\documentclass[11pt,a4paper]{article}
\pdfoutput=1
\usepackage[pdftex]{graphics}
\usepackage{jheppub}
\usepackage{amsmath,amssymb,amsfonts}
\usepackage{enumitem}
\usepackage{multirow}
\usepackage{array,booktabs}
\usepackage{slashed}

\usepackage{accents}
\usepackage{mathrsfs}
\usepackage{mathtools}

\usepackage[export]{adjustbox}
\setlist[itemize]{
    label=\adjustbox{scale=0.7}{$\bullet$}, itemsep=-3pt,topsep=0px
}

\usepackage[usenames,dvipsnames]{xcolor}
\definecolor{labelcolor}{RGB}{194, 175, 116}
\definecolor{rmkcolor}{RGB}{15,120,255}

\usepackage[final]{showlabels} 

\newif\ifToggleMacros
\ToggleMacrostrue  
\ifToggleMacros
\else
\fi

\let\oldc\c
\let\oldi\i

\newcommand{\Fig}[1]{Fig.\,\ref{#1}}

\newcommand{\Eq}[1]{Eq.\,(\ref{#1})}
\newcommand{\Eqs}[2]{Eqs.\,(\ref{#1}) and (\ref{#2})}
\newcommand{\Sec}[1]{Sec.\,\ref{#1}}
\newcommand{\Secs}[2]{Secs.\,\ref{#1} and \ref{#2}}
\newcommand{\App}[1]{App.\,\ref{#1}}
\newcommand{\transition}[1]{\qquad\adjustbox{scale=0.95}{\text{#1}}\qquad}

\DeclareMathOperator{\diag}{diag}
\DeclareMathOperator{\tr}{tr}

\def\mem{\hspace{0.1em}}
\def\hem{\hspace{0.05em}}
\def\nem{\hspace{-0.1em}}
\def\hnem{\hspace{-0.05em}}
\def\hhem{\hspace{0.025em}}
\def\hhnem{\hspace{-0.025em}}
\def\hhhem{\hspace{0.0125em}}

\def\qiq{{\quad\implies\quad}}
\def\iq{{{\implies}\quad}}

\def\minie{{\textstyle\frac{1}{2}}}

\def\a{\alpha}
\def\b{\beta}
\def\c{\gamma}
\def\d{\delta}
\def\e{\epsilon}
\def\ve{\varepsilon}
\def\m{\mu}
\def\n{\nu}
\def\r{\rho}
\def\s{\sigma}
\def\k{\kappa}
\def\l{\lambda}

\def\bpsi{{\smash{\bar{\psi}}\kern0.02em\vphantom{\psi}}}

\def\te{{\tilde{\epsilon}}}
\def\ts{{\tilde{\sigma}}}
\def\ttilde{\tilde{t}}

\def\swedge{{\mem{\wedge}\,}}
\let\oldcap\cap
\renewcommand{\cap}{{\,\oldcap\,}}

\def\da{{\dot{\a}}}
\def\db{{\dot{\b}}}
\def\dc{{\dot{\c}}}
\def\dd{{\dot{\d}}}
\newcommand{\wrap}[1]{{\smash{#1}\vphantom{\b}}}
\def\rambda{\tilde{\lambda}}

\def\id{{\rlap{1} \hskip 1.6pt \adjustbox{scale=1.1}{1}}}

\newcommand{\D}[1]{\mathcal{D}\hnem{#1}\,}

\newcommand{\expval}[1]{
    \big\langle\hem{
        #1
    }\hem\big\rangle
}

\let\oldexp\exp
\renewcommand{\exp}{\oldexp\nem}
\def\Pexp{\mathrm{P}\kern-0.1em\exp}
\newcommand{\wloop}[2]{
    \tr_{#1} \Pexp
    \bigg(\hem{
        \oint_{#2} A
    }\bigg)
}
\def\Pexp{\mathrm{P}\kern-0.1em\exp}
\newcommand{\hloop}[2]{
    \tr_{#1} \Pexp
    \bigg(\hem{
        \oint_{#2} \omega
    }\bigg)
}

\def\M{{\mathcal{M}}}
\def\P{{\mathcal{P}}}
\def\C{{\mathcal{C}}}
\def\S{{\mathcal{S}}}
\def\V{{\mathcal{V}}}
\def\N{{\mathcal{N}}}
\def\Dee{{\mathcal{N}}}
\def\Sig{{\mathcal{V}}}
\def\MX{{\mathcal{X}}}
\def\CX{{\mathcal{K}}}
\def\NX{{\mathcal{Y}}}
\def\HSig{{\mathcal{N}}}
\def\Sh{{\mathcal{C}}}

\DeclareMathOperator{\Link}{link}
\DeclareMathOperator{\Intersect}{int}
\def\LLink{\Link_*}

\def\cen{{\alpha}}
\def\rep{{\hnem\rho}}
\def\fund{{\text{fund}}}

\def\kap{{\hem\star\hhhem}}
\def\ikap{{\hem\star\hhhem}}

\newcommand{\Ad}[2]{{#1}^{-1}\hhnem {#2}\mem {#1}}
\newcommand{\coAd}[2]{{#1}^{-1}\hhnem {#2}\mem {#1}}

\def\mpad{\kern0.525em}
\def\hpad{\kern0.8em}
\def\wpad{\kern0.95em}

\def\Uinv{U_\cen\kern-0.3em\mathrlap{\adjustbox{raise=0.11em}{$^{-1}$}}\kern0.95em}
\def\BF{\text{``$BF\hem$''}}
\def\bpsi{\tilde{\psi}}
\def\sp{{\text{sp}}}
\def\asp{{\widetilde{\text{sp}}}}

\let\oldparagraph\paragraph
\newcounter{alphnum}
\renewcommand{\paragraph}[1]{
    \refstepcounter{alphnum}{%
        \oldparagraph{\scshape\alph{alphnum}. #1}\phantom{}\vskip0.35ex\noindent
    }%
}
\counterwithin*{alphnum}{subsubsection}

\title{
    Generalized Symmetry in Dynamical Gravity
}

\author[a]{Clifford Cheung,}
\author[a]{Maria Derda,}
\author[a]{Joon-Hwi Kim,}
\author[a]{\\Vinicius Nevoa,}
\author[b]{Ira Rothstein,}
\author[a]{Nabha Shah}

\affiliation[a]{Walter Burke Institute for Theoretical Physics,\\ California Institute of Technology, Pasadena, CA 91125}
\affiliation[b]{Department of Physics, Carnegie Mellon University, Pittsburgh, PA 15213}

\abstract{
    We explore generalized symmetry in the context of nonlinear dynamical gravity.  Our basic strategy is to transcribe known results from Yang-Mills theory directly to gravity via the tetrad formalism, which recasts general relativity as a gauge theory of the local Lorentz group.  By analogy, we deduce that gravity exhibits a one-form symmetry implemented by an operator $U_\alpha$ labeled by a center element $\alpha$ of the Lorentz group and associated with a certain area measured in Planck units.
    The corresponding charged line operator $W_\rho$ is the holonomy in a spin representation $\rho$, which is the gravitational analog of a Wilson loop.   The topological linking of $U_\alpha$ and $W_\rho$ has an elegant physical interpretation from classical gravitation: the former materializes an exotic chiral cosmic string defect whose quantized conical deficit angle is measured by the latter.  We verify this claim explicitly in an AdS-Schwarzschild black hole background. Notably, our conclusions imply that the standard model exhibits a new symmetry of nature at scales below the lightest neutrino mass.
    More generally, the absence of global symmetries in quantum gravity suggests that the gravitational one-form symmetry is either gauged or explicitly broken.  The latter mandates the existence of fermions.  Finally, we comment on generalizations to magnetic higher-form or higher-group gravitational symmetries.
}

\begin{document}

\begin{flushright}
    \footnotesize
    CALT-TH 2024-009
\end{flushright}
\maketitle

\newpage
\section{Introduction}
\label{sec:Intro}

Symmetry has long been a vital tool for investigating complex physical systems, particularly at strong coupling.
Historically, most efforts in this expansive subject have focused on conventional symmetries, which act on {\it local} operators.  The standard model of physics exhibits numerous exact and approximate symmetries of this type, for example relating to charge in electromagnetism and chiral symmetry in the strong interactions.

In the past decade, however, the fundamental concept of symmetry has broadened considerably \cite{Alford:1991vr,Alford:1990fc,Alford:1992yx,Bucher:1991bc,Pantev:2005rh,Pantev:2005wj,Pantev:2005zs,Hellerman:2006zs,Nussinov:2009zz,Aharony:2013hda}. As described in the seminal work of \cite{Gaiotto:2014kfa}, it is now understood that the traditional formulation of symmetry is actually the tip of a colossal iceberg.  Rather, there exists a rich patchwork of so-called higher-form symmetries whose distinguishing feature is that they act intrinsically on extended objects described by {\it nonlocal} operators supported on lines, surfaces, and membranes.   Since higher-form symmetries act trivially on local operators, their physical implications are sometimes quite subtle to diagnose.  From this point of view, the standard symmetries found in most quantum field theory textbooks are brusquely relegated to the special case of zero-form symmetry.

The growing body of work on generalized symmetries has revealed new perspectives on a broad spectrum of assorted phenomena in quantum field theory, including phase transitions \cite{Iqbal:2021rkn,Wen:2018zux,Levin:2004mi,Levin:2004js,Hastings:2005xm,Shimizu:2017asf}, anomalies \cite{Choi:2022jqy,Cordova:2022ieu,Gaiotto:2017yup,Wan:2018zql,Delacretaz:2019brr,Hsin:2018vcg,Tanizaki:2017mtm,Cordova:2019bsd,Wan:2018djl,Cordova:2019jnf,Cordova:2019uob,Delmastro:2022pfo}, and symmetry breaking \cite{Kovner:1992pu,Hofman:2018lfz,Lake:2018dqm,Sogabe:2019gif,GarciaEtxebarria:2022jky}. Recent work has even explored new opportunities for physics beyond the standard model, for example in the context of flavor physics \cite{Cordova:2022qtz}, neutrinos \cite{Cordova:2022fhg}, and axions \cite{Hidaka:2020iaz,Hidaka:2020izy,Brennan:2020ehu,Choi:2022fgx,Yokokura:2022alv,Brennan:2023kpw,Choi:2023pdp,Cordova:2023her,Reece:2023iqn,Agrawal:2023sbp}. Such efforts are a welcome development, as they attempt to draw an explicit connection between highly formal developments in mathematical physics and high-energy physics of actual experimental relevance.  That said, the constraints imposed by generalized symmetry on particle physics models tend to be explicable via more conventional means.  This is perhaps not so surprising---these models are easily embedded within renormalizable theories in which all is calculable and there are no surprises to be had or which require explanation.

Gravity, on the other hand, is another story.  Far less is understood about its putative ultraviolet completion.   Consequently, the only truly theory-agnostic approach is to retreat to safely low energies, where gravitational dynamics are described universally by an effective field theory of gravitons on a fixed background, augmented by possible higher-derivative corrections.  For example, see \cite{Donoghue:2022eay,Burgess:2003jk} for a review of this perspective.  
The effective field theory of gravity is clearly a natural target for understanding generalized symmetry in a refreshingly different context.  There has, however,  been relatively little effort in this vein.\footnote{The bulk of work that makes reference to both gravity and generalized symmetries has focused on the implications of swampland conjectures. 
In this picture, one posits a quantum field theory that exhibits certain generalized global symmetries.  The conjectured absence of global symmetries in a theory of quantum gravity 
then imposes constraints on the theory in order to explicitly break or gauge these symmetries. Though interesting, this subject is not the topic of the present work.} Some notable exceptions include interesting recent work studying the higher-form symmetries associated with  parity \cite{McNamara:2022lrw} and topology change \cite{McNamara:2019rup}, as well as generalizations of continuous higher-form symmetries to {\it linearized gravity} \cite{Hinterbichler:2022agn,Benedetti:2021lxj,Benedetti:2023ipt}.

 In this paper, we extend the now well-established insights of higher-form symmetry in gauge theory to the effective theory of {\it nonlinear gravity}  in four-dimensional spacetime.  Our key ingredient is the well-known fact that gravity can itself be recast in gauge theoretic language.   As history would have it, this perspective carries dual meanings.
 On the one hand, gravity is a theory of diffeomorphisms, nonlinearly realized by a self-interacting, massless spin two field.   Since diffeomorphisms are a redundancy, they are on occasion referred to as a gauge symmetry, though colloquially and not in the strict technical sense.
 On the other hand, it is well-known that gravity can also be described by a bona fide gauge theory of local Lorentz transformations, which is the so-called Palatini formalism for the tetrad and spin connection.  Formally, these descriptions are equivalent\footnote{At low energies,  general relativity and the tetradic Palatini formalism are equivalent classically and quantum mechanically since they both reproduce a local effective field theory of a massless spin two particle.  As is well-known, the dynamics of such a theory are uniquely fixed, up to unknown Wilson coefficients.} since gauge symmetry is, after all, pure redundancy and no redundancy is more valid than any other.\footnote{Of course, one can always start from the tetradic Palatini formalism and simply {\it integrate out} the spin connection and {\it gauge fix} the tetrad algebraically, thus reverting to the usual metric description of gravity.  Doing so should yield the same physics, since these redundancies are unphysical.  However, for our analysis it will be far more illuminating to keep the tetrad and spin connection since we will be especially interested in the gravitational interactions of fermions and their worldlines, which play an absolutely essential role in our construction.  Said another way, the pure metric formulation is poorly equipped to describe fermions. 
 }  As
we will see, tetradic Palatini gravity is perfectly suited to our purposes because we can work in lockstep analogy with the familiar approach taken in gauge theory. 
 For concreteness, the bulk of our analysis will be in Euclidean signature, though we will toggle to Lorentzian signature on and off when needed. Our conclusions for gravity are as follows.

First and foremost, our central claim is that tetradic Palatini gravity exhibits an electric one-form symmetry described by the center subgroup $Z(G)$ of the Lorentz group $G$.\footnote{In an abuse of notation,
we will hereafter refer to the gauge group $G$ of the tetradic Palatini formalism
as the ``Lorentz group'' even though we will consider both Euclidean and Lorentzian signatures. 
}
This one-form symmetry depends crucially on the signature and global structure of $G$.  For example, in 
Euclidean signature the center is nontrivial when we consider $Z({\rm SO}(4)\hnem)=\mathbb{Z}_2$ or $Z({\rm Spin}(4)\hnem)=\mathbb{Z}_2 {\mem\times\mem} \mathbb{Z}_2$, while in Lorentzian signature the center is nontrivial for $Z({\rm SL}(2,\mathbb{C})\hnem)=\mathbb{Z}_2$.  In all cases, these center subgroups have a zero-form symmetry action as various parities on Lorentz vector and spinor indices.


Second, we show how the one-form symmetry of gravity is implemented by a topological symmetry operator $U_\cen$.  This object is constructed explicitly in terms of the local degrees of freedom
as the exponential of a certain area operator for a closed surface measured in Planck units and labeled by an element of the center $\cen$.  
The symmetry operator $U_\cen$ acts on a line operator $W_\rep$ known as the spin holonomy, which is simply a Wilson loop for the spin connection computed in a spin representation $\rep$ along a chosen contour.   While $U_\cen$ generates a {\it global} one-form symmetry, it can be implemented as a field transformation that is precisely the form of a local Lorentz transformation, but with nontrivial winding that precludes it from being a genuine local Lorentz transformation.  Using this ``twisted local Lorentz transformation'', we show that $W_\rep$ transforms by a center-valued phase that depends only on the topological linking of the surface and curve which define $U_\cen$ and $W_\rep$, respectively.  We prove the Ward identity for $U_\cen$ and $W_\rep$ using both covariant and canonical approaches.  Notably, this proof is valid to all orders in perturbation theory, at least within the context of the effective field theory description of gravity\footnote{As is well-known, quantum corrections are perfectly well-defined even within a low-energy effective field theory, provided one enforces systematic power counting.  In the effective field theory of gravity, most quantum corrections are ultraviolet sensitive and thus absorbed into incalculable counterterms.  However, there also exist calculable long-distance quantum corrections \cite{Donoghue:2022eay}.} where the topology and dimension of spacetime are preserved.


Thirdly, we show that the interplay of $U_\cen$ and $W_\rep$ has a remarkably simple interpretation in terms of {\it classical gravitation}.  The symmetry operator $U_\cen$ creates a defect in spacetime that is 
a chiral version of a cosmic string defect, and serves as a certain gravitational analog of the Dirac string.   
The tension of $U_\cen$ is quantized so as to induce a $\pi$ deficit angle which is directly measured by the spin holonomy $W_\rep$ as the center-valued linking number.   We then compute the linking number by evaluating $W_\rep$ on various spacetimes, including an AdS-Schwarzschild background.  The topological nature of $U_\cen$ implies that its linking with  $W_\rep$ arises purely from contributions at leading order in the so-called self-force expansion, where $U_\cen$ is treated as a nondynamical background.  Furthermore, this implies that higher order self-force corrections are vanishing, so evidently the classical deficit angle is not quantum corrected at any perturbative order.



Last but not least, we discuss the breaking of the gravitational one-form symmetry.
 As expected, explicit breaking requires a local operator in the representation $\rep$ that renders the spin holonomy $W_\rho$ ``endable,'' thus unspooling its  linking with $U_\cen$.  Physically, this corresponds to the screening of the spin holonomy by spinning particles.
Interestingly, the spin holonomy in the vector representation is automatically screened in pure gravity by orbital angular momentum.  This mirrors the phenomenon in gauge theory where adjoint Wilson lines are screened by the gluon field itself.  On the other hand, holonomies in the spinor representation are endable only by local fermionic operators.  If no such operators exist, then the one-form gravitational symmetry is exact.  Remarkably, this implies the emergence of a hidden symmetry of the real world: below the lightest neutrino mass, there is a gravitational one-form symmetry under which spinor holonomies are charged. 
More generally, in a theory in which the gravitational one-form symmetry is not gauged, the conjectured absence of exact global symmetries in quantum gravity directly implies the existence of fermions.



\section{Gauge Theory}
\label{sec:YM}

In this section, we present a self-contained review of one-form symmetries in gauge theory.  Other treatments can be found in the literature \cite{Gomes:2023ahz, Brennan:2023mmt,Schafer-Nameki:2023jdn,McGreevy:2022oyu,Cordova:2022ruw,witten230b,Bhardwaj:2023kri}. We start with a covariant analysis expressed in the language of path integrals, followed by a treatment in terms of canonical quantization.   Because this section is mostly---though not entirely---a recap of known results from gauge theory,  it may be skipped by readers interested only in our new findings, which pertain to gravity.   However, we note that this gauge theory warm up forms a concrete road map for our parallel analysis of gravity later on.

\subsection{Covariant Formalism}
\label{sec:CovYM}

To begin, let us consider Yang-Mills theory 
for a
gauge group $G$
which is 
a connected matrix Lie group.
We take spacetime to be a Riemannian four-manifold $\M$ of Euclidean signature.
Our discussion will apply irrespective of whether or not the background spacetime is curved, provided it is nondynamical.
As is well-known, this theory admits a first-order formulation in terms of a one-form gauge connection and an auxiliary two-form field,
\begin{align}
    A^a = A^a{}_\m\mem dx^\m
    \transition{and}
    B_a = \frac{1}{2}\mem B_{a\m\n}\mem dx^\m \swedge dx^\n
    \,,
\end{align}
valued in the adjoint and the coadjoint of $G$, respectively, so 
$a,b,\ldots \in \{ 1,\ldots, {\rm dim}(G)\}$.
The action is the integral over $\M$ of the Lagrangian four-form,
\begin{align}
    \label{eq:YM.L}
    L
    = \frac{1}{g^2}\mem B_a \swedge F^a
    - \frac{1}{2g^2} B_a \swedge {*}B^a \transition{where}
    F^a
    = dA^a + \frac{1}{2}\mem f^a{}_{bc}\mem A^b \wedge A^c
    \,,
\end{align}
where $g$ is the gauge coupling.
Throughout, color indices are raised and lowered by the Killing form,
while
the Hodge dual $*$ 
with respect to the background metric
acts on spacetime indices. 
Integrating out the auxiliary two-form field enforces
$B_a = *F_a$,\footnote{
    Here we emphasize to the reader that despite appearances $B$ {\it does not} denote the magnetic field.  Rather, it is the two-form of the {\BF} formulation of Yang-Mills theory, so $\int\nem B = \int\nem *F$ and $\int\nem *B = \int\nem F$ denote electric and magnetic fluxes
    when integrated over a spatial surface, respectively.
} 
and plugging this back in to \Eq{eq:YM.L},
we obtain
$(1/2g^2)\mem F_a \swedge {*}F^a$,
which is the textbook Lagrangian for Yang-Mills theory in a fixed background spacetime.

\subsubsection{Line and Symmetry Operators}
\label{sec:CovYM.symmetry}

In Yang-Mills theory, the one-form symmetry group is identified with the center of the gauge group, 
$Z(G)$.
By definition, the one-form symmetry acts on extended objects rather than local operators. The relevant charged object is the one-dimensional line operator,
\begin{align}
    \label{eq:YM.W}
    W_\rep(\C)
    = \wloop{\rep}{\C}
    \,,
\end{align}
which is a path-ordered Wilson loop along a closed contour $\C$.
Here $\rep$ denotes the irreducible representation in which the trace and exponentiation are defined.\footnote{
    Note that there is no factor of $i$ in the exponential map because we are using an anti-Hermitian convention for the Lie algebra generators.
}

\begin{figure}
    \centering
    \includegraphics[scale=1.05]{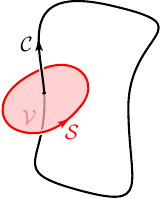}
    \qquad\qquad
    \includegraphics[scale=1.05]{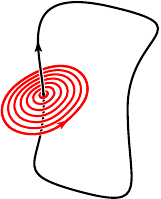}
    \caption{
        {\it Left:}
        The linking number between 
        a one-dimensional contour $\C$ and an exact codimension two surface $\S = \partial\V$ is equal to the number of
        intersections 
        between 
        the coboundary $\V$ and $\C$.
        {\it Right:}
        The coboundary $\V$ 
        defines
        a homotopy for shrinking the surface $\S$ to a point on $\C$.
    }
    \label{fig:CS}
\end{figure}

Meanwhile, the one-form symmetry transformation is implemented by a corresponding symmetry operator, 
\begin{align}
    \label{eq:YM.Uschem}
    U_\cen(\S)
    \,,
\end{align}
which is an instance of the
Gukov-Witten operator \cite{Gukov:2006jk}.
This operator is
supported on a two-dimensional surface $\S$
and
labeled by a center element $\a {\,\in\,} Z(G)$.  We will present a concrete formula for $U_\cen(\S)$ in terms of explicit fields later on.  But for the moment, let us abstractly describe the symmetry operator in terms of the defining property that it generates the following transformation on the line operator,
\begin{align}
    \label{eq:YM.symaction}
    W_\rep(\C)
        \,\,\mapsto\,\,
    \rep(\cen)^{\Link(\C,\S)}\mem
    W_\rep(\C)
    \,,
\end{align}
where
$\rep(\cen)$ is 
the representation of the center element $\cen$
as a complex phase,
and we have defined $\Link(\C,\S)$ to be the linking number between the contour $\C$
and
the surface $\S$.
Crucially,
since the linking number is a topological invariant,
so too is the operator
$U_\cen(\S)$,
in the sense that the surface of
its support $\S$
can be deformed arbitrarily
to yield the same action on the Wilson loop $W_\rep(\C)$
provided it does not degenerate with $\C$.

In this paper, we will always assume that
the two-dimensional support of the symmetry operator
is not only closed but also exact, so the surface $\S = \partial \V$ is the boundary of a three-dimensional volume $\V$.
This is required so that 
the symmetry operator
can be contracted continuously into an infinitesimal two-sphere enclosing the line operator, as depicted in \Fig{fig:CS}.  Intuitively, this deformation corresponds to the physical measurement of the 
electric charge of a body by computing the electric flux flowing through an infinitesimal two-sphere enclosing it.
As a consequence, we see that
\begin{align}
    \label{eq:link=inter}
    \Link(\C,\S)=\Intersect(\C,\V)
    \transition{where}
    \S = \partial\V
    \,.
\end{align}
so the linking number between $\S$ and $\C$
is equal to the intersection number between $\C$ and the coboundary $\V$.

In Maxwell theory, it is well-known that the one-form symmetry is implemented by a shift of the gauge field by a ``flat connection'',
which is {\it closed} wherever it is well-defined and nonsingular, 
but crucially {\it not exact}.
A key fact that we now emphasize is that
this can be realized as a transformation of the fields that takes the form of a gauge transformation for a multivalued---that is, winding---gauge parameter, which is hence is not globally defined.  
For instance, consider the map $A \mapsto A + d\chi$. 
If the zero-form parameter $\chi$ exhibits nontrivial winding,
then $d\chi$ is not, despite its appearance, an exact form. 
For example, we might
choose
$\chi = \ve\mem \phi$,
where $\ve$ is a constant
and
$\phi$ is
the azimuthal angle in cylindrical coordinates.
Crucially,
$d\phi 
= (x\mem dy {\,-\,} y\mem dx)/(x^2{\,+\,}y^2)$
is not exact because its integral around a closed circular loop, $\int_{S^1} d\phi = 2\pi$, is nonzero.

There are two distinct ways to interpret this winding connection.  From the mathematician's perspective, 
$d\phi$ would be described as a closed-but-not-exact one-form defined on the $x$-$y$ plane with the origin excised, which is $\mathbb{R}^2 {\,\setminus\hem} 0$.  However, throughout this paper we will adopt the equivalent physicist's picture, which is to instead 
specify
the exterior derivative of $d\phi$
as a \textit{distributional} two-form in the entire $x$-$y$ plane
\textit{without} excising the origin.  That is, 
by demanding that
Stokes theorem apply, we
deduce
$dd\phi = 2\pi\mem \delta(x)\hem \delta(y)\mem dx\swedge dy$.\footnote{
    For the more mathematically inclined,
    this delta function expression can be thought of as a shorthand for stating a cocycle condition.
    A deformation retract of the triple overlap
    turns into the support of the delta function.
    See also the discussion in \cite{Gukov:2006jk}.
}
In this picture,
under 
the transformation
$A \mapsto A + \ve\mem d\phi$
the field strength 
shifts by
$dd\chi = 2\pi \ve\mem \delta(x)\hem \delta(y)\mem dx\swedge dy$,
which describes a magnetic flux tube, or a Dirac string. 
The total flux $2\pi\ve$ of this Dirac string
can be arbitrary
and is measured by the induced phase on the Wilson loops.   We emphasize that in this more physical picture, the
Maxwell action is always defined over all of space without the excision of any particular support.  Furthermore, 
the shift of the field strength
$F \mapsto F + 2\pi \ve\mem \delta(x)\hem \delta(y)\mem dx\swedge dy$
correctly 
describes the fact
that the one-form symmetry transformation of the gauge field,
$A \mapsto A + \ve\mem d\phi$,
does not leave the Lagrangian invariant on the locus of the Dirac string.\footnote{
    For quantized values of $\varepsilon$, the shift of the gauge field by a flat connection is an invariance of the path integral and thus corresponds to a bona fide gauge transformation.  If $\varepsilon$ is nonquantized, then the shift of the gauge field implements the global one-form symmetry.
}

An exactly analogous construction applies to Yang-Mills theory, which we now describe.  In particular, in this case the one-form symmetry is realized by
\begin{align}
\begin{split}
    \label{eq:YM.fr}
    A^a
        &\,\,\mapsto\,\,
    (\Ad{\Omega}{A})^a
    + (\Omega^{-1} d\Omega)^a
    \transition{and}
    B_a
        \,\,\mapsto\,\,
    (\coAd{\Omega}{B})_a
    \, ,
\end{split}
\end{align}
where $\Omega$ is a zero-form parameter which is valued in the gauge group $G$ and approaches the identity at infinity.\footnote{
    The notations $\Ad{\Omega}{A}$
    and $\coAd{\Omega}{B}$ here
    signify adjoint and coadjoint actions,
    which is validated by the fact that we specialize in
    matrix Lie groups and algebras.
    See \App{app:notations} for a comment.
}
Here we also stipulate the crucial additional condition that $\Omega$ is multivalued and exhibits nontrivial winding.  
In the presence of winding, a global definition of  $\Omega$  requires a collection of multiple charts which define it on {\it subregions} of spacetime, but together yield an atlas for all points.  For any particular subregion chart, the corresponding function will necessarily have a branch cut residing on some volume, which we define to be $\V$.  
The boundary of $\V$ then coincides with $\S$
in this subregion, which is to say $\S {\,=\,} \partial\V$.  So practically, when we define an explicit function for  $\Omega$  in a given subregion we can deduce $\S$ directly from the branch cut hypersurface $\V$.  For subregions outside of this particular chart for  $\Omega$---which in many cases includes asymptotic infinity---we can say nothing until we define another chart for  $\Omega$  in that other patch.

Concretely, we will consider $\Omega$ which exhibits
a discontinuity across $\V$
such that
\begin{align}
    \label{eq:lambda-disc1}
    \lim_{
        \P_{\pm}\to\, \P
    }\hem
        \Omega(\P_{\hnem+})
        \mem 
        \Omega^{-1}\hnem(\P_{\hnem-})
    \,=\,
    \cen
    \transition{where}
    \cen \in Z(G)     
    \,,
\end{align}
where $\P_{\hnem+}$ and $\P_{\hnem-}$ are points
infinitesimally displaced away from the same point on
$\V$,
but in opposite directions.  Since $\Omega$ is multivalued, $d\Omega$ is not exact.  
The seemingly innocuous caveat implies that \Eq{eq:YM.fr} is {\it not} a gauge transformation in the traditional sense,
despite its appearance.
In particular, it does not leave Wilson loops invariant, which is why it corresponds to a global one-form symmetry. 
In some of the literature, the transformation defined in \Eq{eq:YM.fr} is sometimes referred to as a ``large gauge transformation'' \cite{tong2018gauge} in analogy with instanton configurations which support topological winding in a similar fashion.   For the present work we refer to \Eq{eq:YM.fr} as a ``twisted gauge transformation,'' on account of the structural form of \Eqs{eq:YM.fr}{eq:lambda-disc1}.

\begin{figure}[t]
    \centering
    \includegraphics[scale=1.05]{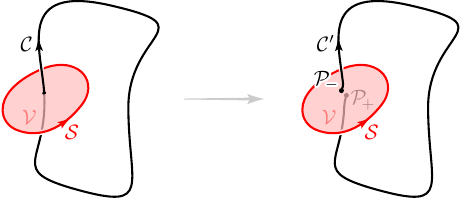}
    \caption{
        Infinitesimal opening of the closed contour $\C$
        at the intersection point $\C{\hem\cap\hem}\V$.
    }
    \label{fig:W'}
\end{figure}

To understand how \Eq{eq:YM.fr}
is equivalent to \Eq{eq:YM.symaction}, 
consider
a Wilson loop for a contour $\C$ that
intersects
with the coboundary $\V$ exactly once, so $\Intersect(\C,\V) {\,=\,} 1$. 
From \Eq{eq:link=inter},
we see that
this also means
$\C$  {\it links} exactly once with $\S$, so
 $\Link(\C,\S) {\,=\,} 1$.
We then find that the Wilson loop transforms as
\begin{align}
\begin{split}
    \label{eq:W'}
    W_\rep(\C)
       \,\,\mapsto\,\,&
        \lim_{\C'\to\mem\C}
        \tr_\rep\nem\bigg[\mem{
            \Pexp
            \bigg(\hem{
                \int_{\C'} 
                    \Ad{\Omega}{A}
                    +
                    \Omega^{-1} d\Omega
            }\bigg)
        }\bigg]
    \,,\\
    =\,\,& 
        \lim_{\C'\to\mem\C}
        \tr_\rep\nem\bigg[\mem{
            \Omega^{-1}\hnem(\P_{\hnem-})
            \,
            \Pexp
            \bigg(\hem{
                \int_{\C'} 
                    A
            }\bigg)
            \,\Omega(\P_{\hnem+})
        }\hem\bigg]
    \,,
\end{split}
\end{align}
where $\C'$ is nearly identical to $\C$ except that it has been infinitesimally ``cut open'' in the vicinity of $\V$
such that $\partial\C' = \P_{\hnem+} - \P_{\hnem-}$,
as depicted in \Fig{fig:W'}.\footnote{
    \App{app:linking}
    contains a detailed accounting  of 
    the various signs and orientations associated with the curves and surfaces shown
    here.
}
Cyclically permuting the terms inside the trace, we find that the Wilson loop transforms as
\begin{align}
\begin{split}  
    \label{eq:W'1}
    W_\rep(\C)
    \,\,\mapsto\,\,
        \lim_{\C'\to\mem\C}
        \tr_\rep\nem\bigg[\mem{
            \cen\,
            \Pexp
            \bigg(\hem{
                \int_{\C'} A
            }\bigg)
        }\bigg]
    \mem=\,
        \rep(\cen)\mem
        W_\rep(\C)
    \,,
\end{split}
\end{align}
where $\rho(\alpha)$ enters with a single power because $\Intersect(\C,\V) {\,=\,} 1$. 
As a result,
we find that
\Eq{eq:W'1}
is precisely the desired transformation of the Wilson loop
for $\Link(\C,\S) {\,=\,} \Intersect(\C,\V)$ $ {\,=\,} 1$.
When generalized to arbitrary linking number, the above 
calculation establishes
the one-form symmetry transformation law for Wilson loops defined in
\Eq{eq:YM.symaction}.

The astute reader will notice that it was essential that the mismatch in the twisted gauge transformation is valued in a center element $\cen {\,\in\,} Z(G)$,
to make the symmetry operator topological.
Otherwise, 
the branch cut $\V$
cannot be arbitrarily chosen, since $W_\rep(\C)$ will transform differently under $\Omega$ depending on 
precisely 
where $\C$ has been cut open to yield $\C'$. In other words, if $\alpha$ were an arbitrary group element, its placement in the Wilson loop would matter and thus the corresponding transformation would not be topological.

\begin{figure}[t]
    \centering
    \includegraphics[width=0.9\linewidth]{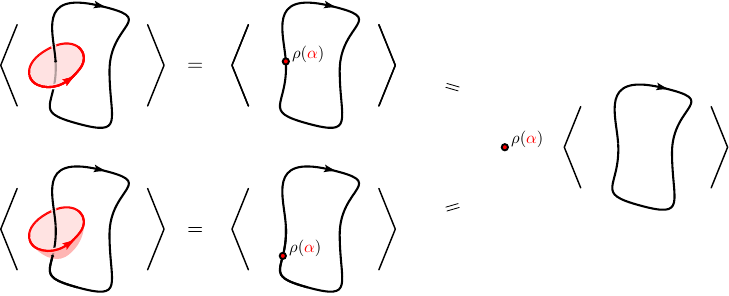}
    \caption{ 
        The one-form symmetry operator essentially inserts $\rep(\cen)$ into the trace of the Wilson loop, where 
        $\cen {\,\in\,} Z(G)$
        is a center element.
        The location of this insertion can be arbitrary
        since $\cen$ commutes with all elements of 
        $G$.
    }
    \label{fig:cuts}
\end{figure}

Before moving on, let us comment on a likely point of confusion.  We have implemented the one-form symmetry transformation using a twisted gauge transformation $\Omega$ that is multivalued with a discontinuity center-valued in $\cen$.  However, we saw earlier that the symmetry operator $U_\cen(\S)$ that implements this transformation should be labeled solely by the center element $\cen$ rather than a whole zero-form parameter $\Omega$.  Why does the twisted gauge transformation depend on $\Omega$ rather than just its twist $\cen$? The resolution to this puzzle is that the naive $\Omega$ dependence in the twisted gauge transformation is spurious.    Since  $U_\cen(\S)$ is a topological surface operator, it only links with one-dimensional objects.  The only such gauge invariant objects are Wilson loops, and we have already shown that the action of the twisted gauge transformation on Wilson loops {\it only} depends on the center element $\cen$, and not the details of $\Omega$.  Hence, different choices of $\Omega$ which have the same twist valued in $\cen$ are physically indistinguishable.  In other words,  symmetry operator is gauge invariant despite the appearance of a reference structure $\Omega$.

\subsubsection{Ward Identity}
\label{sec:CovYM.ward}

Next,
let us now derive the Ward identity which encodes the interplay between the symmetry operator $U_\cen(\S)$ and the line operator $W_\rho(\C)$. Our goal is to prove that
\begin{align}
    \label{eq:Ward}
    \expval{
        U_\cen(\S)\mem W_\rep(\C)
    }
    = \rep(\cen)^{\Link(\C,\S)}
    \expval{
        W_\rep(\C)
    }
    \,,
\end{align}
where the brackets denote the path integral over all fields, so for example
\begin{align}
    \label{eq:YM.piLHS}
    \expval{
        U_\cen(\S) \mem W_\rep(\C) 
    }=
    \int \D{A}\D{B}\,
            e^{-S}\,
            U_\cen(\S) \,
            W_\rep(\C)
    \,.
\end{align}
Obviously, \Eq{eq:Ward} is simply the  transformation law for Wilson loops
in \Eq{eq:YM.symaction}, expressed in the language of
covariant path integrals.

Earlier, we asserted 
that the one-form symmetry transformation in
\Eq{eq:YM.symaction}
is equivalent to the twisted gauge transformation defined in \Eq{eq:YM.fr}.   The latter is implemented by the symmetry operator, which can be written in the explicit form,
\begin{align}
    \label{eq:YM.U}
    U_\cen(\S)
    = \exp\bigg(\mem{
        \frac{2\pi}{g^2}
        \int_{\S} \lambda^a\hem B_a
    }\bigg)
    \transition{where}
    e^{2\pi \lambda} = \cen \in Z(G)
    \,,
\end{align}
where $\lambda^a$ is an adjoint-valued zero-form.  Our claim is that the Ward identity in \Eq{eq:Ward} follows mechanically from the definition of the line and symmetry operators in \Eqs{eq:YM.W}{eq:YM.U}, and furthermore we can see directly how the one-form symmetry transformation arises as a twisted gauge transformation.  
As before, the physical interpretation of \Eq{eq:YM.U} is that it inserts 
a Dirac string or vortex \cite{Reinhardt:2002mb,Engelhardt:1999xw,tHooft:1977nqb} into spacetime.  Note that \Eq{eq:YM.U} is a generalization of the center symmetry operator described in \cite{Gomes:2023ahz}, which utilized temporal winding, and is also a special case of the family of symmetry operators constructed in \cite{Cordova:2022rer}.

The definition of  $U_\cen(\S)$ in \Eq{eq:YM.U} may appear strange since the right-hand side is not manifestly a function of just the center element $\cen$.  Rather, it depends on a color reference $\l$ which has been chosen to exponentiate to $\cen$.  Even worse, $\l$ seems to specify an arbitrary vector in color space that naively violates gauge invariance.  However, exactly like we saw for the twisted gauge transformation, the dependence on $\l$ is actually spurious.  The properties of $U_\cen(\S)$ are dictated entirely by its action on Wilson loops, which we will see depends only on $\cen$. Hence, any distinct choices of $\l$ with the same twist $\cen$ are physically equivalent.\footnote{
    This is similar to what occurs in the case of instantons in gauge theory.  These pure gauge configurations necessarily specify some explicit path in color space, and hence appear naively color breaking.  However, only the topological winding number is the invariant label on these configurations. 
} We will see this borne out explicitly in the subsequent calculation.

\begin{figure}[t]
    \centering
    \includegraphics[scale=0.8]{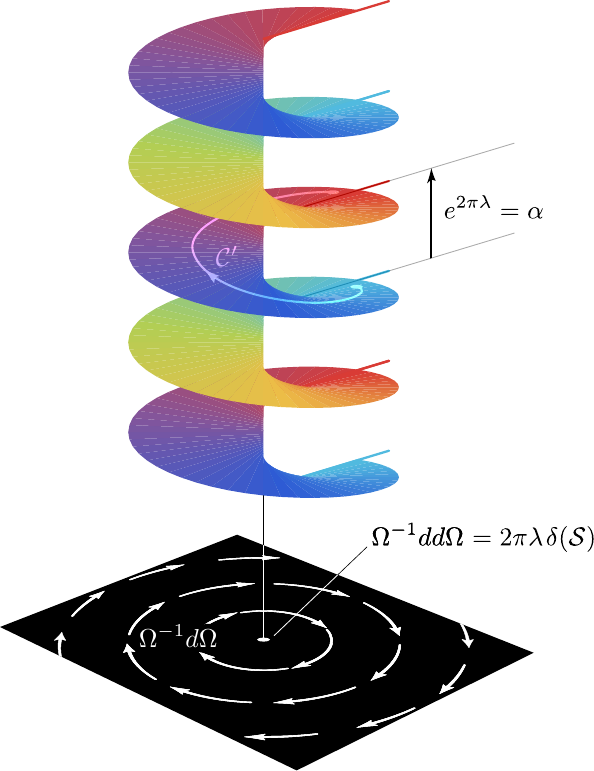}
    \caption{
        The 
        twist
        of the multivalued gauge transformation
        can be characterized by
        the ``curl'' $\Omega^{-1}dd\Omega$ of the flat connection $\Omega^{-1}d\Omega$,
        which localizes along $\S$.
    }
    \label{fig:swirls}
\end{figure}

The proof of the Ward identity is as follows.
To begin, we apply the twisted gauge transformation defined in
\Eq{eq:YM.fr}, under which the field strength becomes
\begin{align}
    \label{eq:YM.Ftransf}
    F^a
        &\,\,\mapsto\,\,
    (\Ad{\Omega}{F})^a
    + (\Omega^{-1} dd\Omega)^a
    \,.
\end{align}
As noted earlier, $dd\Omega=0$ everywhere that $d\Omega$ is well-defined.  However, there are regions of spacetime where it is ill-defined.  Again, this is analogous to $d\phi$ in polar coordinates, which is closed, not exact, and ill-defined at the origin. 
Furthermore, it can be illuminating to consider a distributional interpretation of $dd\phi$ which has nontrivial delta function support at precisely at the origin, so the integral of it over a disc yields $\int_{D^2} dd\phi = \int_{S^1} d\phi=2\pi$. 

Similarly, for the case of 
our twisted gauge transformation,
$dd\Omega$ has localized support on the surface $\S$ so in  particular
\begin{align}
    \label{eq:lambda-disc2}
    (\Omega^{-1} dd\Omega)^a
    = 2\pi\lambda^a\mem \delta({\S})
    \transition{for any $\l^a$ such that}
    e^{2\pi \lambda} = \cen \in Z(G)
    \,.
\end{align}
Here
$\delta(\S)$ is the two-form generalization of the Dirac delta distribution
that peaks on $\S$
\cite{witten230b,Gukov:2006jk},
whose defining feature is that
\begin{align}
    \label{eq:delta2}
    \int_\M \varphi\nem \wedge \delta(\S)
    =
    \int_{\S} \varphi
    \, ,
\end{align}
for any two-form $\varphi$ in $\M$.
 
The equivalence between
\Eq{eq:lambda-disc1}
and
\Eq{eq:lambda-disc2}
can be understood intuitively
from the visualization given in \Fig{fig:swirls}.
The multivaluedness condition in \Eq{eq:lambda-disc1}
implies that the ``gradient'' $\Omega^{-1} d\Omega$
swirls about $\S$,
and 
in turn, its ``curl'' $\Omega^{-1} dd\Omega$  
localizes along $\S$
as a delta function,
describing a Dirac string.
A proof of this equivalence
is left to
\App{app:linking},
which is essentially a colored generalization
of our simpler example $dd\phi = 2\pi\mem \delta(x)\hem \delta(y)\mem dx\swedge dy$.
We will also provide explicit examples
of $\Omega$ 
later on.

With this understanding,
we observe that
that \Eq{eq:YM.U} can be written as
\begin{align}
    U_\cen(\S)
    = \exp\bigg(\mem{
        \frac{1}{g^2}
        \int_{\M} B_a
        \wedge
        2\pi\lambda^a
        \delta(\S)
    }\bigg)
    = \exp\bigg(\mem{
        \frac{1}{g^2}
        \int_{\M} B_a
        \wedge
        (\Omega^{-1} dd\Omega)^a
    }\bigg)
    \,,
\end{align}
from which we can now revisit
the left-hand side of the Ward identity
in \Eq{eq:YM.piLHS}.
The factor involving the action and the symmetry operator
combine to give
\begin{align}
    \label{eq:YM.pi1}
    e^{-S}\,
        U_\cen(\S)
    =
    \exp\bigg({
            -\frac{1}{g^2} \int_\M
                B_a \wedge
                (
                    F
                    {\,-\,}
                    \Omega^{-1}dd\Omega
                )^a
            -\frac{1}{2}\mem
              B_a \swedge {*}B^a
        }\mem\bigg)
    \,.
\end{align}
The twisted gauge transformation of the field strength in \Eq{eq:YM.Ftransf}
eats up the $\Omega^{-1}dd\Omega$ term
and {\it absorbs} the symmetry operator into the action, so
\begin{align}
    \label{eq:YM.eU}
    e^{-S}\, U_\cen(\S) 
        \,\,\mapsto\,\,
    e^{-S}
    \,.
\end{align}
Combing \Eq{eq:YM.symaction} and
\Eq{eq:YM.eU}, we see that the left-hand side of the Ward identity in \Eq{eq:YM.piLHS} transforms to
\begin{align}
\begin{split}
    \expval{
        U_\cen(\S)\mem W_\rep(\C)
    }
    \; \mapsto \;
    \int \D{A}\D{B}\,
        e^{-S}
        \mem
            \rep(\cen)^{\Link(\C,\S)}\mem
            W_\rep(\C)
    \,,
\end{split}
\end{align}
which proves the Ward identity in \Eq{eq:Ward}.

In summary,
\Eqs{eq:YM.fr}{eq:lambda-disc2}
specify 
a change of field variables
that \textit{absorbs} the symmetry operator into the action
as \Eq{eq:YM.eU}.
Note that
\Eq{eq:YM.eU} clearly shows
the action is {\it not invariant} under \Eq{eq:YM.fr}, as the Lagrangian four-form in \Eq{eq:YM.L}
changes on the support of the surface $\S$. 
Hence
we learn again explicitly that
\Eq{eq:YM.fr} is not a typical gauge transformation
in Yang-Mills theory,
which would leave 
the Lagrangian four-form at
all points in spacetime invariant.
Of course, if we excise the region $\S$ from spacetime, then the twisted gauge transformation in \Eq{eq:YM.fr} becomes a bona fide gauge transformation in the resulting punctured manifold,
as mentioned earlier in our discussion of Maxwell theory.

A few more remarks are in order. 
Firstly,
it should be clear
from the above logic that
the twisted gauge transformation
in \Eq{eq:YM.fr}
describes
the action of the one-form symmetry operator $U_\cen(\S)$ on {\it arbitrary} operators.
That is,
if the twisted gauge transformation sends an operator ${\cal O}$ to ${\cal O}_\Omega$,
then there is a corresponding generalized Ward identity,
\begin{align}
\begin{split}
    \label{eq:arbWard}
    \expval{
        U_\cen(\S)\mem
        {\cal O}
    }
    &= 
    \expval{
       {\cal O}_\Omega
   }     
    \,,
\end{split}
\end{align}
reiterating the fact that
the one-form symmetry operator
bridges between different
center-twisted topological sectors
of the gauge bundle.
In fact,
the Ward identity of Wilson loops
in \Eq{eq:Ward} can be regarded as a corollary of \Eq{eq:arbWard}.
Also,
although \Eq{eq:arbWard} applies to local operators,
it should be understood that
any 
such point-supported operator is actually invariant under the one-form symmetry because
the twisted gauge transformation can always be locally untwisted by a conventional gauge transformation.  
Geometrically,
this reflects the fact that a point does not link with a codimension two object.

Secondly,
the derivation of the Ward identity turns out to be very simple in the case of Maxwell theory with an exact contour $\C$.
In this case the Wilson loop can be rewritten as a surface integral of the field strength $F$,
as is familiar from the computation of the Aharonov-Bohm phase of a particle induced by a magnetic flux. 
The twisted gauge transformation of the field strength in \Eq{eq:YM.Ftransf} then creates a localized flux tube on the support of $\S$
whose integral
over the surface
yields the desired linking number.
This is
consistent with the above proof
through
a duality in the linking number computation 
described in \App{app:linking}.

Thirdly, let us elaborate on the group multiplication rule for the symmetry operator, $U_{\cen_1}(\S)\mem U_{\cen_2}(\S) = U_{\cen_1 \cen_2}(\S)$, which is required axiomatically.
Since our symmetry operator implements a twisted gauge transformation, the composition of two such transformations automatically yields
a third, so we know a priori that
the group composition law is valid.
However, establishing this more directly in terms of 
the expression
in \Eq{eq:YM.U} is more subtle.
In particular, suppose $U_{\cen_1}(\S)$ and $U_{\cen_2}(\S)$
are realized with the color reference vectors $\l_1$ and $\l_2$.  The product $U_{\cen_1}(\S)\mem U_{\cen_2}(\S)$ naively exponentiates to symmetry operator with the color reference vector 
$\l_1{\,+\,}\l_2$,
which confusingly is not guaranteed to exponentiate to a center element in general.
However, this puzzle is resolved by realizing that
$\l_1{\,+\,}\l_2$ would correspond to a twisted gauge transformation
with an irrational period. 
As described at length in \Sec{sec:CovYM.symmetry} and in \Fig{fig:cuts},
the periods of the twisted gauge transformations must be center-valued
in order for the symmetry operator to be \textit{topological}.  Otherwise line operators will not transform correctly.

In order to properly implement sequential twisted gauge transformations, 
the representative Lie algebra elements for
$U_{\cen_1}(\S)$ and $U_{\cen_2}(\S)$
can be chosen as
\begin{align}
    \cen_1 = e^{2\pi\mem {\Omega_2}^{-1} \lambda_1 \Omega_2}
    \transition{and}
    \cen_2 = e^{2\pi \lambda_2}
    \,.
\end{align}
These color references merge into a new one,
${\Omega_2}^{-1} \lambda_1\hem \Omega_2 + \lambda_2$, which corresponds to the desired composite twisted gauge transformation,
\begin{align}
\begin{split}    
    \Omega = \Omega_1 \Omega_2
    \qiq
    \Omega^{-1} dd\Omega
    \mem&=\mem
    2\pi
    ({\Omega_2}^{-1} \lambda_1\hem \Omega_2 + \lambda_2)^a
    \mem \delta(\S)
    \,.
\end{split}
\end{align}
%
This establishes the group composition law.
In summary,
the topological nature and gauge invariance of the symmetry operator
together implies that the representative Lie algebra elements for the center elements in the group composition equation
should be chosen in a specific form
such that
the composability and closure of center-twisted gauge transformations
are correctly realized.

Last but not least,
we realize that
the above derivation of the Ward identity
implies a remarkably simple and universal recipe for deducing the symmetry operator directly from the action itself.
In particular, starting from any {\BF}-type Lagrangian
of the form
$B_a \swedge F^a + f(B)$, 
we can define the symmetry operator as the object which is generated by a twisted gauge transformation.
While the one-form symmetry operator is exactly eliminated by a twisted gauge transformation of the action, 
the $f(B)$ term will
only serve as a spectator.
This observation indeed
is the key insight that
will allow us to 
identify
an explicit
a one-form symmetry operator for gravity
in \Sec{sec:Grav}.

\subsubsection{Explicit Examples}
\label{sec:YM.examples}

As summarized in \Eq{eq:arbWard}, the one-form symmetry of Yang-Mills theory is implemented by a twisted gauge transformation $\Omega$ that winds nontrivially with a mismatch valued in the center element $\cen {\,\in\,} Z(G)$.  
In this section,
we will explicitly construct some examples of $\Omega$ and apply them to various classical backgrounds.  The resulting twisted backgrounds will reveal some illuminating physical interpretations for the one-form symmetry operator itself.  
For concreteness, we specialize in gauge group $G=\mathrm{SU}(N)$, whose center is $Z(G)=\mathbb{Z}_N$.

\paragraph{Symmetry Operator as Thin Solenoid}%
Suppose the spacetime is flat Euclidean space $\M = \mathbb{R}^4$,
equipped with Cartesian coordinates $x^\mu = (x,y,z,t)$. 
Consider a twisted gauge transformation,
\begin{align}
    \label{eq:YM.exA}
    \Omega
    = 
    \exp\bigg(\hem{
        \frac{k}{N}\mem
        c\mem \phi
    }\hem\bigg)
    \transition{where}
    e^{2\pi c} = \id
    \,,
\end{align}
where 
$\phi$
is the azimuthal angle
such that $\tan\phi = y/x$,
and $k$ is an integer.
Here 
we have defined
$c$ to be an element of 
the Lie algebra of ${\mathrm{SU}}(N)$ 
that exponentiates to the identity via $e^{2\pi c} = \id$, 
so in the fundamental representation
we might have \cite{Gomes:2023ahz}
\begin{align}
    c^i{}_j
    \mem=\mem i\mem \diag(1,1,\cdots,-N{\mem+\mem}1)
    \,,
\end{align}
where $i,j,\ldots$ denote fundamental indices.

To demonstrate the transformation,
let us consider a {\it trivial background}
corresponding to $A^a {\,=\,} 0$, $B_a {\,=\,} 0$,
where all the Wilson loops are trivial as
$W_\rep(\C) {\,=\,} \dim\rep$,
in particular
$W_\fund(\C) {\,=\,} N$
for the fundamental representation.
However,
applying 
\Eqs{eq:YM.fr}{eq:YM.Ftransf}
on this trivial background
with $\Omega$ in \Eq{eq:YM.exA},
we obtain a {\it nontrivial background},
\begin{align}
\begin{split}    
    \label{eq:YM.exA.cl}
    A^a &= (\Omega^{-1}d\Omega)^a 
    =  \frac{k}{N} 
        \mem c^a\mem
        d\phi
    = \frac{k}{N} 
        \mem c^a\mem
        \frac{x\mem dy {\,-\,} y\mem dx}{x^2+y^2}
 \\
    F^a &= (\Omega^{-1}dd\Omega)^a
    = \frac{k}{N}
        \mem c^a\mem
        dd\phi
    = \frac{k}{N}\mem 
        2\pi c^a\mem
        \delta(x)\hem \delta(y)\mem dx \swedge dy
    \,,
\end{split}
\end{align}
with $B_a=0$ still vanishing.
For instance,
consider a contour $\C$ that loops the $z$-axis once.
Then
the Wilson loop for $\C$,
in the fundamental representation,
is given by
\begin{align}
    \tr_\fund \Pexp
    \bigg({
        \frac{k}{N}\mem c
        \int_0^{2\pi}\kern-0.2em d\phi
    }\hem\bigg)
    &= 
       e^{2\pi i k/N} 
       N
    \,,
\end{align}
in the twisted background described in \Eq{eq:YM.exA.cl}.
This demonstrates how the one-form symmetry transformation on Wilson loops
arise from a twisted gauge transformation.

Interestingly,
from the field strength in \Eq{eq:YM.exA.cl}
we see that
the resulting twisted field configuration 
describes a line of color flux flowing through the $z$-axis for all times $t$. 
Hence, we conclude that the physical interpretation of $\Omega$ is that it spontaneously excites an infinitely thin, straight and static solenoid from the vacuum.  In turn, this implies that the one-form symmetry operator inserts a colored 
Dirac string into spacetime.
With this interpretation,
the Ward identity in \Eq{eq:Ward} can be 
understood as the measurement of the Aharonov-Bohm phase
by the Wilson loop in the background of the Dirac string.
Each time the Wilson loop winds about this thin solenoid, we accrue an additional phase factor of $e^{2\pi i k/N}$,
describing the center twist
represented as a complex phase
in the fundamental representation:
\begin{align}
    \label{eq:YM.exA.mismatch}
    (\mem{
        \Omega(\phi{\,=\,}2\pi)
        \mem 
        \Omega^{-1}\hnem(\phi{\,=\,}0)
    })^i{}_j
    \mem=\mem 
    e^{2\pi i k/N} 
        \mem \delta^i{}_j
    \,.
\end{align}

A few remarks are in order.
Firstly,
it is worth noting that in \Eq{eq:YM.exA} we can shift $k$ by $pN$ for 
$p {\,\in\,} \mathbb{Z}$,
and this realizes various solenoids of different ``strengths'' but all yielding the same monodromy.
This is an instantiation of a comment made earlier, which is that 
different $\Omega$ can realize the same $\cen$.
Note also that a shift by $pN$
can be implemented by a gauge transformation which is not continuously connected to the identity.

Secondly,
since the linking between line and symmetry operators is topological, all of our results must be insensitive to homeomorphic deformations of their corresponding integration surfaces.   For this reason it is an amusing check to consider the twisted gauge transformation for a static but ``wiggly'' Dirac string,
\begin{align}
    \Omega
    = 
    \exp\bigg(\mem{
        \frac{k}{N}\mem
        c\mem
        \phi
    }\mem\bigg)
    \transition{where}
    \tan\phi = \frac{y{\,-\,}Y(z)}{x{\mem-\hem}X(z)}
    \,,
\end{align}
where $X(z)$ and $Y(z)$ describe a static line in space that is not necessarily straight.  Here a simple calculation shows that the discontinuity in the twisted gauge transformation $ \Omega^{-1}dd\Omega$ is proportional to
\begin{align}
        \delta(x{\mem-}X(z))\mem \delta(y{\mem-}Y(z))
        \,
        d(x{\mem-}X(z)) \swedge d(y{\mem-}Y(z))
    \,,
\end{align}
which is a Dirac string that is not straight.  It is obvious that the Aharonov-Bohm phase computed by the Ward identity is not modified by these wiggles.
Going a step further, one can also 
promote the parameterization of the discontinuity to $X(z,t)$ and $Y(z,t)$
corresponding to time-dependent wiggles of the Dirac string.  
This case also accords with the general formula in \Eq{eq:delta-param}.

In principle, the most general possible twisted gauge transformation can
have 
a generator $c$
that varies across spacetime.
This variation in color space 
is perfectly possible
and
should also not alter the monodromies
as long as it is properly derived form a multivalued transformation in accordance with \Eq{eq:lambda-disc2}.

Finally,
let us take stock of the physical interpretation of the above calculation.  The linking number between $U_\cen(\S)$ and $W_\rep(\C)$ is typically interpreted as the center electric flux $U_\cen(\S)$ measured in the presence of the worldline of a colored particle given by $W_\rep(\C)$. 
Interestingly,
here we arrive at a dual, but completely equivalent picture: instead, $W_\rep(\C)$ is the Aharonov-Bohm phase computed for a color Dirac string created by $U_\cen(\S)$.  

Yet,
clearly
we have been cavalier about the global topology of $\S$
while demonstrating this example.
In particular, because the thin solenoid extends off to infinity, we have not actually stipulated whether or how $\S$ ``winds back'' to form a closed  surface. 
However, importantly, $\S$ must be closed in order for $U_\cen(\S)$ to be a topological operator.  So why did the example of the solenoid yield the correct picture, despite the fact that the global structure of $\S$ was not specified?

Physically speaking, this setup yielded a sensible result because 
we effectively zoomed into a local region of $\S$ which links with $\C$ and measured the associated Aharanov-Bohm phase.
That is, in the neighborhood of any point on $\S$, the surface appears as an infinite plane,
and $\Omega$ is simply described by \Eq{eq:YM.exA}.  
As long as the Wilson loop 
does not deviate substantially from this region,
the Aharanov-Bohm phase will be
completely ignorant of 
how the ends of the flux tube
reattach---or possibly even terminate---in some distant region.  This is why we could obtain the correct transformation of the Wilson loop
despite ignoring the global topology of $\S$.


Mathematically speaking,
\Eq{eq:YM.exA}
should be understood as an expression for $\Omega$
in a certain patch
on spacetime, which notably does not include the point at infinity.
The details of ``winding back''
for the closure of $\S$
are contained in the charts for $\Omega$ which cover those other patches,
which we have not defined explicitly.
As a result, 
with the knowledge of $\Omega$ in a single patch,
 Wilson loops are explicitly computable only when restricted to regions in this patch.

\paragraph{Symmetry Operator as Time Monodromy}%
Another interesting example is Yang-Mills theory at finite temperature, described by a compact product manifold with compactified Euclidean time,
\begin{align}
    \M = \M_3 \times \mathrm{S}^1
    \,,\quad
    t
        \,\,\sim\,\,
    t + \beta
    \,.
\end{align}
In addition to the trivial vacuum, there is an infinite set of gauge equivalence classes for the background gauge field.
For example, consider
\begin{align}
    \label{eq:YM.exB.gf}
        A^a &= n\mem \frac{2\pi c^a}{\beta}\mem dt
    \transition{where}
        n \in \mathbb{Z}
    \,.
\end{align}
In this background,
Wilson loops winding about the thermal circle
are trivial,
as
$e^{2\pi n c} = \id$.
Meanwhile, consider the following twisted gauge transformation:
\begin{align}
    \label{eq:YM.exB}
    \Omega
    = 
    \exp\bigg(\mem{
        \frac{2\pi k c}{N}
        \frac{t}{\beta}
    }\mem\bigg)
    \,.
\end{align}
This maps the background considered in \Eq{eq:YM.exB.gf} to
\begin{align}
   \label{eq:YM.exB.gf'}
        A^a &= \bigg({
            n {\,+\,} \frac{k}{N}
        }\bigg)\mem \frac{2\pi c^a}{\beta}\mem dt
    \,,
\end{align}
so the Wilson loops gain a nontrivial phase factor of
$e^{2\pi ik/N}$
per each thermal circle.
Hence, the symmetry operator has induced a monodromy in the time direction.

Note that \Eqs{eq:YM.exB.gf}{eq:YM.exB.gf'}
can be obtained by
identifying the ends of
a flat gauge field configuration in $\M_3$ times an interval
with a twisted boundary condition.
An implementation of this construction in Lorentzian signature can be found in \cite{Gomes:2023ahz}.

An interesting feature of this example
which is absent from the previous one
is that
the Wilson loops do not generally admit a coboundary.
That is, they can be closed but not exact.
However, 
the construction of the one-form symmetry
in terms of a multivalued gauge transformation still applies.

Again, 
in the more rigorous sense
\Eq{eq:YM.exB}
should be taken as the specification of $\Omega$
in a certain patch,
say a ball in $\M_3$ times $\mathrm{S}^1$.
Then the surface support of the symmetry operator
can reside
at a time slice
along the boundary of a large volume
that goes beyond the ball.

\paragraph{Symmetry Operator as Circular Loop}%
In the examples considered thus far, the symmetry operator exhibited support on a surface $\S$ that has infinitely large extent in some direction,
thus always leaving a worry that an explicit global definition of $\S$ as a closed surface is not given.
For completeness, 
we would like to end with an example that explicitly shows 
how the surface support $\S$ can be finite.

Recall earlier how we constructed a static color Dirac string on the $z$-axis.  The corresponding worldsheet extended infinitely in the $t$-$z$ plane, so $\S$ was infinite.  Here we will temper $\S$ onto compact support in two steps.  First, we will roll up the string in its spatial directions, yielding a closed circular loop of finite radius.  Consequently, $\S$ will be spatially compact.  Second, we will pinch off this loop in time by setting the size of this loop to be vanishing except for a finite duration, so $\S$ will be temporally compact as well.

In the first step of this construction, we consider a completely static system in {\it toroidal coordinates}, which foliate three-dimensional space according to a circular ``reference ring'' of radius $a$ in the $x$-$y$ plane. 
In particular, the coordinates $(\tau,\sigma,\phi)$ are defined by
\begin{align}
\begin{split}
    \label{eq:Omega-amp}
    x = \frac{
        a \sinh\tau
    }{\cosh\tau - \cos\sigma}
        \cos\phi
    \,,\quad
    y = \frac{
        a \sinh\tau
    }{\cosh\tau - \cos\sigma}
        \sin\phi
    \,,\quad
    z = \frac{
        a \sin\sigma
    }{\cosh\tau - \cos\sigma}
    \,,
\end{split}
\end{align}
where 
$\phi$
is the azimuthal angle in the $x$-$y$ plane.  
Surfaces of fixed $\tau {\,\geq\,} 0$ label concentric two-tori which enclose the reference ring, while surfaces of fixed $-\pi {\,<\,} \sigma {\,\leq\,} \pi$ label two-spheres which intersect the reference ring.
Also, here we choose the branch cut for $\sigma$
such that the discontinuity at $\sigma=\pm \pi$
develops on the disc enclosed by the reference ring.

Crucially, we can think of $\sigma$ as an angular coordinate that winds like a solenoid about the reference ring.  
Thus
we can let the twisted gauge transformation parameter be
\begin{align}
    \label{eq:Omega-amp}
      \Omega
    = 
    \exp\bigg(\mem{
        \frac{k}{N}\mem
        c\mem
        \sigma
    }\mem\bigg)
    \,,
\end{align}
which clearly induces a rephasing for any holonomy that wraps the reference ring.  
The branch cut resides on a volume $\V$ corresponding to the static disc enclosed by the reference ring.  Its boundary then defines the surface $\S=\partial \V$, which is the reference ring itself, namely a static loop of radius $a$ in the $x$-$y$ plane.
Furthermore, we see that $\Omega$ correctly approaches the identity at spatial infinity, simply because spatial infinity corresponds to $\sigma=0$ in toroidal coordinates.

In the second step, we allow for the radius of the reference ring to change with time.  To allow for this, we define toroidal coordinates for each time slice in which the reference ring has a time-varying radius $a(t)$.  For example, let us define $a(t)$ to smoothly
increase from and decrease to zero
within a time interval $t\in[t_1,t_2]$. 
With this temporal modification, the surface $\S$ has finite support in both time and space.

Let us end with a final remark.
In general, one typically wants to construct
a twisted gauge transformation parameter $\Omega$
for an arbitrarily shaped surface $\S$
in an arbitrary manifold $\M$ with or without boundaries.
How do we know that such an $\Omega$ always exists, given some choice of $\S$ and $\M$?
In all of the examples above
we started with $\Omega$ as an input
and rather determined the surface $\S$ as an output.

Interestingly,
we find that
it is always possible
to find an $\Omega$
with a given $\S$ and $\M$,
on account of a closely analogous question in {\it classical magnetostatics}.
That is, deducing $\Omega$ from $\S$ and $\M$ is mathematically identical to deducing the static magnetic field and potential
of an electric current loop.
To see why, imagine we are experimentalists who construct a loop of electrical line current $\mathbf{J}$, built to specification according to some arbitrary contour.
On account of  Amp\`ere's law, ${\nabla}{\hem\times\mem} \mathbf{H} = \mathbf{J}$, we can then deduce the magnetic field $\mathbf{H}$, or even just measure it. 
In regions away from the current, we can then reconstruct a magnetic scalar potential via $\mathbf{H} {\,=\,} {-\mathbf{\nabla}\Psi}$.
If the manifold $\M$ is not closed, then we can make its boundary $\partial\M$ superconducting to enforce the boundary condition $\mathbf{H}_\perp {\hem=\,} 0$, in which case $\Psi$ can be set to a constant over $\partial\M$ that we fix to zero.
In this analogy, the electric current $\mathbf{J}$, the magnetic field $\mathbf{H}$, and the magnetic potential $\Psi$ each correspond to the Dirac string defined by $\Omega^{-1} d d\Omega$, the twisted gauge connection $\Omega^{-1} d\Omega$, and the ``log'' of the twisted gauge parameter $\Omega$, respectively .

\subsection{Canonical Formalism}
\label{sec:CanYM}

The one-form symmetry of gauge theory can also be understood from the complementary point of view of the Hamiltonian formalism.
To this end, we will study Yang-Mills theory on a spacetime described by a product manifold
$\M = \M_3 {\,\times\mem} \mathbb{R}$
equipped with coordinates $x^\m = (x^i,x^4)$.  In particular, we perform a $3{\,+\,}1$ decomposition in which 
 $x^i$ denotes coordinates of the spatial three-manifold $\M_3$ and $x^4=t$ defines equal-time slices.

\subsubsection{Phase Space}
\label{sec:CanYM.ps}

The first-order formulation of Yang-Mills theory is defined 
in \Eq{eq:YM.L}.  
Writing out all indices explicitly, we obtain
the Lagrangian,
\begin{align}
\begin{split}
    \label{eq:YM.compL}
    \frac{1}{g^2}\mem
    \bigg[\mem{
         \frac{1}{2}\mem
            B_{a\m\n}
            \Big(\hem{
                \partial_\r A^a{}_\s
                +
                \minie\mem
                f^a{}_{bc}\mem A^b{}_\r\mem A^c{}_\s
            }\Big)\mem
            \e^{\m\n\r\s}
        -
        \frac{1}{4}\mem
            B_{a\m\n}  B^{a\m\n}
    }\mem\bigg]
    \,,
\end{split}
\end{align}
where
$\e^{\m\n\r\s}$ denotes the permutation symbol.
Carrying out the $3{\,+\,}1$ decomposition
and integrating out $B_{ai4}$,
we immediately see that
the dynamical coordinates on the phase space are $A^a{}_i$ together with 
their canonical conjugates,
\begin{align}
    \label{eq:Edef}
    E^i{}_a
    = \frac{1}{2}\mem B_{a jk}\mem \e^{ijk}
    \,.
\end{align}
Their 
canonical commutation relations
given by
\begin{align}
\begin{split}
    \label{eq:YM.ccr}
    [ A^a{}_i(x) , A^b{}_j(x') ] &= 0 
    \,,\\
    [ E^i{}_a(x) , A^b{}_j(x') ] &= g^2\, \delta^i{}_j\,\delta^b{}_a\, \delta^{(3)}\hnem(x{\,-\,}x') 
    \,,\\
    [ E^i{}_a(x) , E^j{}_b(x') ] &= 0
    \,,
\end{split}
\end{align}
where $x,x' \in \M_3$ are points in the spatial manifold.
Note that there is no imaginary unit here since we are in Euclidean signature.
The phase space is also equipped with the Gauss constraint and a Hamiltonian,
the details of which are not important for our 
purposes.

\subsubsection{Ward Identity}
\label{sec:CanYM.ward}

In the language of the path integral, a one-form symmetry transformation is implemented through the insertion of a symmetry operator which wraps the line operator.
In the operator formalism, however,
this corresponds to a conjugation of the latter by the former.
To see how 
this
works
in detail, consider a line operator $W_\rep(\C)$, where $\C$ is restricted to an equal-time slice, say at $t{\,=\,}0$. As before, we take the symmetry operator $U_\cen(\S)$ to be 
supported
on an exact surface $\S$ with an associated coboundary $\V$,
so $\S = \partial\V$.

\begin{figure}[t]
    \centering
    \includegraphics[scale=1.33]{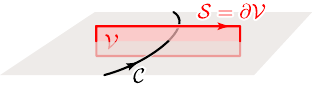}
    \\[1.25\baselineskip]
    \includegraphics[scale=1.33]{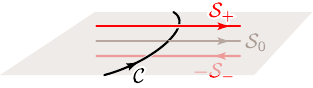}
    \\[1.25\baselineskip]
    \includegraphics[scale=1.33]{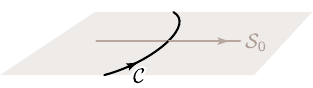}
    \caption{
        Pancaking the symmetry operator
        on a purely spatial line operator $\C$.
        The plane depicts an equal-time slice
        while time flows upwards.
        \textit{Top:}
            The equal-time slice bisects
            the 
            volume $\mathcal{V}$ 
            by a 
            surface $\S_0$.
        \textit{Middle:}
            The 
            surface $\S = \partial\mathcal{V}$
            splits into surfaces $\S_+$ and $-\S_-$
            at the infinitesimal future and past.
        \textit{Bottom:}
            $\S_+$ and $\S_-$ both
            project down to the surface $\S_0$.
    }
    \label{fig:pancake1}
\end{figure}

The geometric set-up is depicted in \Fig{fig:pancake1}.
We assume that
the surface
$\S$ \textit{links} once with 
the purely spatial loop $\C$.  Consequently, 
the coboundary $\V$ is
\textit{intersected} by the loop $\C$ and \textit{bisected} by the spatial slice.
Now imagine continuously squashing or pancaking 
the coboundary
$\V$ along the time direction such that  
its two-dimensional boundary $\S$ infinitesimally hugs the spatial slice.  In this limit,
$\S =\S_+ {\mem\cup\mem} (-\S_-)$ is the union
of two disjoint discs 
$\S_+$ and $-\S_-$
at the infinitesimal future and past
across $t{\,=\,}0$. Once $\V$ has completely collapsed into the spatial slice at $t {\,=\,} 0$, both discs
$\S_\pm$ approach the same surface, which we denote by $\S_0$.
From \Fig{fig:pancake1}, we see that this $\S_0$ 
will be the intersection between $\V$ and $t=0$.
As a result,
the
intersection of $\C$ and $\V$ in the four-manifold $\M$ 
is equivalent to
intersection of
$\C$ and $\S_0$
in the three-manifold $\M_3$ 
as the slice $t{\,=\,}0$.
Therefore,
we have
in general
\begin{align}
    \label{eq:4-to-3}
    \Link(\C,\S)
    \mem= \mem  \Intersect(\C,\V)
    \mem=\mem
    \Intersect_3(\C,\S_0)
    \,,
\end{align}
where $\Intersect_3$
denotes intersection number in $\M_3$.

Now,
we can describe how this
pancaking procedure
boils down the Ward identity
to an equal-time operator equation.
Since $\S =\S_+ {\mem\cup\mem} (-\S_-)$, we see that
the symmetry operator factorizes into
$U_\cen(\S) = 
U_\cen(\S_+)\mem U_\cen(-\S_-)$.\footnote{
    Note that 
    here we have allowed nonclosed surfaces
    for the support of symmetry operators,
    which might be a slight abuse of notation.
}
In turn,
the left-hand side of the Ward identity in \Eq{eq:Ward}
becomes the time-ordered expression
$U_\cen(\S_+)\mem W_\rep(\C)\mem U_\cen(-\S_-)$, which in the 
process of pancaking
limits to the equal-time operator product
$U_\cen(\S_0)\mem W_\rep(\C)\mem \Uinv(\S_0)$.
Using \Eq{eq:4-to-3}, we then find that
the Ward identity 
translates to 
\begin{align}
    \label{eq:operatorWard}
    U_\cen(\S_0)\, W_\rep(\C)\, \Uinv(\S_0)
    \mem=\mem 
    \rho(\alpha)^{\Intersect_3(\C,\S_0)}\, W_\rep(\C)
    \,,
\end{align}
which is an equal-time operator equation.

\Eq{eq:operatorWard}
is the avatar of the Ward identity in the operator formalism,
where all the relevant geometric objects and operations, including the disc $\S_0$ and the closed contour $\C$, are defined within the three-manifold $\M_3$.
The key insight here is that
time ordering in the path integral formalism
turns into
operator ordering in the operator formalism.

\begin{figure}[t]
    \centering
    \includegraphics[
        scale=1.25
    ]{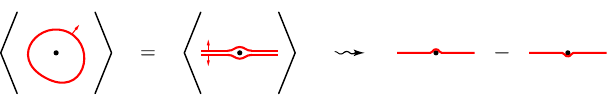}
    \caption{
        Zero-form version of
        the pancaking procedure,
        which the reader may be familiar with.
        Wrapping an operator with the symmetry operator
        translates to an equal-time conjugation,
        where
        operator ordering
        traces back to
        time ordering.
    }
    \label{fig:pancake0}
\end{figure}

Finally,
it is straightforward to 
explicitly evaluate \Eq{eq:operatorWard}.
Applying the $3{\,+\,}1$ decomposition and using \Eq{eq:Edef},
the Hamiltonian formalism avatar of the symmetry operator,
supported on a 
disc
$\S_0 {\,\subset\,} \M_3$
in three dimensions,
is given as
\begin{align}
    \label{eq:YM.Q}
    U_\cen(\S_0)
    = \exp\bigg(\mem{
        \frac{\pi}{g^2}
        \int_{\S_0} dx^j \swedge dx^k\, \e_{ijk}\mem E^i{}_a\mem \lambda^a
    }\bigg)
    \transition{where}
    e^{2\pi\lambda} = \cen
    \,.
\end{align}
To compute 
$U_\cen(\S_0)\mem W_\rep(\C)\mem \Uinv(\S_0)$,
let us first deduce how the conjugation acts on the phase space.
While $E^i{}_a(x)$ is left invariant
because $[E^i{}_a(x) , E^j{}_b(x')] = 0$,
the spatial gauge connection $A^a{}_k(x)$ has a nonvanishing commutator with the exponent of \Eq{eq:YM.Q},
\begin{align}
\begin{split}
    \label{eq:Qaction-calc}
    &
    \frac{\pi}{g^2}\mem \int_{\S_0} 
    dx'^i \swedge dx'^j\, \lambda^b(x')\,
        \e_{ijl}\mem
        [ E^l{}_b(x') , A^a{}_k(x) ]
    \,\\
    &\mem=\mem
    2\pi \int 
    d\s_1\hem d\s_2\,\,
        \frac{\partial X^i}{\partial\s_1} 
        \frac{\partial X^j}{\partial\s_2}
    \, \lambda^a(X)\,
        \e_{ijk}\mem
        \delta^{(3)}\hnem(x-X)
    \,,
\end{split}
\end{align}
where $(\s_1,\s_2) \mapsto X^i(\s_1,\s_2)$ is a parameterization of the surface $\S_0$.
According to \Eq{eq:delta-param},
this
describes the components of the one-form
$2\pi\lambda^a\mem \delta_3(\S_0)$,
where $\delta_3(\S_0)$ is the Dirac delta one-form 
of $\S_0$
defined
in the spatial three-manifold $\M_3$.
Therefore,
we can summarize the action of
the symmetry transformation 
on the phase space variables
as the following:
\begin{align}
\begin{split}
    \label{eq:Qaction}
    U_\cen(\S_0)\,{
        A^a{}_i
    }\,\Uinv(\S_0)
    &= (A^a + 2\pi\lambda^a\mem \delta_3(\S_0))_i
    \,,\\
    U_\cen(\S_0)\,{
        E^i{}_a
    }\,\Uinv(\S_0)
    &= E^i{}_a
    \,.
\end{split}
\end{align}
As a result, we find that the Wilson loop transforms as
\begin{align}
\begin{split}    
    \label{eq:QadjW}
    U_\cen(\S_0)\, W_\rep(\C)\, \Uinv(\S_0)
    \mem=\mem
    \rho(\alpha)^{\Intersect_3(\C,\S_0)}\mem
    W_\rep(\mathcal{C})
    \,,
\end{split}
\end{align}
which 
proves 
the Hamiltonian counterpart of the
Ward identity.
We have used the fact that
the three-dimensional intersection number is
is
$\Intersect_3(\C,\S_0) = \oint_\C \delta_3(\S_0) $.

\section{Gravity}
\label{sec:Grav}

Armed with an understanding of higher-form symmetry in gauge theory, we are now equipped to transcribe all of those results to the context of dynamical gravity.  
As is well-known,
the tetradic Palatini formalism 
is a description of gravity in terms of a  gauge theory of Lorentz transformations. 
Using this framework, we can deduce the higher-form symmetries of gravity by direct analogy.
As before, we start with a covariant analysis 
and then describe
the same physics using the canonical formalism.


\subsection{Covariant Formalism}
\label{sec:CovGrav}

Consider dynamical gravity on a four-dimensional manifold $\M$ with Euclidean signature.  As noted earlier, we work within the regime of validity of an effective field theory of gravity in which the topology and dimensionality of spacetime do not fluctuate.

Our point of departure is the tetradic Palatini formalism,
which formulates gravity as a gauge theory\footnote{Here were refer to gauge theory 
in the restricted sense of a
construction based on principal fiber bundles.  While diffeomorphisms are of course a redundancy of description, they do not define a gauge theory in this restricted sense because
diffeomorphisms inherently shift the base points. } of the four-dimensional Lorentz group $G$
and is inherently first-order.  Here we emphasize again that despite our abuse of nomenclature we will consider both Lorentzian and Euclidean signature.
The degrees of freedom are a one-form spin connection and one-form tetrad field,
\begin{align}
    \omega^{AB} = \omega^{AB}{}_\m\mem dx^\m 
    \transition{and} 
    e^A = e^A{}_{\m}\mem dx^\mu
    \, ,
\end{align}
which transform in the adjoint and fundamental representation of $G$, so  the uppercase indices are $A,B,\ldots \in \{1,2,3,4\}$.

In terms of these fields,
the action for Palatini gravity is given by the integral over $\M$ of the Lagrangian four-form
\begin{align}
    \label{eq:eGrav.L.vec()}
    L
    = \frac{1}{4g^2}\mem 
    \e_{ABCD}\mem
    e^A \swedge e^B \swedge R^{CD}
    - \frac{\Lambda}{24g^2}\mem
    \e_{ABCD}\mem e^A \swedge e^B \swedge e^C \swedge e^D
    \,,
\end{align}
where we have defined the Riemann curvature two-form,
\begin{align}
    \label{eq:Grav.fieldstrength}
    R^A{}_B
    = d\omega^A{}_B + \omega^A{}_C \swedge \omega^C{}_B
    \,,
\end{align}
which is simply the field strength for the spin connection.
The first term in \Eq{eq:eGrav.L.vec()} encodes the Einstein-Hilbert Lagrangian, where we have repackaged Newton's constant into $g= (8\pi G_\text{N})^{1/2}$ to draw a closer analogy with gauge theory.
The second term in  \Eq{eq:eGrav.L.vec()}
defines the cosmological constant $\Lambda$.

It should be reiterated that
both the spin connection and tetrad are taken as {\it independent} degrees of freedom
in the tetradic Palatini formulation.
To see how this approach reproduces conventional
general relativity, let us vary the Lagrangian in \Eq{eq:eGrav.L.vec()} with respect to the spin connection and tetrad to obtain their respective equations of motion,
\begin{align}
    \label{eq:Grav.varL}
            D(e^A \swedge e^B) = 0
                \transition{and} 
            e^{[B}
            \swedge R^{CD]}
            =
            \frac{\Lambda}{3}\mem
            e^B \swedge e^C \swedge e^D
    \,,
\end{align}
where the square brackets on indices denote antisymmetrization.  Here we have defined $D$ as the covariant exterior derivative with respect to the spin connection.

It is not difficult to show that
the first equation of motion in \Eq{eq:Grav.varL} is
\textit{algebraically}
equivalent to 
vanishing of
$De^A = de^A + \omega^A{}_B\swedge e^B$,
which is the definition of torsion.
This fixes the dynamics to be identically torsion-free.
Any coupling of the spin connection to an external source 
generates nonzero torsion precisely 
only
on the support of that source.
Meanwhile, with vanishing torsion
the second equation of motion in \Eq{eq:Grav.varL}
becomes the Einstein field equations for the associated metric,
\begin{align}
    \label{eq:g[e]}
    g_{\m\n}= \delta_{AB}\mem e^A{}_\m\mem e^B{}_\n
    \,.
\end{align}
As emphasized earlier, we work in an effective field theory description of gravity which describes gravitons propagating over a fixed background.  Hence, throughout our analysis the metric and tetrad are implicitly expanded as fluctuations about some choice of background values $\bar g_{\m\n}$ and $\bar e^A{}_\m$, respectively, though it will usually be simpler to manipulate the full field variables rather than their fluctuations.\footnote{In any sensible effective field theory, the background spacetime is nondegenerate and thus the background tetrad $\bar e^A{}_\m$ must be nonzero.  This means the vacuum breaks diffeomorphism invariance and local Lorentz symmetry down to the diagonal, which naively hinders our analysis.  However, our calculation of the Ward identity for the symmetry and line operators utilize the full tetrad field $e^A{}_\m$, which transforms covariantly, so there is no additional complication.  The very same phenomenon occurs in gauge theory, where expanding about a background gauge field $\bar A^a{}_\m$ technically breaks Lorentz invariance and color symmetry down to the diagonal, but of course  with no effect on the Ward identities in the theory.  }

Using integration by parts, it is easy to rewrite the Lagrangian in \Eq{eq:eGrav.L.vec()} so that it does not contain derivatives of the spin connection.  Consequently, 
the spin connection is an auxiliary field that can be eliminated at tree-level by plugging in the classical solution for $\omega^A{}_B$ in terms of $e^A$
using
the torsion-free condition in \Eq{eq:Grav.varL}.
The resulting Lagrangian, 
which depends solely on the tetrad,
is precisely the usual Einstein-Hilbert Lagrangian.
Therefore,
tetradic Palatini gravity is classically
equivalent
to general relativity,
provided there are no sources that couple directly to the spin connection so that torsion is vanishing. As noted earlier, quantum equivalence also follows, since the effective field theory of a massless spin two particle is unique, modulo Wilson coefficients.

Next,
we would like to recall the well-known fact that
tetradic Palatini gravity can be expressed in a way that even more directly parallels the first-order formulation of Yang-Mills theory. 
This fact will play a crucial role in our identification of the one-form symmetry later on.
In particular, 
let us define the Pleba\'nski two-form
\cite{Plebanski:1977zz,Capovilla:1991qb}
by
\begin{align}
    \label{eq:pleb-def}
    B_{AB}   = \frac{1}{2}\mem \e_{ABCD}\mem e^C \swedge e^D \, ,
\end{align}
which is valued in the adjoint of $G$.   Expressed in terms of $B_{AB}$,
the tetradic Palatini Lagrangian in
\Eq{eq:eGrav.L.vec()} becomes
\begin{align}
    \label{eq:eGrav.L}
    L
    = \frac{1}{g^2}\mem 
    B_a \swedge R^a
    - \frac{\Lambda}{6g^2}\mem
    B_a \swedge {\star B}^a
    \,,
\end{align}
which is by construction identical in form to the Yang-Mills Lagrangian in \Eq{eq:YM.L}.
Here the lowercase indices $a,b,\ldots \in \{1,2,3,4,5,6\}$
transform in the adjoint of $G$, and $\star$ 
denotes the Hodge star in the \textit{internal} Lorentz space, so
when the adjoint indices are all converted to
fundamental indices by the 
Lorentz algebra generator $(t_a)^A{}_B$,
we have
\begin{align}
    B_{AB}
    = \frac{1}{2}\mem \e_{ABCD}\mem e^C \swedge e^D
    \transition{and}
    {\star B}^{AB}
    =
    e^A \swedge e^B
    \,.
\end{align}
We emphasize that $\star$ should be distinguished from $\ast$ in \Eq{eq:YM.L}.
Also, we clarify that fundamental indices
$A,B,\cdots$
are raised and lowered by
the Euclidean flat metric $\delta_{AB}$.

In terms of the Pleba\'nski two-form, the Lagrangian in \Eq{eq:eGrav.L},
is clearly of the {\BF}-type Lagrangian of the form $B_a \swedge F^a + f(B)$ which is familiar from gauge theory.
Therefore, we can immediately
identify 
the line operator and
the symmetry operator for the one-form symmetry of gravity by essentially copying the formulae from 
our earlier discussion about
Yang-Mills theory.

So what is the one-form symmetry of dynamical gravity?
In analog with gauge theory, it is defined by the center of $G$, which is the four-dimensional Lorentz group.   As usual, the center depends crucially on the global structure of $G$, which is not specified by
the Lagrangian in \Eq{eq:eGrav.L}.  
Therefore, it is essential to clarify the global structure of $G$ before we can continue further.

To begin, let us consider the case of Euclidean signature.  
Given the well-known Lie algebra isomorphism
$\mathfrak{so}(4)
    \,\cong\,
\mathfrak{su}(2) {\mem\oplus\,} \mathfrak{su}(2)
\,\cong\, \mathfrak{spin}(4)$,
we can choose $G$ to be either\footnote{
    A priori, one can also consider semi-spin groups such as  $\mathrm{SemiSpin(4)}\simeq \mathrm{SU}(2) {\mem\times\,} \mathrm{SO}(3)$, which is a nonstandard quotient.  However, this group does not admit a vector representation so it is not compatible with the tetrad formalism. 
}
\begin{align}
    \label{eq:Gfor4}
        \mathrm{Spin}(4)
        \cong \mathrm{SU}(2) {\mem\times\,} \mathrm{SU}(2)
    \transition{or}
        \mathrm{SO}(4)
        \cong \frac{\mathrm{Spin}(4)}{\mathbb{Z}_2}
    \transition{or}
        \frac{\mathrm{SO}(4)}{\mathbb{Z}_2}
        \cong \frac{\mathrm{Spin}(4)}{\mathbb{Z}_2 {\mem\times\,} \mathbb{Z}_2}
    \,,
\end{align}
whose center subgroups are given by 
\begin{align}
        \mathbb{Z}_2 {\mem\times\,} \mathbb{Z}_2
    \transition{or}
        \mathbb{Z}_2
    \transition{or}
         \id  
    \,,
\end{align}
respectively. The one-form charges will be valued in these center subgroups.
Note that the zero-form symmetry associated with 
the center $\mathbb{Z}_2 {\mem\times\,} \mathbb{Z}_2$ of
$\mathrm{Spin}(4)
\cong \mathrm{SU}(2) {\mem\times\,} \mathrm{SU}(2)$
acts as parity on chiral and antichiral spinor indices,
while
that of the center $\mathbb{Z}_2$
of
$\mathrm{SO}(4)$
acts as parity on vector indices.

Meanwhile, in Lorentzian signature,
possible candidates for the gauge group are
\begin{align}
    \label{eq:GforL4}
        \mathrm{Spin}(3,1)
        \cong \mathrm{SL}(2,\mathbb{C}) 
    \transition{or}
        \mathrm{SO}^+\nem(3,1)
        \cong \frac{\mathrm{Spin}(3,1)}{\mathbb{Z}_2}
    \,,
\end{align}
     where the latter is
    the orthochronous Lorentz group
     and
    the former is its double cover.
    The corresponding center subgroups 
    are 
\begin{align}
        \mathbb{Z}_2 
    \transition{or}
         \id  
    \,,
\end{align}
    respectively, 
   which define allowed one-form symmetry of gravity in Lorentzian signature.  Here the zero-form symmetry associated with the former corresponds to net parity on chiral and antichiral spinor indices, also known as fermion parity.

Any choice of $G$ in \Eq{eq:Gfor4} is viable.  
For the sake of generality, however, in the remainder of our analysis we will be agnostic and take the
gauge group
to be some general $G$
with center subgroup $Z(G)$.

\subsubsection{Line and Symmetry Operators}
\label{sec:CovGrav.symmetry}


We are now equipped to derive explicitly the one-form symmetry of dynamical gravity.
As before, we start with identifying the line operator and its symmetry transformation.

Firstly,
let us recall the direct parallel between
the spin connection in \Eq{eq:eGrav.L} and the gauge connection in \Eq{eq:YM.L}.
As such, it is obvious that the natural line operator in gravity is the spin holonomy, 
\begin{align}
    \label{eq:Grav.W}
    W_\rep(\C)
    = \hloop{\rep}{\C}
    \,,
\end{align}
where $\C$ defines a closed one-dimensional contour and $\rep$ is some spin representation of the Lorentz group $G$.
Far less clear a priori is the identity of the symmetry operator,
\begin{align}
    \label{eq:Grav.Uschem}
    U_\cen(\S)
    \,,
\end{align}
other than that it should be labeled by a center element $\cen \in G$ and defined on an exact two-dimensional surface $\S = \partial\V$ that can topologically link with $\C$.

In perfect analogy with the Wilson loop of gauge theory, we expect that the spin holonomy should transform as
\begin{align}
    \label{eq:Grav.symaction}
    W_\rep(\C)
        \,\,\mapsto\,\,
    \rep(\cen)^{\Link(\C,\S)}\mem
    W_\rep(\C)
    \, ,
\end{align}
under the one-form symmetry of gravity, and this is indeed the case.
As before, ${\Link(\C,\S)}$ is defined to be the linking number between the contour and surface that define the line and symmetry operators.\footnote{
    As noted previously, we work in an effective field theory of gravity in which the topology and dimensionality of spacetime are robust.  Furthermore, the linking number is invariant under any invertible diffeomorphism that is continuously connected to the identity.  To see why, consider a putative family of diffeomorphisms labeled by a parameter $\tau {\,\in\mem} [0,1]$ such that $\tau {\,=\,} 0$ is the identity and $\tau {\,=\,} 1$ unlinks the surfaces.  By continuity, there exists some $\tau$ for which the corresponding diffeomorphism results in surfaces which intersect at a point.  In this case the diffeomorphism is not invertible, since it maps two points, one on each surface, to a single point. 
}
%
%

Mirroring
\Eq{eq:YM.fr} in gauge theory, the one-form symmetry of gravity is 
implemented as a transformation of the fields by a closed but not exact form. In the gravitational context, the appropriate map is a local Lorentz transformation that is multivalued.  In particular, we consider the case where $\S=\partial\V$ and there is a branch cut on the coboundary $\V$ whose discontinuity is center-valued.
Physically,
this twisted Lorentz transformation boosts local laboratories in spacetime
such that
a $\pi$ rotation of frames
is applied after each winding about $\S$.
Specifically, the spin connection and the tetrad will transform as
\begin{align}
    \label{eq:Grav.fr}
      \omega^A{}_B
        &\,\,\mapsto\,\,
    (\Omega^{-1})^A{}_C\mem \omega^C{}_D\mem \Omega^D{}_B
    + (\Omega^{-1})^A{}_C\mem d\Omega^C{}_B
    \transition{and}
    e^A
    \,\,\mapsto\,\,
    (\Omega^{-1})^A{}_B\, e^B
    \,,
\end{align}
from which the transformation of the Pleba\'nski two-form is given by
\begin{align}
    \label{eq:Grav.frB}
    B^A{}_B
        \,\,\mapsto\,\,
    (\Omega^{-1})^A{}_C\hem
    B^C{}_D\mem
    \Omega^D{}_B
    \,,
\end{align}
where $\Omega$ is a multivalued zero-form which defines the twisted Lorentz transformation.  Like in the case of gauge theory, $d\Omega$ is not exact because $\Omega$ carries winding. 
As before, we also impose a discontinuity condition
across the branch cut, given by
 \begin{align}
    \lim_{
        \P_{\pm}\to\, \P
    }\hem
        \Omega(\P_{\hnem+})
        \mem 
        \Omega^{-1}\hnem(\P_{\hnem-})
    \,=\,
    \cen
    \transition{where}
    \cen \in Z(G)     \transition{and} \P {\,\subset\,} \V
    \,,
\end{align}
where $\P_{\hnem+}$ and $\P_{\hnem-}$ 
are infinitesimally split across $\V$.
In turn,
it follows that
the spin holonomy that links once with $\S$ maps to
\begin{align}
\begin{split}
    W_\rep(\C)
    \,\,\mapsto\,\,&
        \lim_{\C'\to\mem\C}
        \tr_\rep\nem\bigg[\mem{
            \Pexp
            \bigg(\hem{
                \int_{\C'} 
                    \Ad{\Omega}{\omega}
                    +
                    \Omega^{-1} d\Omega
            }\bigg)
        }\bigg]
    \,,\\
    =\,\,& 
        \lim_{\C'\to\mem\C}
        \tr_\rep\nem\bigg[\mem{
            \Omega^{-1}\hnem(\P_{\hnem-})
            \,
            \Pexp
            \bigg(\hem{
                \int_{\C'} 
                    \omega
            }\bigg)
            \,\Omega(\P_{\hnem+})
        }\bigg]\\
    =\,\,& 
        \lim_{\C'\to\mem\C}
        \tr_\rep\nem\bigg[\mem{
            \cen\,
            \Pexp
            \bigg(\hem{
                \int_{\C'} \omega
            }\bigg)
        }\bigg]
    \,=\,
        \rep(\cen)\mem
        W_\rep(\C)
    \,,
\end{split}
\end{align}
which exactly instantiates \Eq{eq:Grav.symaction}.
Thus, we conclude that the twisted local Lorentz transformation in \Eq{eq:Grav.fr}
implements the gravitational one-form symmetry.

As in the gauge theory case, we emphasize here that $\Omega$ is not a bona fide gauge transformation when the center twist $\cen$ is nontrivial. In particular, drawing an analogy with
\Eq{eq:lambda-disc2}, we see that the double exterior derivative of $\Omega$ is nonzero,
\begin{align}
    \label{eq:Grav.ddOmega}
    (\Omega^{-1})^A{}_C\mem
    dd\Omega^C{}_B
    = 2\pi\lambda^A{}_B\mem \delta(\S) 
    \transition{for any $\l^A{}_B$ such that}
    e^{2\pi \lambda} = \cen \in Z(G)
    \,,
\end{align}
and in fact has nonvanishing support precisely on $\S$.

Before continuing,
let us address some possible confusions relating to Lorentz and diffeomorphism invariance.
First of all, just like in Yang-Mills theory, we see here that the twisted Lorentz transformation depends on the whole function $\Omega$ rather than just $\cen$.  Naively, this dependence is Lorentz-violating, since $\Omega$ defines a trajectory in the space of Lorentz transformations.
However, just as before, we can see that this dependence is spurious since different choices for $\Omega$ which exhibit the same twist $\cen$ still act indistinguishably on the spin holonomy, and are thus physically equivalent. 

Secondly, in the presence of dynamical gravitation 
there is a further caveat
regarding the {\it diffeomorphism invariance} of the line operator $W_\rep(\C)$ and symmetry operator $U_\cen(\S)$.
Although these objects do not carry dangling indices,
they do depend on a particular choice of 
a contour $\C$ and surface $\S$,
which
define collections of points in spacetime in the very same way that a local operator ${\cal O}(x)$ defines a single point.  
However, any local or quasi-local object such as a point, curve, or surface in spacetime is famously not diffeomorphism invariant, simply because ``$x$'' itself is not diffeomorphism invariant. 
Note that this annoyance is also implicitly present in any discussion of gauge theory Wilson loops in the presence of gravity, which is central to discussions of swampland conjectures.
To address this, one typically appeals to the restoration of diffeomorphism invariance by  ``gravitationally dressing''
\cite{Donnelly:2016rvo,Giddings:2022hba} the operator in question.  
A closely related tactic is to define all positions ``relationally'' in terms of some asymptotic reference, either at spatial infinity or the beginning of time.

However, diffeomorphism invariance is restored here in the very same way as Lorentz invariance.  Since the linking number is itself diffeomorphism invariant, it means that operators related to each other by a diffeomorphism are themselves are physically equivalent.

\subsubsection{Ward Identity}
\label{sec:CovGrav.ward}

Next, let us compute the Ward identity associated with the gravitational one-form symmetry.   Taking inspiration from \Eq{eq:YM.U} in the case of gauge theory, we define the one-form symmetry operator of gravity to be
\begin{align}
    \label{eq:Grav.U}
    U_\cen(\S)
    = \exp\bigg(\mem{
        \frac{\pi}{g^2}
        \int_{\S}\mem \lambda^{AB}\hem B_{AB}
    }\bigg)
    \transition{where}
    e^{2\pi \lambda} = \cen \in Z(G)
    \,,
\end{align}
where $\lambda$ is a zero-form function which is chosen so that at all points in spacetime it 
exponentiates to a center element of the Lorentz group.
%
%
Again,
this formula should be understood as
a realization of the symmetry operator $U_\cen(\S)$
with a representative Lie algebra element $\l$ for $\cen$,
as all choices of $\l$ with the same twist $\cen$ act the same on the spin holonomy and are thus physically equivalent.

Intriguingly,
this surface operator $U_\cen(\S)$
literally computes a certain area-like quantity
associated with $\S$!
In particular, by reverting to tetrad variables and reintroducing Newton's constant,
we find that the symmetry operator is 
\begin{align}  
    \label{eq:Grav.U.alt}
    U_\cen(\S)
    = \exp\bigg(\mem{
        \frac{1}{4 G_\text{N}}
        \int_{\S}\mem
            \frac{1}{2}\,
            {\star\lambda}_{AB}\mem
            (e^A \swedge e^B)
    }\mem\bigg)
    \,,
\end{align}
where 
${\star\lambda}_{AB}
= \frac{1}{2}\mem \e_{ABCD}\mem \lambda^{CD}$
is the Hodge dual of $\lambda^{AB}$.  Here we recognize
 $e^A \swedge e^B$ as
the infinitesimal area element
in the orthonormal frame,
so the exponent 
computes the area smeared with a reference $\star\lambda_{AB}$,
measured in Planck units.
Note that $1/2$ is the canonical normalization factor
for contracting 
antisymmetric tensors.

Additionally, it is 
curious that the exponent in \Eq{eq:Grav.U.alt}, or equivalently in \Eq{eq:Grav.Q},
is tantalizingly similar to the
area operator in loop quantum gravity
which leads to the quantization of tetrahedral volume
\cite{Rovelli:1994ge,Ashtekar:1996eg}.
The only crucial difference is that here
the area is ``dotted'' with a Lorentz generator
$\lambda^{AB}$ that exponentiates to the center,
so in a sense it 
serves as a discrete and topological reincarnation of the area operator in loop quantum gravity.
On the other hand,
$U_\cen(\S)$ is also reminiscent of the Bekenstein-Hawking entropy formula \cite{Bekenstein:1973ur,Hawking:1975vcx}.
Of course,
these could easily be accidents of dimensional analysis on account of the $1/G_{\rm N}$ normalization in the exponent.
But in any case, 
it would be 
fascinating
to see if any of 
these superficial similarities
carry deeper significance. 

Meanwhile,
a virtue of the algebra isomorphism
$\mathfrak{so}(4) \cong \mathfrak{su}(2) {\mem\oplus\,} \mathfrak{su}(2)$
is that we always can split the six generators of the Lorentz group $G$ into three chiral and three antichiral
$\mathrm{SU}(2)$ generators.
Assigning dotted and undotted spinor indices
to each sector,
we see that 
the spin connection decomposes into self-dual and anti-self-dual components,
$\omega^{\da\db}$ and $\omega^{\a\b}$,
while the tetrad is $e^{\da\a}$.
Similarly,
the Pleba\'nski two-form $B_a$
decomposes into
$B^{\da\db} = \e_{\a\b}\mem (e^{\da\a} \swedge e^{\db\b})$
and
$B^{\a\b} = -\te_\wrap{\da\db}\mem (e^{\da\a} \swedge e^{\db\b})$.
See \App{app:notations} for more details.
Meanwhile, 
since the two $\mathfrak{su}(2)$ sectors
commute,
any element of the Lorentz algebra
that exponentiates to a center element can be split into self-dual and anti-self-dual generators
that separately exponentiate to center elements.
Given these facts,
we see that
the symmetry operator decomposes into more primordial building blocks, which are {\it chiral} symmetry operators,
\begin{align}
\begin{split}
    \label{eq:Ublocks} 
    \widetilde U(\S)
    = \exp\bigg({
        -
        \frac{1}{4G_\text{N}}
        \int_{\S}\mem
            \rambda^\db{}_\da\,
            e^{\da\a} \swedge e_\wrap{\a\db}
    }\mem\bigg)
    &\transition{where}
    (e^{2\pi\rambda})^\da{}_\db = -\delta^\da{}_\db
    \,,\\
    U(\S)
    = \exp\bigg({
        -
        \frac{1}{4G_\text{N}}
        \int_{\S}\mem
            \lambda_\a{}^\b\mem
            e_\wrap{\b\da} \swedge e^{\da\a}
    }\mem\bigg)
    &\transition{where}
    (e^{2\pi\l})_\a{}^\b = -\delta_\a{}^\b
    \,.
\end{split}
\end{align}
Here
$\tilde\lambda^\da{}_\db$
and
$\lambda_\a{}^\b$
belong to the self-dual and anti-self-dual Lie algebras.

For the case of Euclidean signature with
$G=\mathrm{Spin}(4)$, 
the chiral operators $\widetilde U(\S)$ and $ U(\S)$
are precisely the one-form symmetry operators corresponding to each factor of the center subgroup $Z(G) = \mathbb{Z}_2\times \mathbb{Z}_2$.  Note that, as zero-form symmetries, each factor of the center acts as parity in the numbers of dotted and undotted spinor indices, respectively.
In Lorentzian signature, however, the one-form symmetry is only nontrivial if $G=\mathrm{SL}(2,\mathbb{C})$, in which case the center subgroup is $Z(G) = \mathbb{Z}_2$.  As a zero-form symmetry this acts as {\it net} parity on spinor indices.   In this case, only the {\it real} combination of the chiral operators,   $\widetilde U(\S)\mem U(\S)$, corresponds to the one-form symmetry operator.

Now let us finally present a derivation of the Ward identity associated with the one-form symmetry of gravity,
\begin{align}
    \label{eq:Grav.Ward}
    \expval{
        U_\cen(\S)\mem W_\rep(\C)
    }
    = \rep(\cen)^{\Link(\C,\S)}
    \expval{
        W_\rep(\C)
    }
    \,. 
\end{align}
Since the tetradic Palatini framework is a first-order formalism, the left-hand side 
is given by a path integral over all configurations of
both the spin connection and the tetrad,
\begin{align}
    \label{eq:Grav.piLHS}
    \expval{
        U_\cen(\S) W_\rep(\C) 
    }
    = \int \D{\omega}\D{e}\,
        e^{-S}\,
        U_\cen(\S)\,
        W_\rep(\C)
    \,.
\end{align}
As before, the symmetry operator $ U_\cen(\S)$
merges with $e^{-S}$
to give
\begin{align}
    \label{eq:Grav.pi1}
    e^{-S}\,
        U_\cen(\S)
    =
    \exp\bigg({
        -\frac{1}{4g^2} \int_\M
            \e_{ABCD}\mem e^A \swedge e^B
            \wedge
            (
                R
                -
                \Omega^{-1}dd\Omega
            )^{CD}
        + \ldots
    }\mem\bigg)
    \,,
\end{align}
in analogy with \Eq{eq:YM.pi1}, and where
the ellipses denote the cosmological constant term.
Meanwhile, we see that the twisted Lorentz transformation in \Eq{eq:Grav.fr} implies 
\begin{align}
    \label{eq:Grav.Ftransf}
    R^A{}_B
        &\,\,\mapsto\,\,
    (\Omega^{-1})^A{}_C\mem R^C{}_D\mem \Omega^D{}_B
    + (\Omega^{-1})^A{}_C\mem
    dd\Omega^C{}_B
    \,,
\end{align}
which is
gravitational analog of
\Eq{eq:YM.Ftransf}.
Applying this to \Eq{eq:Grav.pi1}, we find
\begin{align}
    \label{eq:Grav.eU}
    e^{-S}\, U_\cen(\S) 
        \,\,\mapsto\,\,
    e^{-S}
    \,.
\end{align}
Together with the transformation of the line operator in \Eq{eq:Grav.symaction},
\Eq{eq:Grav.eU}
sends the left-hand side of the Ward identity in \Eq{eq:Grav.piLHS} to 
\begin{align}
\begin{split}
    \expval{
        U_\cen(\S)\mem W_\rep(\C)
    }
    \; \mapsto \;
    \int \D{\omega}\D{e}\,
        e^{-S}\mem
            \rep(\cen)^{\Link(\C,\S)}\mem
            W_\rep(\C)
    \,, 
\end{split}
\end{align}
which is precisely its right-hand side.
This proves
the Ward identity encoding the topological linking of the spin holonomy in \Eq{eq:Grav.W}
with the symmetry operator in \Eq{eq:Grav.U}.

Just like in the case of gauge theory, the one-form symmetry of gravity acts as a twisted Lorentz transformation on {\it any choice of operator}.
Thus,
if an operator transforms under the twisted Lorentz transformation as 
${\cal O} \mapsto {\cal O}_\Omega$,
then the corresponding Ward identity is
\smash{$\expval{
    U_\cen(\S)\mem
    {\cal O}
} = \expval{ {\cal O}_\Omega }$}
in parallel with \Eq{eq:arbWard}.

Before continuing, let us highlight an important point: the one-form symmetry of gravity is independent of our choice of formalism.  In particular, while our derivations have made elaborate use of tetradic Palatini gravity, our final conclusions remain valid independent of this choice.  For example, integrating out the spin connection yields a pure tetrad theory, but this still exhibits the one-form symmetry.   

On the other hand, it is natural to ask about gravity in the pure {\it metric} formulation, where there is no tetrad, spin connection, or local Lorentz symmetry to speak of.  In this case one formulates the dynamics only in terms of the metric, which is manifestly invariant under the twisted Lorentz transformation defined in \Eq{eq:Grav.fr}, since  $g_{\m\n} \mapsto g_{\m \n}$.  Has the one-form symmetry disappeared?  For the spinor holonomy, the answer is yes, but for the simple reason that we cannot even write it down since there is no tetrad field to characterize the gravitational interactions of fermions. This would be analogous to studying Yang-Mills theory with the stipulation that we can only ever use adjoint indices, thus precluding the existence of the very fundamental Wilson loops which are charged under the one-form symmetry.
On the other hand, the vector holonomy and its one-form symmetry properties should presumably have a pure metric description since spin structure should not be necessary.
Note that there has been some interesting recent work 
constructing {\it continuous} one-form symmetries of linearized gravity using the metric alone \cite{Hinterbichler:2022agn,Benedetti:2021lxj,Benedetti:2023ipt}.  It would be very illuminating to see explicitly how those symmetries relate to the ones derived in this paper.

\subsubsection{Chiral Cosmic String}
\label{sec:ChiralCosmicString}

Just as in gauge theory, the one-form symmetry operator in gravity can be interpreted as an insertion of a defect in spacetime. It is then natural to ask, what is the nature of this singular object?  Indeed, how would a relativist interpret such a defect?

To answer this question let us consider empty space, as described by a flat metric.  Next, we apply the twisted Lorentz transformation in \Eq{eq:Grav.fr}, which induces a curvature singularity,
$R^A{}_B = (\Omega^{-1} dd\Omega)^A{}_B = 2\pi \lambda^A{}_B\mem \delta(\S)$, localized on the surface $\S$.
To give a physical interpretation to this twisted geometry, we can straightforwardly reverse engineer the matter source that would directly generate this singularity.
To do this we insert $R^A{}_B = 2\pi \lambda^A{}_B\mem \delta(\S)$ directly into the left-hand side of the Einstein field equations, 
\begin{align}
    \label{eq:stress}
      -\frac{1}{g^2}\mem
    {\star R}_{AB} \swedge e^B=
    \frac{1}{3!}\,
    |e|\,
    T^\m{}_A\,
    \e_{\m\n\r\s}\mem 
    dx^\n \swedge dx^\r \swedge dx^\s
    \,,
\end{align}
where $|e|$ denotes the determinant of $e^A{}_\m$.
From the resulting quantity, we then deduce the stress-energy tensor $T^\m{}_A$ that would be required to generate the corresponding curvature singularity:
\begin{align}
\begin{split}
    \label{eq:stress-string}
    T^\m{}_\k
    &=
        -\frac{1}{8G_\text{N}}\mem
        \frac{1}{|e|}\,
        {\star\l}_{\k\n}\,
        \delta(\S)_{\r\s}\mem
        \e^{\m\n\r\s}
    \,,\\
    &=
        -\frac{1}{4G_\text{N}}\mem
        {\star\lambda}_{\k\n}
    \int d\s_1\hem d\s_2\,\,
        \delta^{(4)}\hnem(x{\mem-\hem}X)\,
        \bigg({
            \frac{\partial X^{\m}}{\partial \s_1}
            \frac{\partial X^{\n}}{\partial \s_2}
            -
            \frac{\partial X^{\n}}{\partial \s_1}
            \frac{\partial X^{\m}}{\partial \s_2}
        }\bigg)
    \,,
\end{split}
\end{align}
where in the last line
we have parameterized the surface $\S$ by the function $X^\m\hnem(\s_1,\s_2)$ in terms of worldsheet coordinates
$(\s_1,\s_2)$
and used a formula given in \Eq{eq:delta-param}.
Also, 
we have freely traded off local Lorentz indices with spacetime indices through the tetrad or its inverse
in these final expressions,
while $T^\m{}_A$ in \Eq{eq:stress}
acts as a source for the tetrad $e^A{}_\m$ in the first-order formulation defined in \Eq{eq:eGrav.L}.
The stress-energy tensor
in \Eq{eq:stress-string}
is localized along a membrane $\S$
and 
describes a defect reminiscent of a \textit{cosmic string},
though it is not literally identical to the Nambu-Goto string.

Interestingly, we can also see that the algebraic Bianchi identity,  $R_{AB} \swedge e^B=0$, actually {\it fails} for this defect configuration, indicating the existence of  \textit{magnetic} stress-energy \cite{cho1991magnetic}.  That is, plugging  $R^A{}_B = 2\pi \lambda^A{}_B\mem \delta(\S)$ into the left-hand side of algebraic Bianchi, we obtain a nonzero expression,
\begin{align}
    \label{eq:dualstress}
     -\frac{1}{g^2}\mem
    R_{AB} \swedge e^B =
    \frac{1}{3!}\,
    |e|\,
    T^\star{}^\m{}_A\,
    \e_{\m\n\r\s}\mem 
    dx^\n \swedge dx^\r \swedge dx^\s
    \, ,
\end{align}
which defines the dual stress-energy tensor $T^\star{}^\m{}_A$. 
Note that the above equation is manifestly Hodge dual to the Einstein field equations in \Eq{eq:stress}.
The fact that the right-hand side of
\Eq{eq:dualstress}
is nonzero is analogous to the failure of the Bianchi identity in Maxwell theory in the presence of
a magnetic monopole. 
For our string geometry,
the components of this dual stress-energy tensor are 
given by
\begin{align}
\begin{split}
    \label{eq:dualstress-string}
    T^\star{}^\m{}_\k
    &=
        -\frac{1}{8G_\text{N}}\mem
        \frac{1}{|e|}\mem
        {\lambda}_{\k\n}\mem
        \delta(\S)_{\r\s}\mem
        \e^{\m\n\r\s}
    \,,
\end{split}
\end{align}
describing a line distribution of NUT charge.
In the meantime,
recall that
the Lorentz generator $\lambda$
that exponentiates to a nontrivial center element
has to be either self-dual or anti-self-dual.
This implies that the electric and magnetic stress-energy tensors in \Eqs{eq:stress-string}{eq:dualstress-string}
are related
either as $T^\m{}_\k = T^\star{}^\m{}_\k$
or $T^\m{}_\k = -T^\star{}^\m{}_\k$,
so this string is
composed of either self-dual or anti-self-dual
matter.
Therefore, we conclude that
the geometry generated by an insertion of the symmetry operator
represents a ``\textit{chiral cosmic string},''
where 
self-dual or anti-self-dual stress-energy
localizes on a two-dimensional surface $\S$.

Meanwhile,
there is another well-known class of line singularities in relativity:
the Misner string 
\cite{Misner:1963flatter,Misner:1965zz,Bonnor:1969ala,sackfield1971physical,dowker1967gravitational,cho1991magnetic},
which is famously the line singularity of the Taub-NUT solution
\cite{Taub:1950ez,Newman:1963yy,Misner:1963flatter,Bonnor:1969ala}.
Does the chiral cosmic string carry a Misner string component?
The answer turns out to be no.
First of all, the Misner string describes a stack of gravitomagnetic \textit{dipoles} \cite{Bonnor:1969ala,sackfield1971physical},
and as such
it can terminate on a set of
gravitomagnetic monopoles, also known as
Taub-NUT black holes.
However, the magnetic stress-energy found in \Eq{eq:dualstress-string}
rather describes 
a line density of distributed gravitomagnetic \textit{monopole}
whose magnitude is equal to the electric stress-energy in \Eq{eq:stress-string}.
Secondly,
the Misner string 
could also be characterized as
a localized source of torsion,
as it
induces a time monodromy
\cite{Misner:1963flatter,dowker1967gravitational,Alfonsi:2020lub}.
However,
it is easy to see that torsion transforms linearly even under twisted local Lorentz transformations,
so 
if torsion was zero in the initial background
it stays identically zero after the insertion of the symmetry operator
as well.
Concretely, one can establish that our string geometry has vanishing torsion
by  plugging in the transformed tetrad and spin connection in \Eq{eq:Grav.fr} 
into $De^A = de^A + \omega^A{}_B \swedge e^B$,
which also confirms that the failure of algebraic Bianchi is solely due to the multivaluedness of the tetrad, since $dde^A \neq 0$.
The absence of torsion
is also clear 
if one recalls the fact that
torsion is induced by local source for the spin connection
in the tetradic Palatini formalism,
while
the symmetry operator  only couples to the tetrad.%
\footnote{
    For this reason,
    the fact that
    the stress-energy tensor in \Eq{eq:stress-string} is not symmetric
    cannot be
    attributed to nonzero torsion.
    It rather traces back to the nonvanishing magnetic stress-energy \cite{cho1991magnetic}.
}
For these reasons,
we conclude that
the singular string geometry generated by the symmetry operator
is of a cosmic kind,
with no Misner string component.

\subsubsection{Linking and Conical Deficit Angle}
\label{sec:ConDef}

We have established that the symmetry operator creates a chiral cosmic string singularity that carries both electric and magnetic gravitational charge.  What is the meaning of the linking of this string with the spin holonomy?
Remarkably, this too has a simple physical interpretation in terms of classical gravitation.
To understand why, let us revisit the path integral computation of the Ward identity, 
\begin{align}
    \expval{
        U_\cen(\S) \mem W_\rep(\C) 
    }
    = \int \D{\omega}\D{e}\mem
        e^{-S}\mem
        U_\cen(\S)\mem
        W_\rep(\C)
    \, ,
\end{align}
but evaluated from using perturbation theory about flat space.  The only dynamical degrees of freedom are the tetrad and spin connection, which encode fluctuations of the physical graviton.   As we will see, from this viewpoint the linking number is computed by an infinite set of perturbative diagrams whose structure offers some nice physical insights.

To apply perturbation theory, we must interpret $ U_\cen(\S)$ and $W_\rep(\C)$ as {\it external sources} for the graviton field.  From \Eq{eq:Grav.W} we see that the spin holonomy $W_\rep(\C)$ couples the graviton to the contour $\C$ with a dimensionless coupling strength.  That is expected because the holonomy describes the coupling of gravity to spin.  Meanwhile, \Eq{eq:Grav.U} implies that the symmetry operator $U_\cen(\S)$ couples the graviton to the surface $\S$ with a coupling $1/g^2$.
Last but not least, from \Eq{eq:eGrav.L}, we see that  in a normalization where the graviton field is dimensionless, graviton vertices scale as $1/g^2$ and graviton propagators scale as $g^2$. 

To classify the various contributions in perturbation theory, let us first consider a tree-level $n$-point correlator of gravitons.  This object has $n-2$ vertices and $2n-3$ propagators, so it scales as $g^{2(n-1)}$.  Loop-level contributions will be higher order in $g^2$, so they are subleading.  Next, we take this $n$-point correlator and attach its external legs to the sources $ U_\cen(\S)$ and $W_\rep(\C)$.  

\begin{figure}[t]
\begin{align*}
    \text{
        Quantum corrected $U$ 
    }:&\qquad
        \includegraphics[scale=1.1,valign=c]{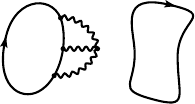}
        \quad\sim\quad
        \mathcal{O}(g^{-2})
    \\[0.3\baselineskip]
    \text{
        Classical holonomy from $U$
    }:&\qquad
        \includegraphics[scale=1.1,valign=c]{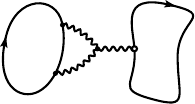}
        \quad\sim\quad
        \mathcal{O}(g^{0})
    \\[0.3\baselineskip]
    \text{
        Quantum holonomy from $U$
    }:&\qquad
        \includegraphics[scale=1.1,valign=c]{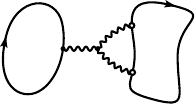}
        \quad\sim\quad
        \mathcal{O}(g^{2})
    \\[0.3\baselineskip]
    \text{
        Quantum corrected $W$
    }:&\qquad
        \includegraphics[scale=1.1,valign=c]{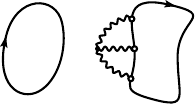}
        \quad\sim\quad
        \mathcal{O}(g^{4})
    \\
    &\qquad
        \kern1.66em
        \mathclap{\adjustbox{scale=0.9}{$
            U_\cen 
        $}}
        \kern5.95em
        \mathclap{\adjustbox{scale=0.9}{$
            W_\rep 
        $}}
\end{align*}
    \caption{
        A schematic depiction of
        Feynman diagrams contributing to the Ward identity
        at various perturbative orders in the gravitational coupling $g$.
        The wavy lines denote gravitons, while the loops of solid lines on the left and right depict the symmetry operator $U_\cen$ and line operator $W_\rep$, respectively.   The first and fourth rows describe quantum corrections to $U_\cen$ and $W_\rep$ separately.  The second row is a contribution to the classical holonomy measured by $W_\rep$, treating $U_\cen$ as a source for the background spacetime.  
        The third row is the one-loop quantum correction to this quantity in that background.
        Since the linking number is dimensionless and topological, it arises purely from the classical holonomy.
    }
    \label{fig:duffs}
\end{figure}

The leading diagrams arise when $n$ external gravitons are connected to $ U_\cen(\S)$, yielding $n$ factors of $1/g^2$. This contribution scales as $1/g^2$ and corresponds to the renormalization of $ U_\cen(\S)$ coming from graviton loops.  Since these contributions do not link with $W_\rep(\C)$, they are unrelated to the topological linking number.  These diagrams are depicted in the first row of \Fig{fig:duffs}.

Meanwhile, the next-to-leading contributions  come from diagrams in which $n-1$ external gravitons are connected to $ U_\cen(\S)$, with the last external graviton connected to $W_\rep(\C)$ as an insertion of the spin connection.  The corresponding diagram scales as a dimensionless constant, since all powers in $g^2$ exactly cancel.  This contribution, 
resumming
up all diagrams for all $n$, is precisely the tree-level one-point function of spin connection computed at all orders in $g^2$ in the presence of the source $U_\cen(\S)$, otherwise known as the classical spin connection in  the corresponding classical gravity problem.  This relationship was understood in the seminal work of Ref.~\cite{Duff:1973zz}, which showed how to perturbatively construct the Schwarzschild metric from an analogous set of diagrams.\footnote{Recently, this connection between perturbative diagrams and classical dynamics has been used to simplify certain contributions to black hole scattering in a recently proposed effective field theory for extreme mass ratio inspirals \cite{Cheung:2023lnj}.}
In any case, we can then compute the next-to-leading contribution to the linking number by inserting this one-point function of the spin connection directly into $W_\rep(\C)$.   See the second row of \Fig{fig:duffs} for a depiction of these contributions.\footnote{Technically, the Feynman diagram contribution in the second row of \Fig{fig:duffs} vanishes since the trace in the spin holonomy acts on a single factor of the antisymmetric spin connection, yielding zero.  However, there are diagrams with multiple insertions of the one-point function of the spin connection into the spin holonomy which enter at the same order in perturbation theory, and they contribute nontrivially to the linking number.} Note that on account of the exponential in $W_\rep(\C)$, it is inserted an infinite number of times.  In perturbation theory, such a calculation would be prohibitively hard.  But crucially, this procedure is literally exactly equivalent to a calculation of the classical spin holonomy $W_\rep(\C)$ evaluated with the spin connection set to its background value sourced by $ U_\cen(\S)$.   

As is well-known, this classical spin holonomy can be viewed as a geometric phase accounting for the precession of a spinning particle as it circumscribes the contour $\C$. 
Hence, the classical spin holonomy $W_\rep(\C)$ precisely measures a certain \textit{conical deficit angle}  induced by $ U_\cen(\S)$. Since $W_\rep(\C)$ rephases by a center element, this conical deficit angle is quantized, and should be viewed intuitively as a $\pi$ phase shift.  The examples in the subsequent section will verify this.  

Of course, there are next-to-next-to-leading contributions and beyond that are ever higher order in $g^2$.  Some of these are just the renormalization of $W_\rep(\C)$, which are depicted in the fourth row of \Fig{fig:duffs}.  The remaining contributions are shown in the third row of \Fig{fig:duffs}, and correspond to quantum loop corrections to the spin holonomy.  
However, we can argue a priori that the topological winding number should not receive quantum corrections.  This follows from dimensional analysis and topology.  Since all quantum corrections enter with additional factors of $g^2$, the corresponding contributions to the dimensionless linking number must involve some other dimensionful parameter. The only other scales available are the relative distances and sizes associated with $\C$ and $\S$.  However, these scales cannot appear, since our results are invariant under topology-preserving deformations.  Consequently,
all quantum corrections must be power divergent and can be absorbed into counterterms at all orders in perturbation theory.
Said another way, the conical deficit angle induced by $ U_\cen(\S)$ should not be renormalized.

\subsubsection{Explicit Examples}
\label{sec:Grav.examples}

To summarize, we have shown that the gravitational one-form symmetry operator implements a twisted Lorentz transformation, which is in turn equivalent to the insertion of a chiral cosmic string defect.  The symmetry operator can be treated as a  matter source which nonlinearly generates the classical one-point function of spin connection, which is then input into the line operator to yield the classical holonomy.  Hence, the classical holonomy  precisely reproduces the Ward identity for the line and symmetry operators.  Let us demonstrate how this works in some simple examples.

\paragraph{Symmetry Operator as Chiral Cosmic String in Flat Space}%
\noindent To begin, let us consider an initial background spacetime
given by flat Euclidean space $\M = \mathbb{R}^4$,
equipped with Cartesian coordinates $x^\m {\,=\,} (x,y,z,t)$ 
and the trivial tetrad
$\delta^A{}_\m = \diag(1,1,1,1)$.
In this background
the classical spin connection coefficients are zero and so the classical holonomy is trivial.

Next, let us induce a twist on this trivial configuration in a way that parallels the example from Yang-Mills theory
described in \Eq{eq:YM.exA}.  For concreteness, we assume that the Lorentz group is $G=\mathrm{SO}(4)$, so the center one-form symmetry is $Z(G) = \mathbb{Z}_2$, under which vector holonomies are charged.
Next, we define a twisted Lorentz transformation that 
generates a string singularity along the surface at $x{\,=\,}y{\,=\,}0$, so
\begin{align}
    \label{eq:Grav.Omega}
    \Omega =    
    \exp\bigg(\mem{
        \frac{k}{2}\mem
        c\mem \phi
    }\hem\bigg)
    \transition{where}
    e^{2\pi c} = \id
    \,,
\end{align}
where $k$ is an integer and
$\phi$
is the azimuthal angle
such that $\tan\phi = y/x$.
For example, the Lorentz reference vector can be chosen to be
\begin{align}   
    \label{eq:Grav.cfund}
    c^A{}_B
    = 
    \left({
    \begin{array}{cccc}   
        \hphantom{\mpad}\mathclap{
            0
        }\hphantom{\mpad}&\hphantom{\mpad}\mathclap{
            0
        }\hphantom{\mpad}&\hphantom{\mpad}\mathclap{
            0
        }\hphantom{\mpad}&\hphantom{\mpad}\mathclap{
            1
        }\hphantom{\mpad}
        \\
        \hphantom{\mpad}\mathclap{
            0
        }\hphantom{\mpad}&\hphantom{\mpad}\mathclap{
            0
        }\hphantom{\mpad}&\hphantom{\mpad}\mathclap{
            1
        }\hphantom{\mpad}&\hphantom{\mpad}\mathclap{
            0
        }\hphantom{\mpad}
        \\
        \hphantom{\mpad}\mathclap{
            0
        }\hphantom{\mpad}&\hphantom{\mpad}\mathclap{
            -1
        }\hphantom{\mpad}&\hphantom{\mpad}\mathclap{
            0
        }\hphantom{\mpad}&\hphantom{\mpad}\mathclap{
            0
        }\hphantom{\mpad}
        \\
        \hphantom{\mpad}\mathclap{
            -1
        }\hphantom{\mpad}&\hphantom{\mpad}\mathclap{
            0
        }\hphantom{\mpad}&\hphantom{\mpad}\mathclap{
            0
        }\hphantom{\mpad}&\hphantom{\mpad}\mathclap{
            0
        }\hphantom{\mpad}
    \end{array}
    }\right)
    \,,
\end{align}
which is shown in the vector representation of the Lorentz group.
Then,
according to \Eqs{eq:Grav.fr}{eq:Grav.Ftransf},
the spin connection and curvature transform to
\begin{align}
\begin{split}
    \label{eq:Grav.ex1-ewr}
        \omega^A{}_B &= (\Omega^{-1}d\Omega)^A{}_B
    =  \frac{k}{2} 
        \mem c^A{}_B\mem
        d\phi
    = \frac{k}{2} 
        \mem c^A{}_B\mem
        \frac{x\mem dy {\,-\,} y\mem dx}{x^2+y^2}
 \\
    R^A{}_B &= (\Omega^{-1}dd\Omega)^A{}_B
    = \frac{k}{2}
        \mem c^A{}_B\mem
        dd\phi
    = \frac{k}{2}\mem  
        2\pi c^A{}_B\mem
        \delta(x)\hem \delta(y)\mem dx \swedge dy
    \,,
\end{split}
\end{align}
where the latter
exhibits the chiral cosmic string defect whose worldsheet is in the $x$-$y$ plane.
We emphasize here that
the orientation of the Lorentz transformation $c^A{}_B$ is completely independent of
the direction that the string actually spans in spacetime.  In \Eq{eq:Grav.cfund}, we have arbitrarily chosen $c^A{}_B$ to act on the
$x$-$t$ and $y$-$z$ planes.  This is an essentially random choice---this construction exists for any $c^A{}_B$ that exponentiates properly to the identity, as stipulated in \Eq{eq:Grav.cfund}.  As we saw earlier, while this naively chooses a Lorentz violating reference vector, it is spurious because the reference drops out of the Ward identity for the symmetry operator.
Lastly, note that the intrinsic chirality of this construction is evident if we write these expressions in spinor notation, where $c^A{}_B$ splits into 
$\tilde{c}^\da{}_\db = -i\hem (\s_1)^\da{}_\db$
and $c_\a{}^\b = 0$.  Hence, this defect interacts with antichiral fermions, but not chiral fermions.  It is for this reason that we refer to the string itself as chiral.

Plugging in the background spin connection above into the holonomy, we obtain
\begin{align}
    \tr_{\text{vec}} \Pexp
    \bigg(\mem{
        \frac{k}{2}\mem c
        \int_0^{2\pi}\kern-0.2em d\phi
    }\hem\bigg)
    &= 4\hem (-1)^{k} 
    \,,
\end{align}
where the trace and exponentiation are performed in the vector representation.
From the factor of $(-1)^{k}$,
we immediately see that
this holonomy flips sign depending on the parity of $k$, which defines the number of windings.

It is not difficult to understand that
this sign factor
from the Wilson loop
describes
a conical deficit angle
quantized in the units of $\pi$.
Suppose a Lorentz vector $v^A$ is initialized to a value $v_0^A$ 
at a point on $\phi {\,=\,} 0$
and then parallel-transported
around the string.
The spin connection enters into the equation for parallel transport, 
$\dot{v}^A = -\omega^A{}_{B\r}\mem v^B\mem \dot{x}^\r$,
whose solution is
$v^A = (\Omega^{-1})^A{}_B\mem v^B_{0}$.
That is, the parallel transport of this vector is precisely implemented by a twisted Lorentz transformation.
As a result,
during a round trip
the vector experiences 
a succession of smooth rotations
that accumulates to
a net rotation of $e^{\pi k c} = \pm \id$,
corresponding to a deficit angle of $k\pi$.\footnote{
    One might ask whether the same result trivially follows 
    from solving $\dot{v}^\m = -\Gamma^\m{}_{\n\r}\mem v^\n\mem \dot{x}^\r$,
    which describes the
    parallel transport of a vector
    in terms of spacetime indices.
    The answer is no, since this equation computes the holonomy of the {\it Christoffel symbol}, which is a functional of the metric rather than the spin connection.  
    So as an operator 
    in the first-order formalism, the Christoffel holonomy    
    is simply not equal to the spin holonomy defined in \Eq{eq:Grav.W}. 
    There is no contradiction here:
    to implement the symmetry transformation of the Christoffel holonomy, one should determine the analog of the twisted Lorentz transformation for spacetime indices,
    which may be 
    possible
    in a metric-affine formulation of gravity.
}
This exactly describes how a conical deficit angle is measured
in classical gravitation.  For even $k$ the holonomy is trivial, while for odd $k$ the vector experiences a net $\pi$ rotation. 
 Note that this rotation is always {\it orientation-preserving}, since we consider the proper Lorentz group throughout.  Consequently, our results are consistent with \cite{McNamara:2022lrw}, which elegantly argues for the impossibility of {\it orientation-changing} defects.  Furthermore, we see that the chiral cosmic string differs from the traditional cosmic string, because the angle deficit is accumulated around an arbitrary axis $c^A{}_B$ which is totally unrelated to the actual orientation of the string in spacetime.

One can repeat this exercise for the case of spinor holonomy, which requires a Lorentz group $G=\textrm{Spin}(4)$, whose center one-form symmetry is $Z(G) = \mathbb{Z}_2 \times \mathbb{Z}_2$.  Like before, we apply a twisted Lorentz transformation of the form
of \Eq{eq:Grav.Omega}, except in the spinor representation.  In this case,
as noted earlier
\Eq{eq:Grav.cfund} translates to
\begin{align}   
    c_\a{}^\b = 0 
    \transition{and}  
    \tilde{c}^\da{}_\db = -i\hem (\s_1)^\da{}_\db
    \,.
\end{align}
It is then easy to compute the holonomies in the chiral and antichiral spinor representations,
\begin{align}
\begin{split}
    \tr_\sp
    \Pexp
    \bigg(\mem{
        \frac{k}{2}\mem c
        \int_0^{2\pi}\kern-0.2em d\phi
    }\hem\bigg)
    &= 2  \\
    \tr_\asp
    \Pexp
    \bigg(\mem{
        \frac{k}{2}\mem \tilde c
        \int_0^{2\pi}\kern-0.2em d\phi
    }\hem\bigg)
    &= 2\hem (-1)^{k} 
    \,,
\end{split}
\end{align}
so the antichiral spin holonomy is rephased by $(-1)^{k}$, while the chiral spin holonomy is invariant.
Thus the chiral cosmic string 
detects the parity of antichiral spinor indices.  Like before, these spin holonomies can be deduced from the parallel transport of fermions around a loop, described by
$\dot\psi_\alpha = -\omega_\alpha{}^\beta{}_\r\mem \psi_\beta\mem \dot{x}^\r$ 
and
$\dot\bpsi^{\dot\alpha} = \tilde\omega^\da{}_\wrap{\db\r}\mem \bpsi^\db\mem \dot{x}^\r$.  

\pagebreak

\paragraph{Symmetry Operator as Cosmic String in Black Hole Geometry}%
%
As a more general example,
we next consider the one-form symmetry operator in
 curved spacetime.
Conveniently, we can exploit well-known expressions from classical gravity in order to compute at all orders in perturbation theory in the gravitational constant.

Our starting point is
an
AdS-Schwarzschild background
in Boyer-Lindquist coordinates $x^\mu = (r,\theta,\phi,t)$.
The line element is given by
\begin{align}
\label{eq:Sch.line}
    ds^2
    = \frac{1}{f(r)^2}\mem dr^2 + r^2\mem d\theta^2 + r^2 \sin^2\theta\mem d\phi^2
    + f(r)^2\mem dt^2
    \,,
\end{align}
where we have denoted
\begin{align}
    \label{eq:Sch.f(r)}
     f(r) 
     = \sqrt{\mem{
        1 {\,-\,} \frac{2G_\text{N}M}{r} {\,+\,} \frac{r^2}{l^2}
     }\mem}
     \,,
\end{align}
where $M$ is the mass of the black hole
and $l$ is the AdS radius.

The goal of this analysis is to compute the classical spin holonomy in the presence or absence of the symmetry operator in order to verify 
the validity of the Ward identity 
of the one-form symmetry.
To be concrete, let us consider the spin holonomy for a circular loop $\C$ in the plane defined by $\theta=\pi/2$ and with constant radius $r=r_0$:
\begin{align}
    \label{eq:Sch.W_exp}
    {{W_\text{{\nem}vec}(\C)}}
    =  \tr_\text{vec} \Pexp
    \bigg(\hem{
        \int_0^{2\pi}  \omega_\phi(
            r_0
            ,
            \pi/2
            ,
            \phi
            ,
            0
        )\mem d\phi
    }\mem\bigg)
    \,.
\end{align}
The Ward identity for the gravitational one-form symmetry implies that this spin holonomy should
change its value
depending on whether or not we insert the symmetry operator. 


Firstly,
consider a pure AdS-Schwarzschild background in the absence of the symmetry operator.
Here the spin connection relevant to the holonomy in \Eq{eq:Sch.W_exp} is
\begin{align}
    \label{eq:Sch.w}
    \omega^A{}_{B\phi}(r,\pi/2,\phi,t)
    =
    \left(\,\,{
    \begin{array}{cccc}   
        \hphantom{\hpad}\mathclap{
            0
        }\hphantom{\hpad}&\hphantom{\hpad}\mathclap{
            0
        }\hphantom{\hpad}&\hphantom{\hpad}\mathclap{
            f(r)
        }\hphantom{\hpad}&\hphantom{\hpad}\mathclap{
            0
        }\hphantom{\hpad}
        \\
        \hphantom{\hpad}\mathclap{
            0
        }\hphantom{\hpad}&\hphantom{\hpad}\mathclap{
            0
        }\hphantom{\hpad}&\hphantom{\hpad}\mathclap{
            0
        }\hphantom{\hpad}&\hphantom{\hpad}\mathclap{
            0
        }\hphantom{\hpad}
        \\
        \hphantom{\hpad}\mathclap{
            -f(r)
        }\hphantom{\hpad}&\hphantom{\hpad}\mathclap{
            0
        }\hphantom{\hpad}&\hphantom{\hpad}\mathclap{
            0
        }\hphantom{\hpad}&\hphantom{\hpad}\mathclap{
            0
        }\hphantom{\hpad}
        \\
        \hphantom{\hpad}\mathclap{
            0
        }\hphantom{\hpad}&\hphantom{\hpad}\mathclap{
            0
        }\hphantom{\hpad}&\hphantom{\hpad}\mathclap{
            0
        }\hphantom{\hpad}&\hphantom{\hpad}\mathclap{
            0
        }\hphantom{\hpad}
    \end{array}
    }\nem\right)
    \,.
\end{align}
This is independent of $\phi$,
so the path ordering can be dropped.
Straightforward calculation shows that
the Wilson loop in the vector representation is given by
(see also \cite{Modanese:1993zh,Fredsted:2001rt})
\begin{align}
    \label{eq:Sch.W}
    {{W\hnem(\C)}}
    = 2 + 2 \cos(\hhem{2\pi f(r_0)}\hnem)
    \,.
\end{align}
%
In fact, 
the calculation can be done ``symbolically'' by
observing that
the spin connection component in \Eq{eq:Sch.w}
splits into self-dual and anti-self-dual parts as
\begin{align}
    \omega^A{}{}_{B\phi}(r,\pi/2,\phi,t)
    =
    \frac{(c_+ {\mem+\,} c_-)^A{}_B}{2}\mem f(r)
    \,,
\end{align}
where we have defined
\begin{align}
    \label{eq:Sch.cpm}
    (c_+)^A{}_B
    = 
    \left({
    \begin{array}{cccc}   
        \hphantom{\mpad}\mathclap{
            0
        }\hphantom{\mpad}&\hphantom{\mpad}\mathclap{
            0
        }\hphantom{\mpad}&\hphantom{\mpad}\mathclap{
            1
        }\hphantom{\mpad}&\hphantom{\mpad}\mathclap{
            0
        }\hphantom{\mpad}
        \\
        \hphantom{\mpad}\mathclap{
            0
        }\hphantom{\mpad}&\hphantom{\mpad}\mathclap{
            0
        }\hphantom{\mpad}&\hphantom{\mpad}\mathclap{
            0
        }\hphantom{\mpad}&\hphantom{\mpad}\mathclap{
            -1
        }\hphantom{\mpad}
        \\
        \hphantom{\mpad}\mathclap{
            -1
        }\hphantom{\mpad}&\hphantom{\mpad}\mathclap{
            0
        }\hphantom{\mpad}&\hphantom{\mpad}\mathclap{
            0
        }\hphantom{\mpad}&\hphantom{\mpad}\mathclap{
            0
        }\hphantom{\mpad}
        \\
        \hphantom{\mpad}\mathclap{
            0
        }\hphantom{\mpad}&\hphantom{\mpad}\mathclap{
            1
        }\hphantom{\mpad}&\hphantom{\mpad}\mathclap{
            0
        }\hphantom{\mpad}&\hphantom{\mpad}\mathclap{
            0
        }\hphantom{\mpad}
    \end{array}
    }\right)
    \transition{and}
    (c_-)^A{}_B
    = 
    \left({
    \begin{array}{cccc}   
        \hphantom{\mpad}\mathclap{
            0
        }\hphantom{\mpad}&\hphantom{\mpad}\mathclap{
            0
        }\hphantom{\mpad}&\hphantom{\mpad}\mathclap{
            1
        }\hphantom{\mpad}&\hphantom{\mpad}\mathclap{
            0
        }\hphantom{\mpad}
        \\
        \hphantom{\mpad}\mathclap{
            0
        }\hphantom{\mpad}&\hphantom{\mpad}\mathclap{
            0
        }\hphantom{\mpad}&\hphantom{\mpad}\mathclap{
            0
        }\hphantom{\mpad}&\hphantom{\mpad}\mathclap{
            1
        }\hphantom{\mpad}
        \\
        \hphantom{\mpad}\mathclap{
            -1
        }\hphantom{\mpad}&\hphantom{\mpad}\mathclap{
            0
        }\hphantom{\mpad}&\hphantom{\mpad}\mathclap{
            0
        }\hphantom{\mpad}&\hphantom{\mpad}\mathclap{
            0
        }\hphantom{\mpad}
        \\
        \hphantom{\mpad}\mathclap{
            0
        }\hphantom{\mpad}&\hphantom{\mpad}\mathclap{
            -1
        }\hphantom{\mpad}&\hphantom{\mpad}\mathclap{
            0
        }\hphantom{\mpad}&\hphantom{\mpad}\mathclap{
            0
        }\hphantom{\mpad}
    \end{array}
    }\right)
    \,.
\end{align}
The calculation is simplified once one observes that
these Lorentz generators commute.

\begin{figure}
    \centering
    \includegraphics[scale=1.2,valign=c]{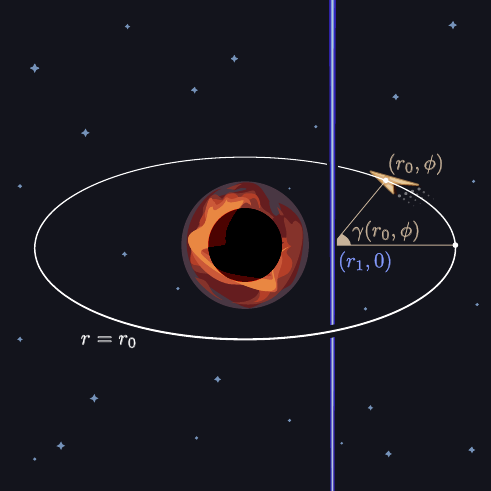}
    \vskip5pt
    \caption{
        A spaceship is in circular orbit about a black hole adjacent to a chiral cosmic string.
        On board, a spinning top
        experiences an additional $\pi$ rotation of its spin per round trip, as compared to without the string.
    }
    \label{fig:spaceship}
\end{figure}

Next,
we compute the same spin holonomy
but with the insertion of a symmetry operator $U_\cen(\S)$.
As depicted in \Fig{fig:spaceship},
we take $\S$ to define the worldsheet of a chiral cosmic string spanning the azimuthal direction. 
Without loss of generality, let us take the string to intersect the plane $\theta=\pi/2$
at $(r,\phi) = (r_1,0)$.
The associated twisted Lorentz transformation is,
for example,
\begin{align}
    \label{eq:Sch.Omega}
    \Omega
    = 
    \exp\bigg(\mem{
        \frac{c}{2}\mem
        \gamma(r,\phi)
    }\hem\bigg)
    \transition{where}
    e^{2\pi c} = \id
    \,,
\end{align}
which is simply an instance of \Eq{eq:Grav.Omega}
with $k = 1$ winding.
Here $\gamma(r,\phi)$ describes the ``apparent'' azimuthal angle measured from the string at $(r,\phi) = (r_1,0)$,
so concretely,
\begin{align}
    \tan
    \gamma(r,\phi)
    = 
        \frac{r \sin\phi}{r \cos\phi {\,-\,} r_1}
    \,.
\end{align}
In principle, we can choose any generator for $c$
to demonstrate the transformation of the Wilson loop.
For an explicit check, let us work out a simple case
in which the multivalued transformation parameter $\Omega$
commutes with
the spin connection component in \Eq{eq:Sch.w}.
Namely, we take
the Lorentz generator in \Eq{eq:Sch.Omega} to be
the self-dual generator in \Eq{eq:Sch.cpm}, so
$c^A{}_B = (c_+)^A{}_B$.
By explicit
calculation we find that
the spin connection transforms according to
\begin{align}
\begin{split}
    \label{eq:Sch.w_transf}
    \omega^A{}_{B\phi}(r,\pi/2,\phi,t)
    = \frac{(c_+ {\mem+\,} c_-)^A{}_B}{2}\mem f(r)
        \,\,\,\mapsto\,\,\,
    \frac{(c_+ {\mem+\,} c_-)^A{}_B}{2}\mem f(r)
    +
        \frac{(c_+)^A{}_B}{2}\mem \frac{\partial\gamma}{\partial\phi}(r,\phi)
    \,,
\end{split}
\end{align}
where we have used that $(c_+)^A{}_B$ commutes with $(c_-)^A{}_B$.
By construction 
the
path ordering of this transformed spin connection 
is trivial
and can be dropped.
Finally, we find that the transformed spin holonomy is
\begin{align}
\begin{split}
    \label{eq:Sch.Wnew}
    {{W_\text{{\nem}vec}(\C)}}
        \,\,\,\mapsto\,\,\,
    &
        \tr \exp\bigg(\mem{
            \frac{c_+ {\mem+\,} c_-}{2}\mem
            2\pi f(r_0)
            + 
            \frac{c_+}{2}\mem 
            2\pi n
        }\mem\bigg)
    \,,\\
    &=
        2 \cos
            n\pi
        +
        2 \cos\nem \Big(\mem{
            2\pi f(r_0) {\,+\,}
            n\pi
        }\mem\Big)
    \,,\\
    &=
        \Big(\mem{
            2
            {\,+\,}
            2 \cos(\hhem{2\pi f(r_0)}\hnem)
        }\Big)
        \cdot (-1)^n
    \,,\\
    &=
        {{W_\text{{\nem}vec}(\C)}}
        \cdot (-1)^n
    \,,
\end{split}
\end{align}
where $n$ is linking number
measuring the number of times
$\C$ winds about $\S$:
\begin{align}
    n
    \,=
    \frac{1}{2\pi}
    \int_0^{2\pi}
        \frac{\partial\gamma}{\partial\phi}(r,\phi)
        \mem d\phi
    \,=\mem
    \Bigg\{
    \begin{aligned}[c]
        \,\,
            1
        &\quad
            \text{if} \quad  r_1 < r_0
        \,,\\
        \,\,
            0
        &\quad
            \text{if} \quad  r_1 > r_0
        \,.
    \end{aligned}
\end{align}
Note how each term in \Eq{eq:Sch.W} gets
``twisted'' to
each term in \Eq{eq:Sch.Wnew},
by the addition of $n\pi$ inside the argument of cosines.
In conclusion, we find that the twisted local Lorentz transformation, which corresponds to an insertion of the operator $U_\cen(\S)$, 
flips the sign of the spin holonomy 
when the contour $\C$ and the surface $\S$ are linked.
This verifies the Ward identity for the one-form symmetry 
in a curved background.

More generally, it is possible to choose the twisted Lorentz transformation such that the twisted spin connection does not commute with itself at different points.
In this case the path ordering in the spin holonomy is much more difficult to compute directly and so we resort to numerical methods.  
%
%
For example,
let us consider an arbitrary self-dual generator parametrized by
a unit three-vector
$(n_1,n_2,n_3)$:
\begin{align}
\label{eq:Sch.lambda2}
    c^A{}_B
    \,= 
    \left({
    \begin{array}{cccc}   
        \hphantom{\wpad}\mathclap{
            0
        }\hphantom{\wpad}&\hphantom{\wpad}\mathclap{
            n_3
        }\hphantom{\wpad}&\hphantom{\wpad}\mathclap{
            -n_2
        }\hphantom{\wpad}&\hphantom{\wpad}\mathclap{
            n_1
        }\hphantom{\wpad}
        \\
        \hphantom{\wpad}\mathclap{
            -n_3
        }\hphantom{\wpad}&\hphantom{\wpad}\mathclap{
            0
        }\hphantom{\wpad}&\hphantom{\wpad}\mathclap{
            n_1
        }\hphantom{\wpad}&\hphantom{\wpad}\mathclap{
            n_2
        }\hphantom{\wpad}
        \\
        \hphantom{\wpad}\mathclap{
            n_2
        }\hphantom{\wpad}&\hphantom{\wpad}\mathclap{
            -n_1
        }\hphantom{\wpad}&\hphantom{\wpad}\mathclap{
            0
        }\hphantom{\wpad}&\hphantom{\wpad}\mathclap{
            n_3
        }\hphantom{\wpad}
        \\
        \hphantom{\wpad}\mathclap{
            -n_1
        }\hphantom{\wpad}&\hphantom{\wpad}\mathclap{
            -n_2
        }\hphantom{\wpad}&\hphantom{\wpad}\mathclap{
            -n_3
        }\hphantom{\wpad}&\hphantom{\wpad}\mathclap{
            0
        }\hphantom{\wpad}
    \end{array}
    }\right)
    =\,
        n_1\mem (t_1^+)^A{}_B
        +
        n_2\mem (t_2^+)^A{}_B
        +
        n_3\mem (t_3^+)^A{}_B
    \,.
\end{align} 
In this case the twisted spin connection is given by
\begin{align}
    & 
    \left({
    \begin{array}{cccc}   
        \hphantom{\mpad}\mathclap{
            0
        }\hphantom{\mpad}&\hphantom{\mpad}\mathclap{
            0
        }\hphantom{\mpad}&\hphantom{\mpad}\mathclap{
            1
        }\hphantom{\mpad}&\hphantom{\mpad}\mathclap{
            0
        }\hphantom{\mpad}
        \\
        \hphantom{\mpad}\mathclap{
            0
        }\hphantom{\mpad}&\hphantom{\mpad}\mathclap{
            0
        }\hphantom{\mpad}&\hphantom{\mpad}\mathclap{
            0
        }\hphantom{\mpad}&\hphantom{\mpad}\mathclap{
            1
        }\hphantom{\mpad}
        \\
        \hphantom{\mpad}\mathclap{
            -1
        }\hphantom{\mpad}&\hphantom{\mpad}\mathclap{
            0
        }\hphantom{\mpad}&\hphantom{\mpad}\mathclap{
            0
        }\hphantom{\mpad}&\hphantom{\mpad}\mathclap{
            0
        }\hphantom{\mpad}
        \\
        \hphantom{\mpad}\mathclap{
            0
        }\hphantom{\mpad}&\hphantom{\mpad}\mathclap{
            -1
        }\hphantom{\mpad}&\hphantom{\mpad}\mathclap{
            0
        }\hphantom{\mpad}&\hphantom{\mpad}\mathclap{
            0
        }\hphantom{\mpad}
    \end{array}
    }\right)
    \frac{f(r)}{2}
    -
    \left({
    \begin{aligned}[c]
        {}&{}
        \Big(\mem{
            n_2n_1\mem \texttt{s}^2
            - n_3\mem \texttt{s}\texttt{c}
        }\mem\Big)
        \mem (t_1^+)^A{}_B
        \\
        {}+
        {}&{}
        \Big(\mem{
            \tfrac{1}{2}
            - (1{\mem-\,}{n_2}^2)\mem \texttt{s}^2
        }\mem\Big)
        \mem (t_2^+)^A{}_B
        \\
        {}+
        {}&{}
        \Big(\mem{
            n_2n_3\mem \texttt{s}^2
            +
            n_1\mem \texttt{s}\texttt{c}
        }\mem\Big)
        \mem (t_3^+)^A{}_B
    \end{aligned}
    }\right)
    f(r)
    + \frac{c^A{}_B}{2}\mem
        \frac{\partial\gamma}{\partial\phi}(r,\phi)
    \,,
\end{align}
where 
$\texttt{s} = \sin(\gamma(r,\phi)/2)$,
$\texttt{c} = \cos(\gamma(r,\phi)/2)$.
By evaluating the path-ordered exponential numerically, we can verify explicitly the final line of \Eq{eq:Sch.Wnew} for this more general case.

In the above examples
we have always assumed that
the holonomy contour and the chiral cosmic string are separated from the black hole horizon.
Since the divergence of the line element in \eqref{eq:Sch.line}
at the horizon is a coordinate singularity,
the loop and the string can actually be placed across or inside 
the horizon.
It would be amusing to compute the holonomy in these more exotic configurations
using coordinate systems
in which the metric is regular on the horizon.


\subsection{Canonical Formalism}
\label{sec:CanGrav}

Last but not least, let us now analyze the one-form symmetry of gravity from the point of view of the Hamiltonian formalism.
The relevant framework
is the nonchiral or real version of the
Ashtekar \cite{ashtekar1986new} formulation,
as studied in
\cite{Buffenoir:2004vx,Alexandrov:2008fs,Celada:2012ua,Celada:2016bf}.


\subsubsection{Phase Space}
\label{sec:CanGrav.ps}

In \Eq{eq:eGrav.L}
we described the Lagrangian for tetradic Palatini gravity, 
expressed in terms of the Pleba\'nski two-form field.  With all indices written out explicitly, this  Lagrangian reads
\begin{align}
\begin{split}
    \label{eq:eGrav.compL}
    \frac{1}{g^2}\mem
    \bigg[\mem{
         \frac{1}{2}\mem
            B_{a\m\n}
            \Big({
                \partial_\r\mem \omega^a{}_\s
                + 
                \minie\mem
                f^a{}_{bc}\mem \omega^b{}_\r\mem \omega^c{}_\s
            }\Big)\mem
            \e^{\m\n\r\s}
        -
        \frac{\Lambda}{6}\mem
            B_{a\m\n}\hem  {\star B}^a{}_{\r\s}
            \mem\e^{\m\n\r\s}
    }\mem\bigg]
    \,.
\end{split}
\end{align}
Carrying out the $3{\,+\,}1$ decomposition,
we immediately find that 
the dynamical degrees of freedom
coordinatizing the
phase space 
are $\omega^a{}_i$
together with their canonical conjugates,
\begin{align}
    \label{eq:Edef-re}
    E^i{}_a
    = \frac{1}{2}\mem B_{a jk}\mem \e^{ijk}
    \,.
\end{align}
Upon quantization, the canonical commutation relations are then
\begin{align}
\begin{split}
    \label{eq:Grav.ccr}
    [ \omega^a{}_i(x) , \omega^b{}_j(x') ] &= 0 
    \,,\\
    [ E^i{}_a(x) , \omega^b{}_j(x') ] &= g^2\, \delta^i{}_j\,\delta^b{}_a\, \delta^{(3)}\hnem(x{\,-\,}x') 
    \,,\\
    [ E^i{}_a(x) , E^j{}_b(x') ] &= 0
    \,,
\end{split}
\end{align}
where $x,x' \in \M_3$ are points in the spatial manifold.
As expected,
\Eqs{eq:Edef-re}{eq:Grav.ccr}
exactly mirror \Eqs{eq:Edef}{eq:YM.ccr} from gauge theory, which is indeed the entire point of the Ashtekar formulation \cite{ashtekar1986new,Ashtekar:2020xll}.

We restrict our focus to
these canonical commutation relations,
as they are the only crucial element
for proving the one-form symmetry in Hamiltonian framework
as we learned in \Sec{sec:CanYM}.
It is known that
integrating out the nondynamical degrees of freedom
eventually leads to
a well-posed constrained Hamiltonian system
equipped with several constraints
\cite{Buffenoir:2004vx,Alexandrov:2008fs,Celada:2012ua,Celada:2016bf},
such as
the Gauss constraints for local Lorentz transformations
as well as
the Hamiltonian and diffeomorphism constraints
familiar from the ADM \cite{adm} analysis.%
\footnote{
    In particular, one can start from the Pleba\'nski Lagrangian \cite{Plebanski:1977zz},
    where the Pleba\'nski two-form in \Eq{eq:pleb-def}
    is taken as a fundamental degree of freedom
    rather than a composite field constructed from  the tetrad.
    In this case, the $3{\,+\,}1$ decomposition
    naturally gives
    the phase space $(\omega^a{}_i , E^i{}_a)$,
    where the variable $E^i{}_a$ is not composite.
}

\subsubsection{Ward Identity}
\label{sec:CanGrav.ward}

Following the logic of gauge theory, we now show how the one-form symmetry of gravity is implemented in the Hamiltonian formulation. 
Like before, we consider $W_\rep(\C)$ oriented in a spatial slice at $t{\,=\,}0$ and linked with $U_\cen(\S)$.
Again, we pancake $\S$ onto the spatial slice so that it becomes the disjoint union of two discs, defined by $\S =\S_+ {\mem\cup\mem} (-\S_-)$.  Here
$\S_+$ and $-\S_-$ are two halves of a squashed sphere which straddle the spatial slice,
as depicted in \Fig{fig:pancake1}.  These discs are eventually projected down 
to a surface $\S_0$ in the time slice $t{\,=\,}0$.

The symmetry operator factorizes into 
$U_\cen(\S) = 
U_\cen(\S_+)\mem U_\cen(-\S_-)$, so the symmetry transformation becomes the equal-time operator equation,
\begin{align}
    \label{eq:pancakedWard}
    U_\cen(\S_0)\, W_\rep(\C)\, \Uinv(\S_0)
    \mem=\mem 
    \rho(\alpha)^{\Intersect_3(\C,\S_0)}\, W_\rep(\C)
    \,,
\end{align}
where $\S_0$ and $\C$ all belong to the spatial three-dimensional manifold $\M_3$.
Here $U_\cen(\S_0)$ is the avatar of the symmetry operator
in the Hamiltonian framework,
taking a two-dimensional surface $\S_0$ with boundary
as its support.
Applying the $3{\,+\,}1$  decomposition described in the previous section, 
the symmetry operator becomes
\begin{align}
    \label{eq:Grav.Q}
    U_\cen(\S_0)
    = \exp\bigg(\mem{
        \frac{\pi}{g^2}
        \int_{\S_0} dx^j \swedge dx^k\, \e_{ijk}\mem E^i{}_a\mem \lambda^a
    }\bigg)
    \transition{where}
    e^{2\pi\lambda} = \cen
    \,,
\end{align}
which exhibits the expected dependence on $E^i{}_a$ 
as
the variable conjugate to the spin connection $\omega^a{}_i$.
Using the canonical commutation relations in \Eq{eq:Grav.ccr},
the left-hand side of
\Eq{eq:pancakedWard} 
can be evaluated as
\begin{align}
\begin{split}    
    \label{eq:QadjW-re}
    U_\cen(\S_0)\mem W_\rep(\C)\mem \Uinv(\S_0)
    \mem&=\mem
    \tr_\rep \Pexp \bigg(\mem{
            \oint_\C \omega
            +
            2\pi\lambda\, \delta_3(\S_0)
    }\bigg)
    \mem=\mem
    \rho(\alpha)^{\Intersect_3(\C,\S_0)}\mem
    W_\rep(\mathcal{C})
    \,,
\end{split}
\end{align}
thus establishing the Hamiltonian version of the Ward identity,
introduced
in \Eq{eq:operatorWard}
and reproduced in \Eq{eq:pancakedWard}. 
Note the near isomorphism between
\Eqs{eq:Grav.Q}{eq:QadjW-re} for gravity
and
\Eqs{eq:YM.Q}{eq:QadjW} for gauge theory.
Indeed, this parallel between
Yang-Mills and gravity
has been the central insight in our construction of the gravitational one-form symmetry. 
It also
strongly
aligns with
the broader philosophy of the
Ashtekar formulation, which is that Yang-Mills and gravity are formulated in essentially the same phase space.

We have taken as our starting point the definition of the spin holonomy in \Eq{eq:Grav.W} and the symmetry operator in \Eq{eq:Grav.U}
and shown how these form the ingredients of a gravitational one-form symmetry.
However, it is amusing that the reverse logic is actually possible. 
Starting from the spin holonomy, we can actually {\it deduce} the symmetry operator from first principles,
as the line and charge operators 
are necessarily integrals of
conjugate phase space variables.
Specifically,
this is required in order for their commutation relations to yield delta functions that eventually integrate to become field-independent linking numbers. 
Since $E^i{}_a$ is conjugate to the spin connection $\omega^a{}_i$, we see that \Eq{eq:Grav.Q} 
and its covariant counterpart
were actually inevitable.
Namely, this provides an alternative argument for the definition of the symmetry operator in \Eq{eq:Grav.U}.

\subsection{Symmetry Breaking}
\label{sec:Breaking}

As is well-known, global higher-form symmetries can be broken, either explicitly or spontaneously.  
In analogy with Yang-Mills theory, the one-form symmetry of gravity is explicitly broken in the presence of matter fields that transform nontrivially under the Lorentz group.  For example, if the theory includes a local operator in the spin representation $\rep$, then it is possible to define a Lorentz invariant line holonomy  $W_{\rep}(\C)$ for a contour $\C$ that is not closed, but rather terminates on this local operator.  The spin holonomy is then ``endable,'' so it can be unlinked topologically from the symmetry operator $U_\cen(\S)$, and the one-form symmetry is explicitly broken.  The physical interpretation of this phenomenon is that the spin holonomy is screened by spinning particles.

\begin{figure}[t]
    \centering
    \begin{align*}
        {\renewcommand{\arraystretch}{1.6}
        \renewcommand{\arraycolsep}{1.0em}
        \begin{array}{cc}
            \includegraphics[valign=c,scale=1.1]{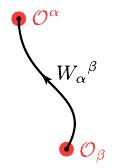}
            &
            \includegraphics[valign=c,scale=1.1]{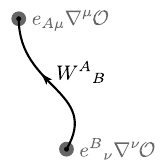}
        \end{array}
        }
    \end{align*}
    \caption{
        Spinor holonomy is screened by fermions.
        Vector holonomy is screened by orbital angular momentum.
    }
    \label{fig:screen}
\end{figure}

Interestingly, this implies that the gravitational one-form symmetry is explicitly broken by particles with spin.  
For example, if the Lorentz group is $G={\rm Spin}(4)$, then a holonomy in the spinor representation can only end on a local operator ${\cal O }_\alpha$ with a free spinor index.  Hence, the corresponding holonomy is endable if there exist fermions in the spectrum.
The case of $G= {\rm SO}(4)$ is more subtle, however. 
A holonomy in the vector representation of the Lorentz group must end on an operator with a free Lorentz vector index,
which can be 
any operator carrying orbital angular momentum
dotted with a tetrad 
such as
$e^A{}_\m \nabla^\m\mathcal{O}$.\footnote{
    A more subtle question arises in formulations of gravity with differences in field content.
    For example, in Pleba\'nski theory
    \cite{Plebanski:1977zz},
    the associated two-form field $B$ is a fundamental degree of freedom.
    The tetrad is sculpted from $B$ through a constraint 
    \cite{Buffenoir:2004vx,Celada:2016bf},
    while the metric can be expressed in terms of $B$ in closed form \cite{Urbantke:1984eb,Capovilla:1991qb}.
    Consequently, there is no field in the pure gravity sector that carries a single vector index, so it is unclear whether the vector spin holonomy is endable in this case.
}  Consequently, the vector holonomy is analogous to the adjoint Wilson loop of gauge theory, which is automatically screened by dynamical gluons.  See \Fig{fig:screen} for a depiction of this phenomenon.

The above logic has implications for the standard model of physics.  Since fermions exist, we know that the Lorentz group is ${\rm SL}(2,\mathbb{C})$, which implies that there is a $\mathbb{Z}_2$ one-form gravitational symmetry under which spinor holonomies are charged.
Presuming the lightest neutrino is not massless, this one-form symmetry is unbroken below that scale.  This implies a new, albeit subtle, exact symmetry in the known laws of physics.


Notably, the explicit breaking of higher-form symmetry is strongly suggested by the so-called swampland conjectures.  In particular, there is strong evidence that all global symmetries are necessarily broken at some scale in a consistent theory of quantum gravity \cite{Misner:1957mt,Polchinski:2003bq,Banks:2010zn,Harlow:2018tng}.  A well-known avatar of this is the weak gravity conjecture \cite{Arkani-Hamed:2006emk,Harlow:2022ich}, which states that a $U(1)$ gauge theory must exhibit a state whose charge exceeds its mass in Planck units.  The weak gravity conjecture quantitatively forbids the strict global symmetry limit of vanishing charge in a $U(1)$ gauge theory.  

The swampland conjectures imply that any  higher-form symmetry should be either gauged or explicitly broken.  In the case of gauge theory coupled to gravity, the latter scenario requires the existence of a tower of charged states, as described  by the so-called completeness conjectures \cite{Heidenreich:2021xpr}.
Applying the same logic to dynamical gravity, we expect that something similar applies to the gravitational one-form symmetry.  One option is that this symmetry is gauged, for example as would occur if the Lorentz group is $\mathrm{Spin}(4)/{\mathbb{Z}_2}\times {\mathbb{Z}_2}$. Alternatively, if the Lorentz group is $\mathrm{Spin}(4)$, then the one-form symmetry is not gauged and must be  explicitly broken, thus implying the existence of fermions in the spectrum.

Finally, let us speculate briefly on the possibility of phases in gravity.  Taking inspiration from gauge theory, it is natural to 
wonder whether the expectation value of the spin holonomy is an order parameter for symmetry breaking.  In the case of gauge theory, it is well-known that an area versus perimeter law scaling of the Wilson loop expectation value is a diagnostic of whether or not a theory is confining.  However, the analogous construction in gravity is far murkier.  In particular, 
 as we noted earlier, the diffeomorphism invariance of dynamical gravity suggests that the contour of the spin holonomy---and any contour, actually---must be defined relationally with respect to some invariant boundary data.  So to be an  order parameter, the spin holonomy must presumably be computed for a contour that circumscribes the boundary.

Even ignoring these subtleties, the fact that the effective field theory description of gravity is intrinsically weakly coupled suggests that confinement is not in play.
More generally it is very unclear whether a low-energy effective theory of gravity on a fixed background could even access different phases, or what that would even mean.  One speculation is that this might have something to do with degenerate configurations of the metric and their corresponding domain walls \cite{Jacobson:1992xy,Romano:1993bj,Bengtsson:1997wr,Volovik:1999vb}.  Another possibility is that a putative gravitational phase diagram might delineate various choices of compactification or of asymptotic behavior of the metric.  Indeed, it is easy to see that the spin holonomy is highly sensitive to the cosmological constant.
For these reasons, it would be interesting to explicitly compute the expectation value of the spin holonomy in various examples. A number of existing works have calculated the spin holonomy in various contexts \cite{Modanese:1993zh,Modanese:1991nh,Fredsted:2001rt,Alawadhi:2021uie,Jacobson:1992ya,Donoghue:2016vck,Brandhuber:2008tf}. 


\section{Future Directions}
\label{sec:Future}

In this paper, we have initiated an exploration of generalized symmetry in the context of dynamical gravity.  Taking our cues from the one-form symmetries of Yang-Mills theory, we have considered gravity in the tetradic Palatini formalism, which is a gauge theory of the local Lorentz group. We have argued that the gravitational one-form symmetry is defined by the center of the Lorentz group.  The  object which is charged under this symmetry is the spin holonomy $W_\rep(\C)$.  The one-form symmetry transformation is implemented by an operator $U_\cen(\S)$, which has dual interpretations, both as a twisted Lorentz transformation but also as a chiral cosmic string defect carrying both electric and magnetic gravitational charge. The topological linking of the line and symmetry operators corresponds to the measurement of a quantized conical deficit angle by the spin holonomy.  In the standard model, this implies the existence of a new symmetry below the mass of the lightest neutrino.
The present work leaves numerous avenues for future exploration, which we now describe.

First and foremost are a number of very simple extensions of this work which should be relatively straightforward.  These include the question of generalization to higher spacetime dimension, which offers a richer spectrum of one-form symmetry groups.
For example, the center group is maximally $Z({\rm Spin}(4)\hnem) = \mathbb{Z}_2 {\mem\times\mem} \mathbb{Z}_2$ in four dimensions, but this grows to $Z({\rm Spin}(6)\hnem) = \mathbb{Z}_4$ in six dimensions.  Another concrete direction is the inclusion of gravitational higher-curvature corrections.  These contributions will clearly preserve the one-form symmetry, whose corresponding symmetry operator will be equal to the surface integral of the canonical conjugate of the curvature in the effective field theory.

Secondly, while the present work has focused on gravitational one-form symmetry of electric type, it is natural to ask whether one can derive an analogous magnetic construction.  For gauge theories,
the electric and magnetic symmetry operators are 
straightforwardly
related by 
the spacetime Hodge duality on the field strength
\cite{Schafer-Nameki:2023jdn},
which is computed
with respect to a \textit{background} volume form.
For gravity,
the analogous procedure necessarily introduces metric dependence,
since the Levi-Civita permutation symbol is a {\it density}.
This additional metric dependence implies that the putative magnetic symmetry operator is
no longer the proper integral of a form.     It is unclear whether this technical obstruction is insurmountable or merely peculiar.  Meanwhile, recent work on cobordism classes in quantum gravity \cite{McNamara:2019rup} appears to construct certain magnetic gravitational defects.  Perhaps a direct link can be drawn between those results and the approach taken here.


A third topic of future study is
higher-group symmetry,
which describes a certain nonabelian structure built from the fusion of  multiple higher-form symmetries   \cite{Kapustin:2013uxa,Benini:2018reh,Cordova:2018cvg,Cordova:2020tij}.
A well-studied example of this is axion-Yang-Mills theory, which exhibits a two-form symmetry for the axion, together with the one-form symmetry of the gauge theory \cite{Brennan:2020ehu}.
These symmetries fuse to yield a two-group structure, which in fact bounds the scale of axion strings from below by the lightest particle in the fundamental of color.
Acquainted with this remarkable fact, 
it would be interesting to investigate
if gravity can also exhibit a higher-group symmetry.
It seems quite likely that a similar two-group symmetry will appear in gravity coupled to an axion, in which case we should expect that the axion string scale is  bounded by the mass of the lightest fermion.

\begin{figure}[t]
    \centering
    \begin{align*}
        {\renewcommand{\arraystretch}{1.6}
        \renewcommand{\arraycolsep}{0.6em}
        \begin{array}{ccc}
            \includegraphics[valign=c,width=0.3\linewidth]{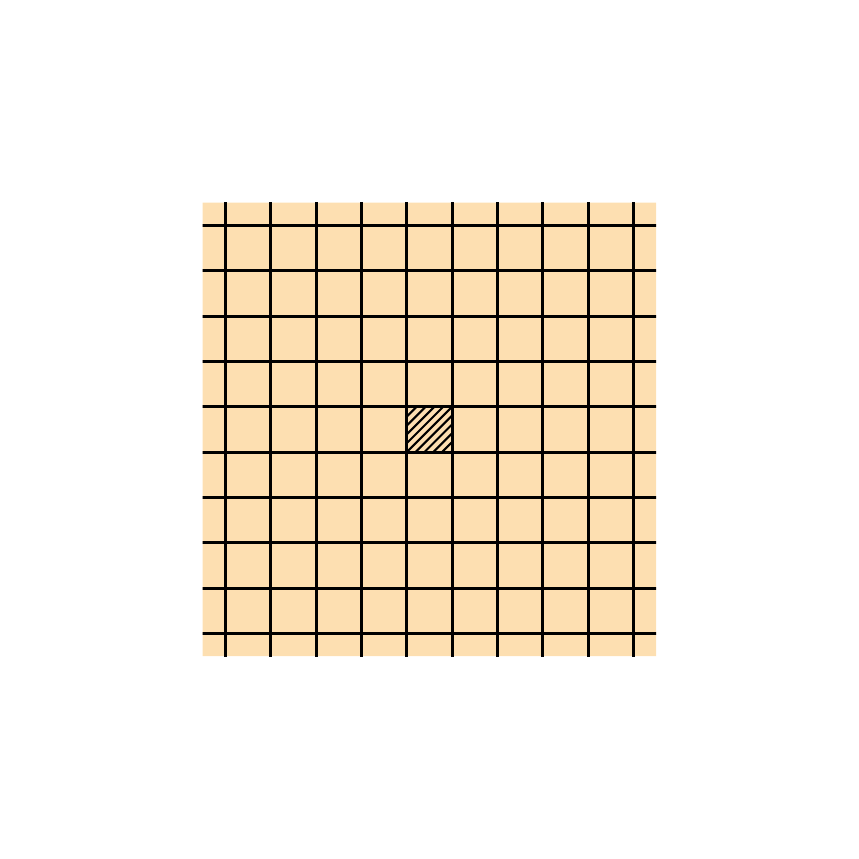}
            &
            \xrightarrow[\displaystyle\text{
            }]{\footnotesize\text{
                \,
                twisted diffeo
                \,
            }}
            &
            \includegraphics[valign=c,width=0.3\linewidth]{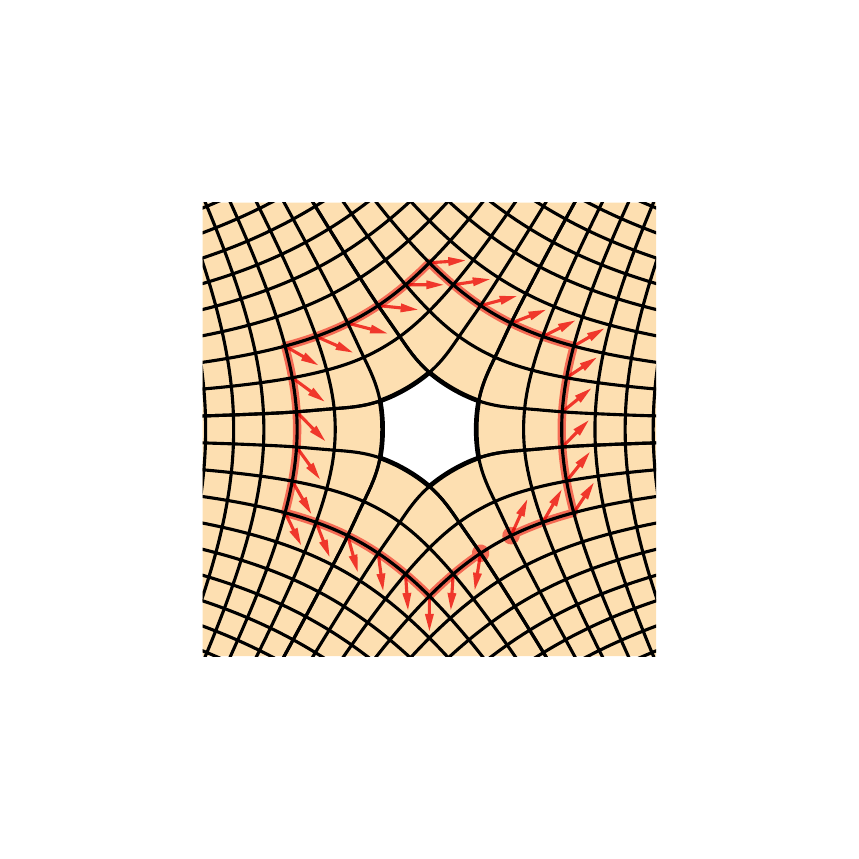}
            \\
            \adjustbox{scale=0.9,valign=c}{
                ``Empty Space''
            }
            &
            &
            \adjustbox{scale=0.9,valign=c}{
                ``Cosmic String''
            }
        \end{array}
        }
    \end{align*}
    \caption{
        A diffeomorphism with a multivalued Jacobian
        creates a cosmic string from empty space,
        analogous to disinclinations in lattice systems.
    }
    \label{fig:cosmic}
\end{figure}

Fourthly, while our approach of treating gravity as a {\BF}-type theory yields a simple route to the one-form symmetry, there is  the question of how this framework is explicitly realized in other physically equivalent formulations of gravity \cite{krasnov2020formulations} in the literature.
%
For example,
it is known that
in the presence of a nonzero cosmological constant
one can integrate out the tetrad altogether, 
yielding a theory of gravity described purely in terms of the spin connection \cite{Capovilla:1991kx,Krasnov:2021zen}.
It would be interesting to understand the emergence of the gravitational one-form symmetry in this ``pure connection'' formalism given that all of our results were derived in the presence of a cosmological constant.  Another open question is the fate of the one-form symmetry in gravitational formulations with subtly different field content,
as referred to in a footnote in \Sec{sec:Breaking}.

A fifth area of study concerns the question of what topological symmetry can  teach us about classical gravitation.
As a theory of spacetime geometry,
gravity boasts a rich array of classical vacuum solutions,
each showcasing distinctive \textit{singularity structures} that are themselves a focal point of study
\cite{penrose1965gravitational}.
In particular, it could be very illuminating to 
initiate
a systematic analysis and classification of gravitational singularities 
from the perspective of topological operators and
their algebra.
For example, even in the absence of spin holonomy, the
the one-form symmetry operator can link with  ``holes'' in spacetime.
It would be interesting to study whether there is
physical information
encoded in
such a linking, for example regarding the singularity of
a black hole.
Moreover, 
it is conceivable that known spacetime singularities in the literature have an alternative interpretation as symmetry operators, like we discovered for the chiral cosmic string.

\begin{figure}[t]
    \centering
    \begin{align*}
        {\renewcommand{\arraystretch}{1.6}
        \renewcommand{\arraycolsep}{0.6em}
        \begin{array}{ccc}
            \includegraphics[valign=c,width=0.36\linewidth]{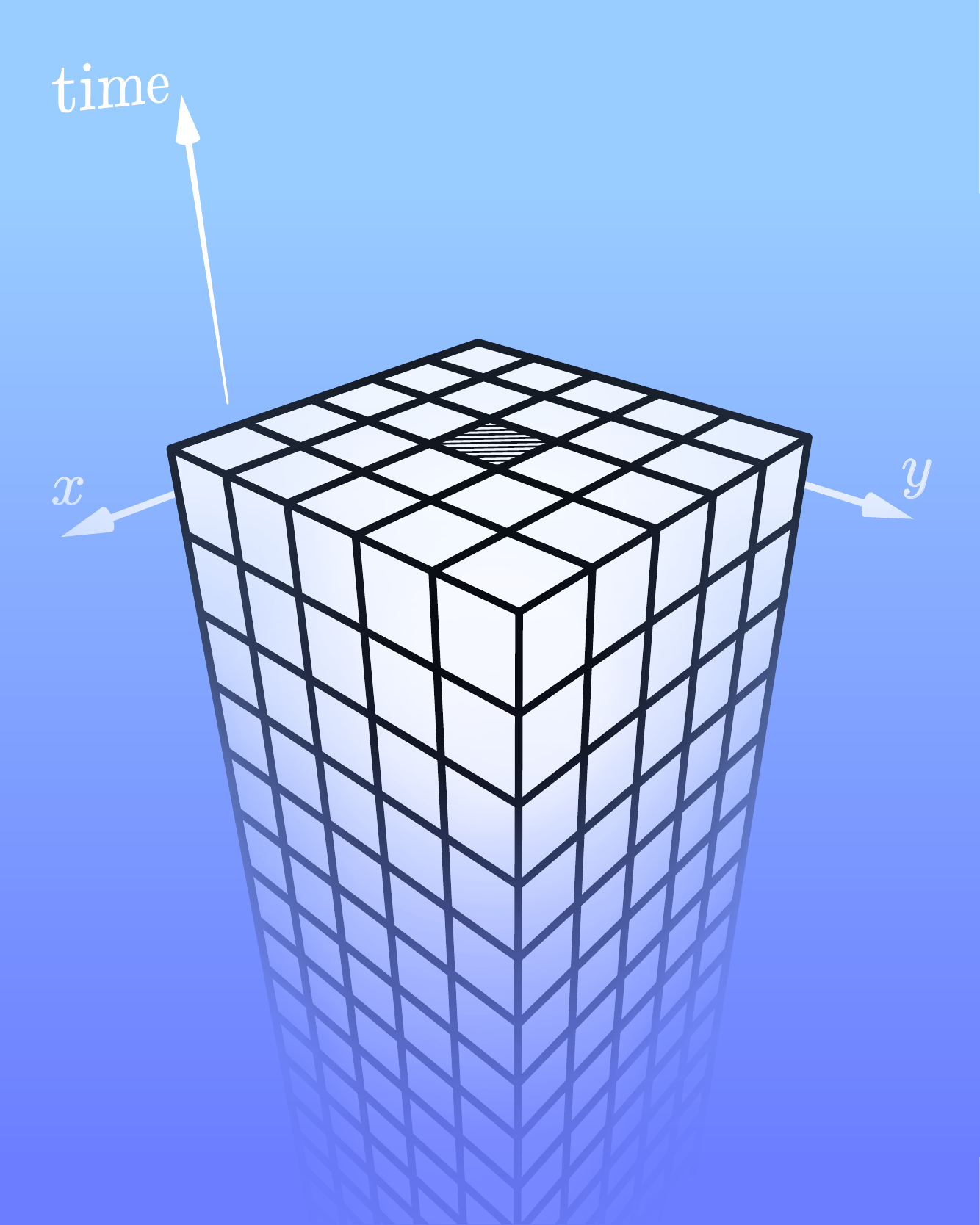}
            &
            \xrightarrow[\displaystyle\text{
            }]{
                \text{
                \,
               twisted diffeo
                \,
            }}
            &
            \includegraphics[valign=c,width=0.36\linewidth]{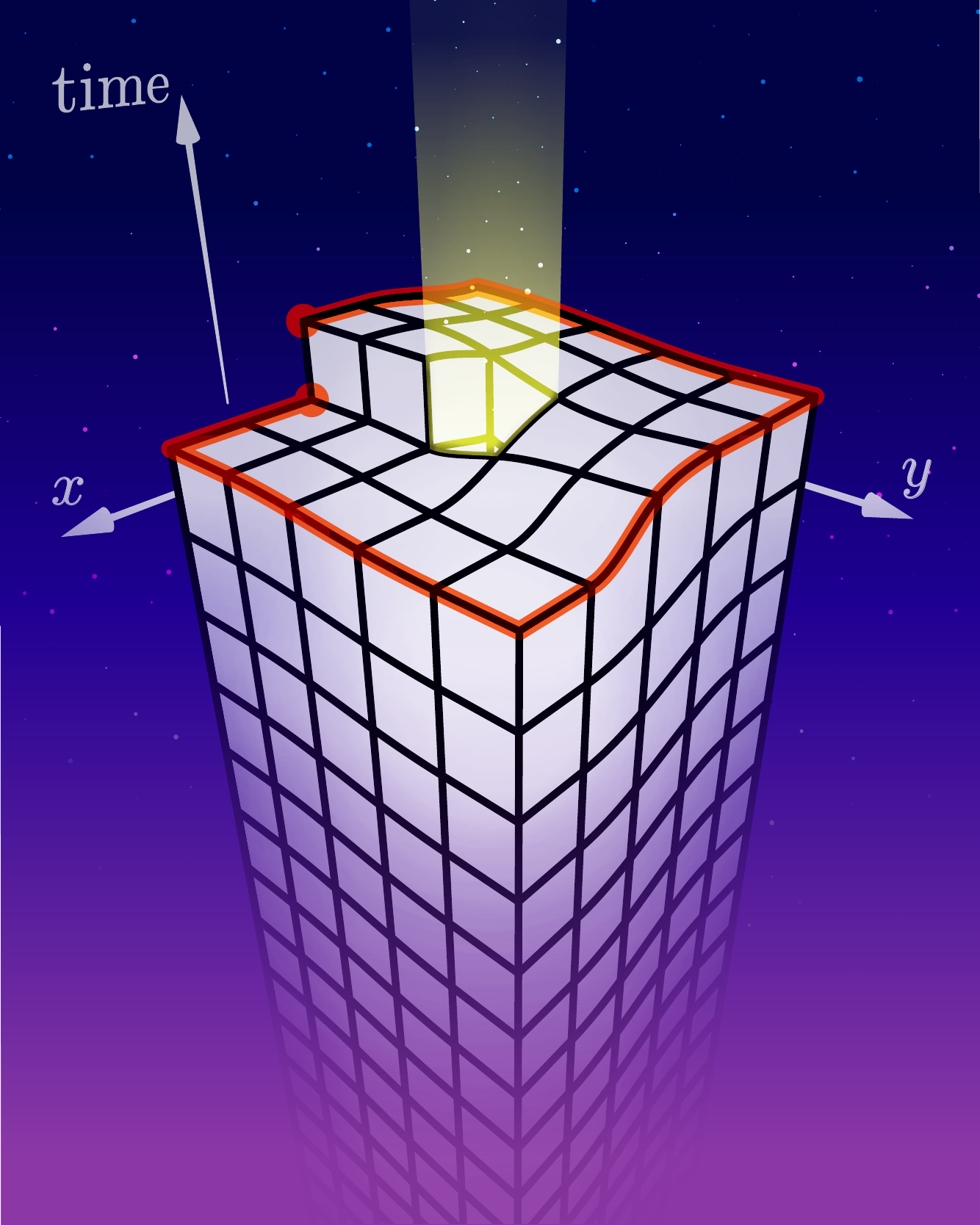}
            \\
            \adjustbox{scale=0.9,valign=c}{
                ``Empty Space''
            }
            &
            &
            \adjustbox{scale=0.9,valign=c}{
                ``Misner String''
            }
        \end{array}
        }
    \end{align*}
    \caption{
        A multivalued diffeomorphism 
        creates a Misner string from empty space,
        whose time monodromy
        is analogous to the Burgers vector of screw dislocations in lattice systems.
    }
    \label{fig:misner}
\end{figure}

Last but not least,
it would be interesting to
see if the 
theoretical framework developed in this paper
can be applied
to
lattice systems
such as crystals
with impurities
\cite{chaikin1995principles,Grozdanov:2018ewh,Pace:2023kyi}. Indeed, as described in
\cite{Fiziev:1995te,Kleinert:1996yi,Kleinert:2008zzb}, the physics of lattice systems
has a formulation that is strikingly
reminiscent of gravity.
In this picture,
lattice disinclinations
are analogous to cosmic strings, as depicted in \Fig{fig:cosmic}.
Furthermore,
the analog of ``translation holonomy'' is manifested by
the Burgers vector,
which measures
the net drift 
in the ``lattice frame''
per round trip about a defect.
Hence, as depicted in \Fig{fig:misner},
we observe that
lattice dislocations 
are analogous to
the Misner string%
---%
a Dirac string of 
time monodromy flux
endable on Taub-NUT charges
\cite{Misner:1963flatter,Misner:1965zz,Bonnor:1969ala,sackfield1971physical,dowker1967gravitational,cho1991magnetic,Alfonsi:2020lub}.
These two types of lattice defects
are described in terms of
\textit{multivalued} coordinate transformations
\cite{Fiziev:1995te,Kleinert:1996yi,Kleinert:2008zzb},
which are
the diffeomorphism analogs of the twisted gauge transformations
that played such a crucial role in our construction of a gravitational one-form symmetry.
It would be interesting
if 
this convergence between
gravity and lattice systems
could cross-pollinate new insights 
across these fields.

\noindent\vphantom{.}

\noindent\textbf{Acknowledgements.}
We are very much grateful to
    Clay C\'ordova,
    Kurt Hinterbichler,
    Ted Jacobson,
    Mrunmay Jagadale,
    Austin Joyce,
    Anton Kapustin,
    Jake McNamara,
    Julio Parra-Martinez,
    Matthew Reece,
    and Shu-Heng Shao
 for many insightful discussions and detailed comments on our draft.
C.C., M.D., J.-H.K., V.N., and N.S. are supported by the Department of Energy (Grant No.~DE-SC0011632) and by the Walter Burke Institute for Theoretical Physics. J.-H.K. is also supported in part by Ilju Academy and Culture Foundation.  I.Z.R. is supported by the Department of Energy (Grant No.~DE-FG02-04ER41338 and FG02-06ER41449).

\appendix

\section{Notations and Conventions}
\label{app:notations}

Our results rely on numerous notational conventions and terminologies.  For completeness, let us briefly summarize our nomenclature here.

\paragraph{Manifold and Index Conventions}%
Throughout our paper,
the symbol $\M$ denotes the full four-dimensional manifold of spacetime.  Within it resides
zero-, one-, two-, and three-dimensional 
submanifolds, which we denote by
$\P$, $\C$, $\S$, and $\V$,
respectively.  We often define $\S$ to be exact, so it is the boundary of a corresponding coboundary manifold $\V$ such that $\S = \partial\V$.  On the other hand, we take $\C$ to be closed, so it has no boundary and thus $\partial \C=0$.
In the Hamiltonian formalism, we perform a $3{\,+\,}1$ decomposition which  generates quantities associated with a spatial three-dimensional submanifold $\M_3$, like the three-dimensional intersection number  $\Intersect_3$.

Spacetime indices are 
$\m,\n,\r,\s,\ldots \in \{1,2,3,4\}$, while spatial indices are 
$i,j,k,l,\ldots \in \{1,2,3\}$.
Adjoint indices of the Lorentz group are $a,b,c,d,\ldots \in 
\{1,2,3,4,5,6\}$ while vector indices are
$A,B,C,D,\cdots \in \{1,2,3,4\}$.
Undotted and dotted spinor indices are 
$\a,\b,\c,\d,\ldots \in \{0,1\}$
and
$\da,\db,\dc,\dd,\ldots \in \{0,1\}$,
respectively.  

All epsilon tensors $\e^{\m\n\r\s}$ and $\e_{\m\n\r\s}$ are pure permutation symbols valued in $\{+1,-1,0\}$, sans dressing by metric-dependent determinant factors, so their indices are never partially raised or lowered.
In particular, we employ the sign and normalization conventions,
\begin{align}
    \e^{1234} = +1
    \,,\quad
    dx^\m \swedge dx^\n \swedge dx^\r \swedge dx^\s
    = \e^{\m\n\r\s}\mem d^4x
    \,,\quad
    \e_{\m\n\r\s}\mem \e^{\m\n\r\s}
    = +4!
    \,.    
\end{align}
On the other hand, $\e^{ABCD}$ and $\e_{ABCD}$ have Lorentz vector indices which are raised and lowered by the flat Euclidean metric
$\delta_{AB} = \diag(+1,+1,+1,+1)$.

\paragraph{Lorentz Algebra Conventions}%
Let us now summarize
our conventions 
for the Lorentz Lie algebra.
First of all,
we employ an anti-Hermitian convention for Lie algebra generators $t_a$
so that $[t_a, t_b] = f^c{}_{ab}\mem t_c$
without the imaginary unit.
%
The adjoint and coadjoint actions on Lie algebra elements and their duals act as
$
    (t_a)^i{}_j\mem X^a
    \mapsto
    (\Omega^{-1})^i{}_k\mem (t_a)^k{}_l\mem \Omega^l{}_j\mem X^a
$ 
and 
$
    Y_a\mem (t^a)^i{}_j
    \mapsto
    Y_a\mem
    (\Omega^{-1})^i{}_k\mem (t^a)^k{}_l\mem \Omega^l{}_j
$, 
respectively,
so that $Y_a\mem X^a \propto Y_a\mem (t^a\hhem t_b)^i{}_i\mem X^b$ is invariant.
In the main text, we have simply denoted these as
$X^a \mapsto (\Ad{\Omega}{X})^a$
and
$Y_a \mapsto (\coAd{\Omega}{Y})_a$.

The generators of the Lorentz Lie algebra $\mathfrak{so}(4)$
are six anti-symmetric matrices
$(t_a)^{AB} = (t_a)^{[AB]}$,
while those of 
the dual Lie algebra 
$\mathfrak{so}(4)^*$
are
$(t^a)_\wrap{AB} = (t^a)_\wrap{[AB]}$.
These are normalized according to
\begin{align}
    \delta^a{}_b
    =
    \minie\mem
    (t^a)_{AB}\mem (t_b)^{AB}
    \transition{and}
    (t_a)^{AB}\mem
    (t^a)_{CD}
    = 2\mem \delta^{[A}{}_C\mem \delta^{B]}{}_D
    \,.
\end{align}
Accordingly,
raised and lowered adjoint indices are
related
to fundamental indices by
\begin{align}
\begin{split}
    X^{AB} = (t_a)^{AB}\mem X^a
    &\,,\quad
    X^a = 
    \minie\mem
    (t^a)_{AB}\mem X^{AB}
    \,,\\
    Y_{AB} = Y_a\mem (t^a)_{AB}
    &\,,\quad
    Y_a = 
    \minie\mem
    Y_{AB}\mem (t_a)^{AB}
    \,.
\end{split}
\end{align}
Note that
the pairing between $\mathfrak{so}(4)$
and $\mathfrak{so}(4)^*$ is given by
\begin{align}
    Y_a\mem X^a
    = 
    \minie\mem
    Y_{AB}\mem X^{AB}
    \,.
\end{align}
Finally, the structure constants are
\begin{align}
    (t_a)^{AB}\mem
    f^a{}_{bc}\mem X_1^b\mem X_2^c
    = X_1^{AC}\mem \delta_{CD}\mem X_2^{DB}
    - X_2^{AC}\mem \delta_{CD}\mem X_1^{DB}
    \,,
\end{align}
when
expressed in the adjoint and fundamental representations.

Interestingly, two metrics can be endowed to the Lie algebra $\mathfrak{so}(4)$.
Firstly, we have
the usual positive-definite Killing form that universally exists in any dimensions:
\begin{align}
    \delta_{ab}
    = 
    \minie\mem
    \delta_{AC}\mem \delta_{BD}\mem
    (t_a)^{AB}\mem (t_b)^{CD}
    \,,\quad
    \delta^{ab}
    = 
    \minie\mem
    (t^a)_{AB}\mem (t^b)_{CD}\mem
    \delta^{AC}\mem \delta^{BD}
    \,,\quad
    \delta^{ac}\mem \delta_{cb} = \delta^a{}_b
    \,.
\end{align}
Indices are raised and lowered with this metric,
which is consistent with
the usual practice of
raising and lowering fundamental indices with the flat Euclidean metric.

On the other hand,
there is also a metric specific to four dimensions:
\begin{align}
\textstyle
    \kap_{ab}
    = \frac{1}{4}\mem \e_{ABCD}\mem
    (t_a)^{AB}\mem (t_b)^{CD}
    \,,\quad
    \ikap^{ab}
    = \frac{1}{4}\mem (t^a)_{AB}\mem (t^b)_{CD}\mem 
    \e^{ABCD}
    \,,\quad
    \ikap^{ac}\hem \kap_{cb} = \delta^a{}_b
    \,.
\end{align}
Evidently, this implements the Hodge dual.
For instance,
\begin{align}
\begin{split}    
    (\kap_{ab}\mem X^b)\mem (t^a)_{AB}
    &= 
    \minie\mem \e_{ABCD}\mem X^{CD}
    = {\star X}_{AB}
    \,,\\
    (t_a)^{AB}\mem
    (\ikap^{ab} Y_b)
    &= 
    \minie\mem
    \e^{ABCD}\mem Y_{CD}
    = {\star Y}^{AB}
    \,.
\end{split}
\end{align}
Note that
this metric naturally appears in the Palatini Lagrangian in \Eq{eq:eGrav.L}.\footnote{
    In fact,
    Pleba\'nski gravity
    without the Immirzi constant
    can be described solely in terms of this split-signature 
    Killing form.
}
It has $(3,3)$ split signature,
which connects to the decomposition
$\mathfrak{so}(4) \cong \mathfrak{su}(2) {\mem\oplus\,} \mathfrak{su}(2)$.
Note also the identities
\begin{align}
    \ikap^{ad} \delta_{dc}\mem
    \ikap^{ce} \delta_{eb}
    = \delta^a{}_b
    \,,\quad
    f^a{}_{bc}\mem \ikap^{be}\hem \ikap^{cf}
    =
    f^a{}_{bc}\mem \delta^{be}\hem \delta^{cf}
    \,,
\end{align}
which hold
because
the Hodge star squares to the identity
in Euclidean signature.



Next, we discuss the spinor representations.
In accordance with the isomorphism
$\mathfrak{so}(4) \cong \mathfrak{su}(2) {\mem\oplus\,} \mathfrak{su}(2)$,
the six Lorentz generators split into
two sets of $\mathfrak{su}(2)$ generators,
which are symmetric $2{\,\times\,}2$ matrices
normalized according to
\begin{align}
\begin{split}
        \delta^a{}_b
        &= (\ttilde^a)_\wrap{\da\db}\mem (\ttilde_b)^{\da\db}
        + (t^a)_{\a\b}\mem (t_b)^{\a\b}
    \,,\\
        (\ttilde_a)^{\da\db}\mem (\ttilde^a)_\wrap{\dc\dd}
        &= \delta^{(\da}{}_\wrap{\dc}\mem \delta^{\db)}{}_\wrap{\dd}
        \,,\quad
        (t_a)^{\a\b}\mem (t^a)_\wrap{\c\d}
        = \delta_\wrap{\c}{}^{(\a}\mem \delta_\wrap{\d}{}^{\b)}
    \,.
\end{split}
\end{align}
In accordance with the above,
adjoint indices are unpacked 
to spinor indices as
\begin{align}
\begin{split}
    \tilde{X}^{\da\db} = (\ttilde_a)^{\da\db} X^a
    \,,\quad
    X^a = 
    (\ttilde^a)_\wrap{\da\db}\mem \tilde{X}^{\da\db}
    &\,,\quad
    \tilde{Y}_\wrap{\da\db} = Y_a\mem (\ttilde^a)_\wrap{\da\db}
    \,,\quad
    Y_a = 
    \tilde{Y}_\wrap{\da\db}\mem (\ttilde_a)^{\da\db}
    \,,\\
    X^{\a\b} = (t_a)^{\a\b} X^a
    \,,\quad
    X^a = 
    (t^a)_\wrap{\a\b}\mem X^{\a\b}
    &\,,\quad
    Y_\wrap{\a\b} = Y_a\mem (t^a)_\wrap{\a\b}
    \,,\quad
    Y_a = 
    Y_\wrap{\a\b}\mem (t_a)^{\a\b}
    \,.
\end{split}
\end{align}
For example, it follows that
\begin{align}
    X^a = (\ttilde^a)_\wrap{\da\db}\mem \tilde{X}^{\da\db} + (t^a)_\wrap{\a\b}\mem X^{\a\b}
    \,,\quad
    Y_a\mem X^a
    = 
    \tilde{Y}_\wrap{\da\db}\mem \tilde{X}^{\da\db}
    +
    Y_{\a\b}\mem X^{\a\b}
    \,.
\end{align}
Also,
the structure constants
are given as
\begin{align}
\begin{split}
    (t_a)^{\da\db}\mem
    f^a{}_{bc}\mem X_1^b\mem X_2^c
    &= (\tilde X_1)^{\da\dc}\mem \te_\wrap{\dc\dd}\mem (\tilde X_2)^{\dd\db}
    - (\tilde X_2)^{\da\dc}\mem \te_\wrap{\dc\dd}\mem (\tilde X_1)^{\dd\db}
    \,,\\
    (t_a)^{\a\b}\mem
    f^a{}_{bc}\mem X_1^b\mem X_2^c
    &= (X_1)^{\a\b}\mem \e_{\b\d}\mem (X_2)^{\d\c}
    - (X_2)^{\a\b}\mem \e_{\b\d}\mem (X_1)^{\d\c}
    \,.
\end{split}
\end{align}
The dotted generators describe self-dual (right-handed) rotations
while the undotted generators describe anti-self-dual (left-handed) rotations:
\begin{align}
\begin{split}
    \kap_{ab}\mem (\ttilde^b)_\wrap{\da\db}
    = + \te_\wrap{\da\dc}\hem \te_\wrap{\db\dd}\mem
        (\ttilde_a)^{\dc\dd}
    \,,\quad
    \kap_{ab}\mem (t^b)_\wrap{\a\b}
    = - \e_{\a\c}\hem \e_{\b\d}\mem
        (t_a)^{\c\d}
    \,.
\end{split}
\end{align}
In fact,
self-dual and anti-self-dual projectors arise as
\begin{align}
\begin{split}
    \label{eq:sig-completeness}
    (\ttilde^a)_\wrap{\da\db}\mem (\ttilde^b)_\wrap{\dc\dd}\mem
    \te^{\da\dc}\mem \te^{\db\dd}
    = \minie\mem (\hhem{
        \delta^{ab} + \ikap^{ab}
    }\hem)
    \,,\quad
    (t^a)_\wrap{\a\b}\mem (t^b)_\wrap{\c\d}\mem
    \e^{\a\c}\mem \e^{\b\d}
    = \minie\mem (\hhem{
        \delta^{ab} - \ikap^{ab}
    }\hem)
    \,.
\end{split}
\end{align}
Having stated our conventions
abstractly, let us be slightly more concrete and note that the self-dual and anti-self-dual generators are
\begin{align}
    (\ttilde^{AB})_\wrap{\da\db}
    = 
        \minie\mem
        (\te\mem \ts^{[A} \s^{B]})_\wrap{\da\db}
    \,,\quad
    (t^{AB})_{\a\b}
    = 
        \minie\mem
        (\s^{[A} \ts^{B]} \e)_{\a\b}
    \,,
\end{align}
where an explicit representation of the Euclidean sigma matrices is given as
\begin{align}
\begin{split}
    v_A\mem (\s^A)_{\a\da}
    =
    \bigg(
    {\renewcommand{\arraycolsep}{0.35em}
    \begin{array}{cc}
        iv_4 + v_3 
        &
        v_1 -iv_2
        \\
        v_1 +iv_2
        &
        iv_4 - v_3
    \end{array}}
    \bigg)
    \,,\quad
    v_A\mem (\ts^A)^{\da\a}
    =
    \bigg(
    {\renewcommand{\arraycolsep}{0.35em}
    \begin{array}{cc}
        iv_4 - v_3 
        &
        -v_1 +iv_2
        \\
        -v_1 -iv_2
        &
        iv_4 + v_3
    \end{array}}
    \bigg)
    \,,
\end{split}
\end{align}
with the convention
$\te^{01} = \e^{01} = \te_{10} = \e_{10} = +1$
for the epsilon tensors.

Finally,
we end with a demonstration of index conversions
in the context of tetradic Palatini gravity.
Let the spinor indices be raised and lowered as
\begin{align}
    \psi^\a = \e^{\a\b} \psi_\b
    \,,\quad
    \psi_\a = \e_{\a\b}\mem \psi^\b
    \,,\quad
    \tilde{\psi}^\da = \te^{\da\db} \tilde{\psi}_\db
    \,,\quad
    \tilde{\psi}_\da = \te_\wrap{\da\db}\mem \tilde{\psi}^\db
    \,.
\end{align}
The spin connection splits into
self-dual and anti-self-dual parts as
\begin{align}
    \omega^a
    = (\ttilde^a)_\wrap{\da\db}\mem \tilde{\omega}^{\da\db} + (t^a)_\wrap{\a\b}\mem \omega^{\a\b}
    \,.
\end{align}
The self-dual and anti-self-dual parts
of the field strength are given by
\begin{align}
    \tilde{R}^\da{}_\db
    = d\tilde{\omega}^\da{}_\db - \tilde{\omega}^\da{}_\dc \swedge \tilde{\omega}^\dc{}_\db
    \,,\quad
    R_\a{}^\b
    = d\omega_\a{}^\b + \omega_\a{}^\c \swedge \omega_\c{}^\b
    \,.
\end{align}
Meanwhile,
it is convenient to convert the vector index of the tetrad to spinor indices
with 
a customary
factor of $-1/2$,
which originates from the $-2$ of
$(\ts_A)^{\da\a}\mem (\s^A)_\wrap{\b\db} = -2\mem \delta_\wrap{\b}{}^\a\mem \delta^\da{}_\wrap{\db}$:
\begin{align}
    e^{\da\a} = -\frac{1}{2}\mem (\ts_A)^{\da\a}\mem e^A
    \,.
\end{align}
Then the spinor components of the
Pleba\'nski two-form
defined in \Eq{eq:pleb-def}
are given by
\begin{align}
\begin{split}
    \tilde{B}^{\da\db}
    = \e_\wrap{\a\b}\mem e^{\da\a} \swedge e^{\db\b}
    \,,\quad
    B^{\a\b}
    = -\te_\wrap{\da\db}\mem e^{\da\a} \swedge e^{\db\b}
    \,.
\end{split}
\end{align}
In turn, the Lagrangian four-form
in \Eq{eq:eGrav.L}
boils down to
\begin{align}
    \label{eq:eGrav.L.spinor()}
    -
    \frac{1}{g^2}\mem 
    \Big(\mem{
        e^{\da\a} \hnem\swedge e_\wrap{\a\db}  \swedge \tilde{R}^\db{}_\da
        +
        e_\wrap{\b\da} \hem\swedge e^{\da\a} \swedge R_\a{}^\b
    }\mem\Big)
    + \frac{\Lambda}{3g^2}\mem
    e_\wrap{\a\da} \swedge e^{\da\b} \hnem\swedge e_\wrap{\b\db} \swedge e^{\db\a}
    \,.
\end{align}
The above expressions
are consistent with the chiral symmetry operators in \Eq{eq:Ublocks}.


\section{Geometry of Linking}
\label{app:linking}

In this appendix, we expand on the technical details of the
Dirac delta form and topological numbers in general spacetime dimension.
Here we will indicate the dimensionality of each manifold  with a subscript.
All manifolds will be assumed to be orientable.

\paragraph{Definitions}%
%
Consider a $p$-dimensional submanifold $\N_p$
in a $d$-dimensional manifold
$\mathcal{M}_d$.
We define a differential form analog of Dirac's delta distribution as
\begin{align}
    \label{eq:delta-def}
    \int_{\N_p} \alpha^{(p)}
    \,\,\,=\,\, 
        \int_{\mathcal{M}_d} 
            \alpha^{(p)}\nem
            \wedge
            \delta(\N_p)
    \,,
\end{align}
where $\alpha^{(p)}$ is an arbitrary smooth $p$-form in $\mathcal{M}_d$.
Evidently,
$\delta(\N_p)$ is a differential $(d{\,-\,}p)$-form
distribution
that peaks on the submanifold $\N_p$
while vanishing elsewhere.

More explicitly, 
suppose $\N_p$ is
parameterized by coordinates $\s^1,\s^2,\cdots,\s^p$ as
$x^\m = X^\m\hnem(\s)$,
where $x^\m$ denote coordinates of $\mathcal{M}_d$.
Then $\delta(\N_p)$ is given by
\begin{align}
\begin{split}
    \label{eq:delta-param}
    \frac{1}{(d{\,-\,}p)!}
    \bigg[\,{
    \int d^p\s\,\,
        \delta^{(d)}\hnem(x{\,-\,}X(\s))\,
        \frac{\partial X^{\l_1}}{\partial \s^1}
        \cdots
        \frac{\partial X^{\l_p}}{\partial \s^p}
    }\,\bigg]\mem
    \e_{\l_1\cdots\l_p\m_1\cdots\m_{d-p}}\mem
    dx^{\m_1} \swedge \cdots \swedge dx^{\m_{d-p}}
    \,,
\end{split}
\end{align}
where 
$\delta^{(d)}\hnem(x{\,-\,}X(\s))$ is the ordinary Dirac delta function.
This is higher-form generalization of Dirac delta function will be referred to as a ``\textit{Dirac delta form}.''
The ordinary Dirac delta corresponds to $p=0$.
It is also interesting to note that
$\delta(\N_d)$
for a top-dimensional submanifold $\N_d$
is its characteristic function,
e.g.,
$\delta(\M_d) = 1$.

\begin{figure}[t]
    \centering
    \includegraphics[scale=1.5,valign=c]{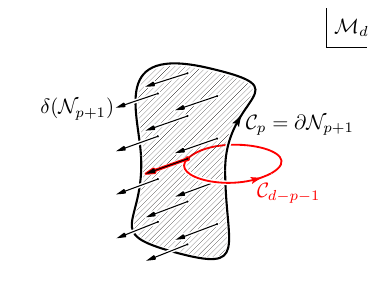}
    \caption{
        Linking between two submanifolds
        in $d$ dimensions.
    }
    \label{fig:deltaform}
\end{figure}

In terms of Dirac delta forms,
the \textit{intersection number} between
two orientable submanifolds
$\N_p$ and $\N_{d-p}$
in $\M_d$
can be defined in a coordinate-free fashion:
\begin{align}
    \label{eq:int-def}
    \Intersect(\N_p,\N_{d-p})
    \,= \int_{\M_d} \delta(\N_p) \swedge \delta(\N_{d-p})
    \,.
\end{align}
Note that the integrand 
$\delta(\N_p) \swedge \delta(\N_{d-p})$
is a top form that localizes at the intersection points.
In fact, it could be argued that
$\delta(\N_p) \swedge \delta(\N_{d-p})
= \delta(\N_p \cap \N_{d-p})$.
It is instructive to check if
\Eq{eq:int-def} defines the intersection number
in an expected way
with specific values of $p$ and $d$,
such as $p=0$, or $d=3$.
Also,
it is easy to see that
\begin{align}
    \label{eq:int-swap}
    \Intersect(\N_p,\N_{d-p})
    =
    (-1)^{p(d-p)}\, \Intersect(\N_{d-p},\N_p)
    \,.
\end{align}
Given \Eq{eq:int-def},
we can easily define the \textit{linking number}
between two \textit{closed} orientable submanifolds
$\C_p$ and $\C_{d-p-1}$
provided that 
one of them is exact.
Without loss of generality,
suppose
$\mathcal{C}_{p} = \partial\Dee_{p+1}$.
Then we define
\begin{align}
    \label{eq:lk*}
    \LLink(\partial\Dee_{p+1},\C_{d-p-1})
    \,=\, \Intersect(\Dee_{p+1},\C_{d-p-1})
    \,= \int_{\C_{d-p-1}} \delta(\Dee_{p+1})
    \,,
\end{align}
which counts how many times 
$\C_{d-p}$ penetrates the coboundary $\Dee_{p+1}$
as depicted in \Fig{fig:deltaform}.

Note that the above definition necessarily puts the two arguments of $\LLink$ in an unequal footing:
the first argument must be exact.
However,
as practiced in the main text,
it can be convenient to use a more handy notation
for the linking number
such that
the exact argument can be placed in any slot.
With hindsight, we define such a notation ``$\Link$''
to satisfy the following relations:
\begin{align}
\begin{split}
    \label{eq:lkdef}
    \Link(\partial\Dee_{d-p} , \C_p)
    &=
    \LLink(\partial\Dee_{d-p} , \C_p)
    \,,\\
    \Link(\C_{d-p-1} , \C_p)
    &= (-1)^{dp+1}\mem \Link(\C_p , \C_{d-p-1})
    \,.
\end{split}
\end{align}
The first equation states that 
$\Link$ and $\LLink$
give the same number when the first argument is exact.
The second equation 
allows the arguments of $\Link$ to be freely rearranged,
given the implicit assumption that
one of the arguments is exact.
The factor $(-1)^{dp+1}$
is necessitated from
consistency with \Eq{eq:link-duality}.
Note also that the following holds:
\begin{align}
    \label{eq:drop2nd}
    \Link(\C_{d-p-1} , \partial\Dee_{p+1})
    &= 
        (-1)^{d-p}\mem \Intersect(\C_{d-p-1} , \Dee_{p+1})
    \,.
\end{align}

\paragraph{Identities}%
Using either
the coordinate-free/axiomatic definition 
of the Dirac delta form
in \Eq{eq:delta-def}
or the explicit definition in \Eq{eq:delta-param},
one can derive various identities
of Dirac delta forms
and also
intersection and linking numbers.
Some simple ones are
\begin{align}
    \label{eq:bd}
    d\mem \delta(\N_p)
    &=
        (-1)^p\, \delta(\partial\N_p)
    \,,\\
    \label{eq:link-duality}
    \Link(\partial\Dee_{p+1},\partial\Dee_{d-p})
    &=
        (-1)^{dp+1}\mem
        \Link(\partial\Dee_{d-p},\partial\Dee_{p+1})
    \,.
\end{align}
The first identity
encodes Stokes theorem,
while the second identity
describes a duality in the linking number computation.

A few more identities are found
in the context of Hamiltonian formalism.
Suppose the spacetime is a product manifold
$\M_d = \MX_{d-1} {\mem\times\,} \mathbb{R}$
with coordinates 
$x^\m = (x^i, t)$,
where the last\footnote{
    This matches the convention chosen in \Eq{eq:delta-def}
    where the Dirac delta form is appended from the right.
} coordinate $x^{\m=d} {\,=\,} t$
labels hypersurfaces foliating $\M_d$.
Then we
consider
submanifolds $\CX_p$ and $\NX_{d-p-1}$
of $\MX_{d-1}$
that can possibly intersect,
with the former being closed:
$\partial\CX_p = 0$.
The corresponding relevant objects in spacetime
are 
$\C_p$ and $\HSig_{d-p}$, defined as
the following.
First,
we consider
an embedding of the closed submanifold $\CX_p$
in $\M_d$:
\begin{align}
    \C_p = \CX_p {\mem\times\mem} \{ t_0 \}
    \,.
\end{align}
Second, we consider
a timelike
$(d{\,-\,}p)$-dimensional submanifold $\HSig_{d-p}$
of $\M_d$
defined as the following,
where $\mathcal{I} = [t_-,t_+]$ is an interval in $\mathbb{R}$
such that $t_- < t_0 < t_+$:
\begin{align}
    \label{eq:Sigreduce1}
    (-1)^{d-p}\mem
    \HSig_{d-p} 
    = 
    \NX_{d-p-1} {\mem\times\mem} \mathcal{I}
    \,.
\end{align}
The customary orientation of $\HSig_{d-p}$
by a sign factor $(-1)^{d-p}$
stipulates that its boundary is given in the form
\begin{align}
    \label{eq:Sigreduce2}
    \partial\HSig_{d-p} 
    \,=\,
        \Big(\hem{
            -\NX_{d-p-1} {\mem\times\mem} \{ t_+ \}
        }\hem\Big)
    {\mem\,\cup\,\mem}
        \Big(\,{
            \NX_{d-p-1} {\mem\times\mem} \{ t_- \}
        }\hem\Big)
    {\mem\,\cup\,\mem}
        \mem
        \Gamma
    \,,
\end{align}
for a timelike submanifold $\Gamma$
in $\M_d$.
Specifically,
the consistency between
\Eqs{eq:Sigreduce1}{eq:Sigreduce2}
can be established
by considering
the Stokes theorem relating $\HSig_{d-p}$ and $\partial\HSig_{d-p}$,
where a factor of $(-1)^{d-p-1}$
arising from rearranging the order of the time differential $dt$
combines with the minus sign in \Eq{eq:Sigreduce2}
to give $(-1)^{d-p}$.

With these definitions, 
it eventually follows that
\begin{align}
    \label{eq:intred}
        (-1)^{d-p}\mem
        \Link(\partial\HSig_{d-p},\C_p)
        \mem=\mem
        \Intersect(\NX_{d-p-1} {\mem\times\mem} \mathcal{I},\C_p)
        \mem=\mem
        (-1)^p\mem
        \Intersect_{d-1}(\NX_{d-p-1},\CX_p)
    \,,
\end{align}
where 
$\Intersect_{d-1}(\NX_{d-p-1},\CX_p)$
is the intersection number between
$\NX_{d-p-1}$ and $\CX_p$
in $\MX_{d-1}$:
\begin{align}
    \Intersect_{d-1}(\NX_{d-p-1},\CX_p)
    \,\,=\mem \int_{\MX_{d-1}}
        \delta_{d-1}(\NX_{d-p-1})
        \swedge
        \delta_{d-1}(\CX_p)
    \,.
\end{align}
To derive \Eq{eq:intred},
one may 
work in component terms with
\Eq{eq:delta-param}.
One will find that
the factor of $(-1)^p$
arises from an index rearrangement,
\begin{align}
    \label{eq:0rearrange}
    \e_{i_1\cdots i_{d-p-1} d\mem j_1\cdots j_p}
    = (-1)^p\mem \e_{i_1\cdots i_{d-p-1} j_1\cdots j_p d}
    = (-1)^p\mem \e_{i_1\cdots i_{d-p-1} j_1\cdots j_p}
    \,,
\end{align}
where we work with the convention such that $\e_{12\cdots d} = +1$.

\paragraph{Example: Abelian \textit{p}-Form Symmetry in \textit{d} Dimensions}%
Having derived various sign factors,
we can
describe
a universal sign convention
for higher-form symmetries
of general rank $p$ 
in Euclidean signature spacetimes of
general $d$ dimensions.
Let us take
an abelian $BF$ theory
as a prototypical model,\footnote{
    For example,
    it is an amusing check to
    consider the case of
    $d{\,=\,}1$, $p{\,=\,}0$
    with a Lagrangian one-form $P\mem dX - H(P)\mem dt$,
    which describes a point particle.
    In particular,
    it readily follows that
    operators
    $U_\e(
    [t_1,t_2]
    ) =
    e^{-\e\hem P(t_2)}\hem e^{\e\hem P(t_1)} 
    $
    and $Q(t) = P(t)$
    implement
    translations
    of $X$, so for example
    $e^{-\e P}\hhem X\mem e^{\e P} = X + \e$.
}
whose action reads
\begin{align}
    \label{eq:aBF}
    S
    \,\,= \int_{\M_d}
        B^{(d-p-1)}\nem \swedge F^{(p+1)}
        + f(B^{(d-p-1)})
    \,,
\end{align}
where $F^{(p+1)} = dA^{(p)}$.
Here the superscripts
denote the ranks,
while $f(B^{(d-p-1)})$ is a $d$-form functional of
$B^{(d-p-1)}$.

First, let us describe the $p$-form symmetry in the covariant formalism.
For simplicity,
consider the Wilson loop for an exact support:
\begin{align}
    W_q(\partial\Dee_{p+1})
    = 
    \exp
    \bigg(\hem{
        q
        \int_{\Dee_{p+1}} F^{(p+1)}
    }\bigg)
    \,.
\end{align}
We want this to be transformed as
\begin{align}
\begin{split}
    W_q(\partial\Dee_{p+1})
        \mem\,\,\mapsto\,\,\,
        \exp\Big(\mem{
            q\ve\mem
            \Intersect(\Sh_{d-p-1},\Dee_{p+1})
        }\hem\Big)
        \,
        W_q(\partial\Dee_{p+1})
    \,,
\end{split}
\end{align}
when the symmetry operator is inserted along
an exact submanifold
$\Sh_{d-p-1} = \partial\HSig_{d-p}$.
Clearly,
this
can be \textit{undone} by a field redefinition
that shifts the field strength as
$
    F^{(p+1)}
        \mapsto
    F^{(p+1)}
    - \ve\, \delta(\Sh_{d-p-1})
$,%
\footnote{
    Concretely, 
    from \Eq{eq:bd}
    it can be realized as
    $
        A^{(p)}
            \mapsto
        A^{(p)} - (-1)^{d-p}\mem \e\, \delta(\HSig_{d-p})
    $, $
        B^{(d-p-1)}
            \mapsto
        B^{(d-p-1)}
    $.
}
which, in turn,
transforms the action in \Eq{eq:aBF} as
\begin{align}
\begin{split}    
    -S
    \,\,\,\mapsto\,\,
    -S
    \,+\,
    \ve \int_{\M_d}
        \kern-0.1em
        B^{(d-p-1)}\nem \swedge \delta(\Sh_{d-p-1})
    \,,
\end{split}
\end{align}
from which the symmetry operator is 
identified:
\begin{align}
    \label{eq:apform.U}
    U_\ve(\Sh_{d-p-1})
    = \exp\bigg(\mem{
        \ve
        \int_{\Sh_{d-p-1}}\kern-0.6em
            B^{(d-p-1)}
    }\mem\bigg)
    \,.
\end{align}
Namely,
the symmetry operator
is derived
as the term that
is generated from the action term in the path integral after a field transformation that
pushes the
symmetry charge
back into the defect operator.
Crucially, the $BF$ structure
leverages between the both sides of the Ward identity,
which reads
\begin{align}
    \label{eq:apWard-int}
    \expval{
        U_\ve(\Sh_{d-p-1})\mem W_q(\partial\Dee_{p+1})
    }
    =
    \exp\Big(\mem{
        q\ve\mem
        \Intersect(\Sh_{d-p-1},\Dee_{p+1})
    }\hem\Big)
    \,
    \expval{
        W_q(\partial\Dee_{p+1})
    }
    \,.
\end{align}
Finally,
denoting $\C_p = \partial\Dee_{p+1}$,
we can state the Ward identity in terms of a linking number:
\begin{align}
    \label{eq:apWard}
    \expval{
        U_\ve(\Sh_{d-p-1})\mem W_q(\C_p)
    }
    =
    \exp\Big(\mem{
        (-1)^{d-p}\mem
        q\ve\mem
        \Link(\Sh_{d-p-1},\C_p)
    }\hem\Big)
    \,
    \expval{
        W_q(\C_p)
    }
    \,,
\end{align}
which follows through 
using
\Eq{eq:drop2nd}
(or 
Eqs.\,\eqref{eq:int-swap},
\eqref{eq:lk*},
and \eqref{eq:link-duality}
with $\LLink$).
 
Note that
the Ward identities in the main text
are reproduced when
one takes
$d{\,=\,}4$ and $p{\,=\,}1$
and makes the following identifications:
\begin{align}
    \C_p
        \,\leftrightarrow\,
    \C
    \,,\quad
    \C_{d-p-1}
        \,\leftrightarrow\,
    \S
    \,,\quad
    \N_{d-p}
        \,\leftrightarrow\,
    \V
    \,.
\end{align}

Next,
we can also double check
from the angle of Hamiltonian formalism.
\Eq{eq:Sigreduce2}
as a convetion for
pancaking the coboundary $\HSig_{d-p}$
of the symmetry operator
down to a submanifold $\NX_{d-p-1}$ in $\MX_{d-1}$
stipulates that
the Ward identity
in \Eq{eq:apWard}
boils down to a conjugation of the defect operator as
\begin{align}
    \label{eq:apWard-pancake}
    e^{-\ve Q(\NX_{d-p-1})} 
        \,
    W_q(\CX_p)
        \,
    e^{\ve Q(\NX_{d-p-1})}
    =
    \exp\Big(\mem{
        (-1)^{p}\mem
        q\ve\mem
        \Intersect_{d-1}(\NX_{d-p-1},\CX_p)
    }\hem\Big)
    \,
        W_q(\C_p)
    \,,
\end{align}
provided that the charge operator is defined as
\begin{align}
    \label{eq:ap-Q}
        Q(\NX_{d-p-1})
    = 
        \int_{\NX_{d-p-1}}\kern-0.6em
            B^{(d-p-1)}
\end{align}
as an object that lives in the $d{\,-\,}1$ dimensions.
Note that the identity in
\Eq{eq:intred}
has been used to deduce \Eq{eq:apWard-pancake}
from \Eq{eq:apWard}.
It is straightforward to verify \Eq{eq:apWard-pancake}
from the canonical commutation relations:
\begin{align}
\begin{split}
    [ 
        A_{i_1\cdots i_p}
        ,
        B_{k_1\cdots k_{d-p-1}}
    ]
    \,\e^{k_1\cdots k_{d-p-1} j_1\cdots j_p}
    \mem&=\mem
    (d{\,-\,}p{\,-\,}1)!\mem p!\mem
    (-1)^p\,
    \delta^{j_1}{}_\wrap{[i_1}
    {\cdots\,}
    \delta^{j_p}{}_\wrap{i_p]}
    \,,\\
    \iq
    [
        A_{i_1\cdots i_p}
        ,
        Q(\NX_{d-p-1})
    ]
    \mem&=\mem
    (-1)^p\mem
    (\delta_{d-1}(\NX_{d-p-1}))_{i_1\cdots i_p}
    \,,
\end{split}
\end{align}
where the factor $(-1)^p$
precisely arises from the same index rearrangement
as in \Eq{eq:0rearrange},
provided
the conventions $\e_{12\cdots d} = +1$
and $t = x^d$.

The above discussion
corresponds to
the calculation carried out in \Secs{sec:CanYM.ward}{sec:CanGrav.ward},
so one can check consistency by taking
$d{\,=\,}4$ and $p{\,=\,}1$.
Yet, note that the spatial surface ``$\S_0$'' there 
corresponds to $\NX_{d-p-1}$ here
with a flip of orientation:
\begin{align}
    \CX_p
        \,\leftrightarrow\,
    \C
    \,,\quad
    \NX_{d-p-1}
        \,\leftrightarrow\,
    -\S_0
    \,.
\end{align}
This single flip is because
the main text attempts
to avoid
any possible distractions
from
convention-dependent
minus signs.
%
%
%



\paragraph{Proof of Equivalence between \Eqs{eq:lambda-disc1}{eq:lambda-disc2}}%
Before ending, we should show that
\Eqs{eq:lambda-disc1}{eq:lambda-disc2},
used in the main text,
are equivalent statements.
In $d$-dimensional spacetime, the claim reads
\begin{align}
\begin{split}
    \label{eq:claim}
    &
        \lim_{
            \P_{\pm}\to\, \P
        }\hem
            \Omega(\P_{\hnem+})
            \mem
            \Omega^{-1}\hnem(\P_{\hnem-})
        = \cen
        \in Z(G)
    \\
    &\iff\quad
    \exists\, \l^a
    \transition{s.t.}
        (\Omega^{-1}dd\Omega)^a
        = 
        \lambda^a\mem \delta({\partial\Sig_{d-1}})
    \transition{and}
        e^{2\pi\l} = \smash{\cen^{(-1)^d}}
    \,,
\end{split}
\end{align}
where 
$\P_{\hnem+}$ and $\P_{\hnem-}$
are points infinitesimally deviating from a point on $\Sig_{d-1}$ from above and below.
The notion of above and below
is well-defined,
as $\Sig_{d-1}$ is
oriented and codimension one.
As shown below,
the identities stated in 
\Eqs{eq:int-swap}{eq:link-duality}
play a role
in the proof.

First of all,
the fact that
$\Omega$ as a multivalued function is everywhere smooth off the surface $\partial\Sig_{d-1}$
implies
that
$dd\Omega$ localizes on $\partial\Sig_{d-1}$
and vice versa,
which is in turn
equivalent to the existence of 
an algebra-valued zero-form $\lambda^a$ such that
$(\Omega^{-1}dd\Omega)^a
= 2\pi\lambda^a\mem \delta({\partial\Sig_{d-1}})$.

\begin{figure}[t]
    \centering
    \includegraphics[scale=1.185]{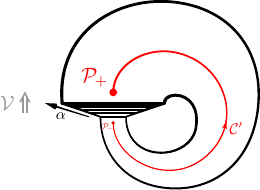}
    \caption{
        The center periodicity condition on the multivalued group parameter $\Omega$,
        in terms of the discontinuity
        $
        \lim_{
            \P_{\pm}\to\, \P
        }\hem
            (\hhhem\mem{
                \Omega(\P_{\hnem+})
                \mem
                \Omega^{-1}\hnem(\P_{\hnem-})
            }\hem)
        = \cen
        $
        across a branch cut $\Sig$.
    }
    \label{fig:stairs}
\end{figure}

Thus, it remains to verify that
the 
center
periodicity
condition
on $\Omega$
is equivalent to
$e^{2\pi\l} = \cen^{(-1)^d}$.
It suffices to work with
an exact one-dimensional loop $\C_1 = \partial\Dee_2$
in the vicinity of $\partial\Sig_{d-1}$.
Suppose 
$\Intersect(\Sig_{d-1},\partial\Dee_2) = -1$.
Then the periodicity condition implies
\begin{align}
\begin{split}
    \label{eq:claimproof-1}
    \lim_{
        \C'_1 \to\, \C_1
    }\hem
    \Pexp
    \bigg(\hem{
        \oint_{\C'_1} \Omega^{-1}d\Omega
    }\bigg)
    = \lim_{
        \C'_1 \to\, \C_1
    }\hem
        \Omega^{-1}\hnem(\P_{\hnem-})
        \mem
        \Omega(\P_{\hnem+})
    = \alpha
    \,,
\end{split}
\end{align}
where $\C'_1$ is the infinitesimal opening of $\C_1$ described before.
Meanwhile, the left-hand side can be computed as a surface-ordered integral, thanks to the nonabelian Stokes theorem:
\begin{align}
    \label{eq:claimproof-2}
    \Pexp
    \bigg(\hem{
        \oint_{\partial\Dee_2} \Omega^{-1}d\Omega
    }\bigg)
    =
    \Pexp
    \bigg(\hem{
        \int_{\Dee_2} \Omega^{-1}dd\Omega
    }\bigg)
    &=
        \exp
        \bigg(\hem{
            \int_{\M_d}
                \Omega^{-1}dd\Omega
                \swedge
                \delta(\Dee_2)
        }\bigg)
    \,.
\end{align}
Note that
the last equality has dropped the surface ordering because the integrand localizes at a single point.
Plugging in
$(\Omega^{-1}dd\Omega)^a
= 
2\pi\lambda^a\mem \delta({\partial\Sig})$,
the right-hand side
boils down to
$\exp\,(2\pi\lambda\mem \Intersect(\partial\Sig_{d-1},\Dee_2))$,
which equals
$\exp\,((-1)^d\mem 2\pi\l)$
given 
$\Intersect(\Sig_{d-1},\partial\Dee_2) = -1$
by \Eqs{eq:int-swap}{eq:link-duality}.
Therefore, 
demanding that 
the outcomes of
\Eqs{eq:claimproof-1}{eq:claimproof-2}
are the same,
we find
$\alpha = \exp\,((-1)^d\mem 2\pi\l)$,
i.e.,
$e^{2\pi\l} = \cen^{(-1)^d}$.
Lastly, 
recalling the fact that $\cen {\,\in\,} Z(G)$,
it can be seen that
supposing 
generic curves with
arbitrary linking numbers
does not impose a further condition on $\l^a$.
This concludes the proof.

To make it certain that 
all the signs have worked properly,
it is instructive to double check
by considering the abelian version of the statement:
\begin{align}
    \label{eq:claim-a}
        \lim_{
            \P_{\pm}\to\, \P
        }\hem
            \chi(\P_{\hnem+})
            {\,-\,}
            \chi(\P_{\hnem-})
        = (-1)^d\mem 2\pi\lambda
    \quad\iff\quad
        dd\chi
        = 2\pi\lambda\, \delta({\partial\Sig_{d-1}})
    \,.
\end{align}
Here $\lambda$ is a constant.
The periodicity condition in the left-hand side is equivalent to
\begin{align}
    d\chi + (-1)^d\mem 2\pi\lambda\, \delta(\Sig_{d-1}) = df
    \,,
\end{align}
for some
single-valued zero-form
$f$;
i.e., 
$d\chi$
is ``gauge equivalent'' to $(-1)^{d-1}\mem 2\pi\lambda\, \delta(\Sig_{d-1})$.
Thus
it follows that
$dd\chi = (-1)^{d-1}\mem 2\pi\lambda\, d\mem \delta(\Sig_{d-1})$,
which equals $2\pi\lambda\, \delta(\partial\Sig_{d-1})$
by \Eq{eq:bd}.

Note that
the points $\P_{\hnem+}$ and $\P_{\hnem-}$ in \Fig{fig:W'}
are correctly deviating from above and below
as stated in \Eq{eq:claim}
($-\Link(\C,\S) = \Link(\S,\C) = \Intersect(\V,\C) = -1$),
but the 
three-dimensional
visualization 
seemingly
appears the opposite
due to the
relative orientation with the
hidden fourth dimension.

\let\c\oldc
\let\i\oldi
\bibliography{references.bib}

\end{document}